\PassOptionsToPackage{table,hyperref,dvipsnames}{xcolor}

\documentclass[examplefnt,biber]{nowfnt}

\usepackage[utf8]{inputenc}
\usepackage{adjustbox}
\usepackage{booktabs}
\usepackage{subcaption}
\usepackage{tikz}
\usepackage{multirow}
\usepackage{tcolorbox}
\usepackage{pifont}
\usepackage[ruled,vlined]{algorithm2e}
\usepackage{bbm}
\usepackage{pgfplots}
\usepackage{soul}
\usepackage{amsmath}
\usepackage{amssymb}
\usepackage{xspace}
\usepackage{multicol}
\usepackage{selectp}
\usepackage{tabularray}
\usepackage{subcaption}
\usepackage{rotating}
\usepackage{makecell}
\usepackage{cleveref}
\usepackage{thmtools}

\declaretheorem[name=RecSys re-interpretation]{reinterpretation}

\usetikzlibrary{positioning,fit,shapes.geometric,backgrounds,calc}
\pgfplotsset{compat=1.18}

\newcommand{\xmark}{\ding{55}}%
\newcommand{\cmark}{\ding{51}}

\definecolor{forestgreen(web)}{rgb}{0.13, 0.55, 0.13}
\definecolor{lightgray}{rgb}{0.83, 0.83, 0.83}

\makeatletter
\renewcommand\nowfnt@bookauthortitlepage{%
  \newpage
  \thispagestyle{\nowfnt@titlepagestyle}%
  \begingroup
  \raggedleft
  {%
   \renewcommand{\baselinestretch}{1.05}%
   \nowfnt@font@title
   \if@nowfntjournal
     \textcolor{\nowfnt@journalcolor}{\nowfnt@font@coverjournalname\nowfnt@journaltitle}\par
   \fi
   \nowfnt@articletitle
   \ifdefempty{\now@subtitle}{}{\par\bigskip\nowfnt@font@subtitle\now@subtitle}%
   \par
  }%
  \vspace{2.5ex}%
  \noindent\rule{\textwidth}{2pt}%
  \par\vspace{1ex}%
  \if@nowfntwrapper
   \begingroup\fontsize{13}{15}\sffamily\bfseries
   \nowfnt@editedby\par\vspace{1ex}\endgroup
  \fi
    \ifbool{authortwocolumn}%
   {\nowfnt@printauthorsbooktitletwocolumn}%
   {\nowfnt@printauthorsbooktitle}%
  \par\vfill
  \par\endgroup
}
\makeatother

\articledatabox{\mbox{}}

\title{On the Impact of Graph Neural Networks in Recommender Systems}
\subtitle{A Topological Perspective}
\maintitleauthorlist{
Daniele Malitesta \\
LUISS Guido Carli University\\
dmalitesta@luiss.it
\and
Alberto Carlo Maria Mancino\\
Politecnico di Bari\\
alberto.mancino@poliba.it
\and
Claudio Pomo \\
Politecnico di Bari \\
claudio.pomo@poliba.it
\and
Vito Walter Anelli\\
Politecnico di Bari\\
vitowalter.anelli@poliba.it
\and
Alejandro Bellogín\\
Universidad Autónoma de Madrid\\
alejandro.bellogin@uam.es
\and
Tommaso Di Noia\\
Politecnico di Bari\\
tommaso.dinoia@poliba.it
}

\issuesetup
{%
 copyrightowner={D.~Malitesta, A. C. M.~Mancino, C.~Pomo, V. W.~Anelli, A.~Bellogín, and T.~Di Noia},
 volume        = xx,
 issue         = xx,
 pubyear       = 2026,
 isbn          = xxx-x-xxxxx-xxx-x,
 eisbn         = xxx-x-xxxxx-xxx-x,
 doi           = 10.1561/XXXXXXXXX,
 firstpage     = 1, 
 lastpage      = 18
 }

\usepackage{mwe}

\author[1]{Malitesta,Daniele}
\author[2]{Mancino,Alberto Carlo Maria}
\author[2]{Pomo,Claudio}
\author[2]{Anelli,Vito Walter}
\author[3]{Bellogín,Alejandro}
\author[2]{Di Noia,Tommaso}

\affil[1]{LUISS Guido Carli University; dmalitesta@luiss.it}
\affil[2]{Politecnico di Bari; name.surname@poliba.it}
\affil[3]{Universidad Autónoma de Madrid; alejandro.bellogin@uam.es}

\begin{document}

\makeabstracttitle

\begin{abstract}
In recommender systems, user–item interactions can be modeled as a bipartite graph, where user and item nodes are connected by undirected edges. This graph-based view has motivated the rapid adoption of graph neural networks (GNNs), which often outperform collaborative filtering (CF) methods such as latent factor models, deep neural networks, and generative strategies. Yet, despite their empirical success, the reasons why GNNs offer systematic advantages over other CF approaches remain only partially understood.

This monograph advances a topology-centered perspective on GNN-based recommendation. We argue that a comprehensive understanding of these models’ performance should consider the structural properties of user–item graphs and their interaction with GNN architectural design. To support this view, we introduce a formal taxonomy that distills common modeling patterns across eleven representative GNN-based recommendation approaches and consolidates them into a unified conceptual pipeline.

We further formalize thirteen classical and topological characteristics of recommendation datasets and reinterpret them through the lens of graph machine learning. Using these definitions, we analyze the considered GNN-based recommender architectures to assess how and to what extent they encode such properties. Building on this analysis, we derive an explanatory framework that links measurable dataset characteristics to model behavior and performance.

Taken together, this monograph re-frames GNN-based recommendation through its topological underpinnings and outlines open theoretical, data-centric, and evaluation challenges for the next generation of topology-aware recommender systems.

\end{abstract}

\chapter{Introduction}

Recommender systems (RSs) have undergone a substantial shift with the introduction of graph neural networks  (GNNs)~\citep{DBLP:journals/corr/abs-2104-13478, DBLP:journals/tnn/ScarselliGTHM09, DBLP:series/synthesis/2020Hamilton}, which model users and items as nodes in a bipartite, undirected graph.
Unlike classical collaborative filtering (CF) approaches, which learn representations solely from direct user-item interactions, graph-based models propagate information across multi-hop neighborhoods, capturing high-order connectivity patterns that can enrich the learned embeddings and improve accuracy.
This shift has led to the development of graph CF methods, which redefine traditional CF paradigms by integrating graph representation learning techniques~\citep{DBLP:journals/csur/WuSZXC23, DBLP:conf/wsdm/GaoW0022, DBLP:conf/sigir/0001DWLZ020, DBLP:conf/cikm/MaoZXLWH21, DBLP:journals/tkde/YuXCCHY24}.
These foundational efforts established the viability of graph-based CF and motivated a large body of research seeking to refine GNN architectures for recommender systems.

\begin{sidewaysfigure*}
\centering
\includegraphics[width=\textwidth]{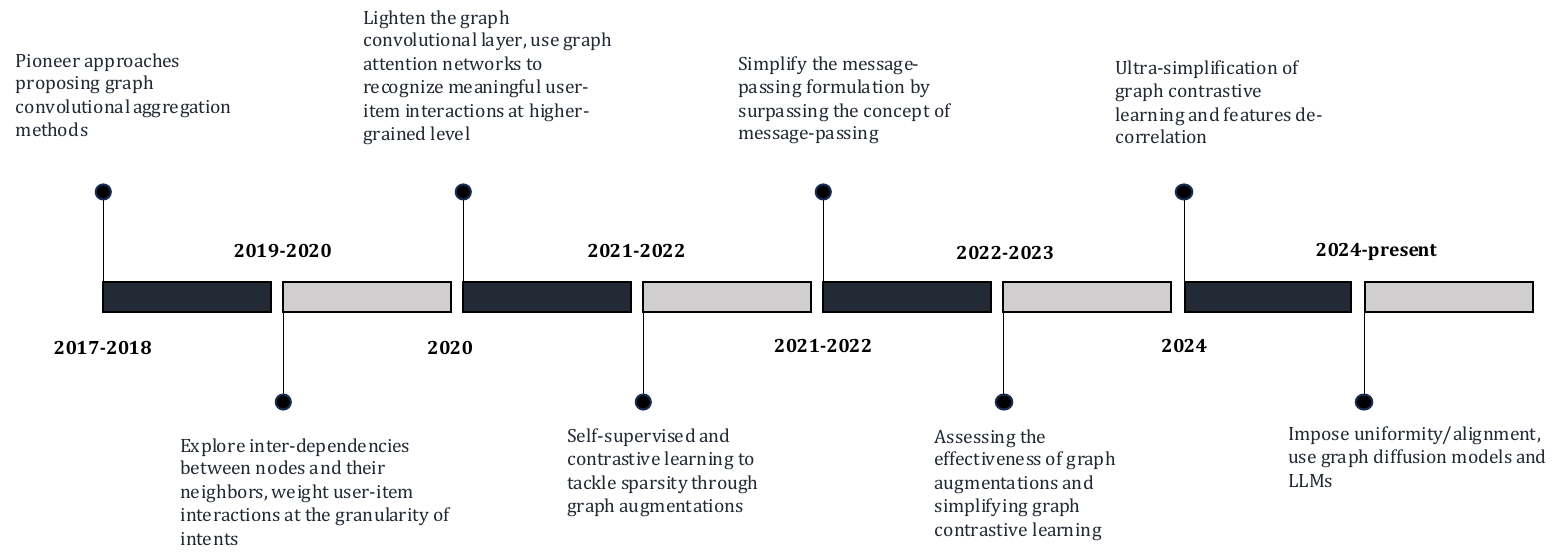}
\caption{A non-exhaustive timeline of representative approaches in graph collaborative filtering (CF) by grouping solutions depending on the adopted graph representation learning strategies.}
\label{fig:timeline}
\end{sidewaysfigure*}

Early works in this direction adapted general-purpose graph convolutional layers such as GraphSAGE~\citep{DBLP:conf/nips/HamiltonYL17} and GCN~\citep{DBLP:conf/iclr/KipfW17} to recommendation, often by placing message-passing layers on top of latent-factor models~\citep{DBLP:journals/corr/BergKW17, DBLP:conf/kdd/YingHCEHL18, DBLP:conf/sigir/Wang0WFC19}. 
These foundational efforts established the viability of graph-based CF and motivated a large body of research seeking to refine GNN architectures for recommender systems. Subsequent advancements explored ways to simplify message passing~\citep{DBLP:conf/icml/WuSZFYW19}, remove non-linearities~\citep{DBLP:conf/sigir/0001DWLZ020, DBLP:conf/aaai/ChenWHZW20}, and incorporate edge importance through reweighting schemes~\citep{DBLP:conf/cikm/AnelliDNSFMP22, DBLP:conf/recsys/MancinoFBMNS23} or attention mechanisms, as in GAT~\citep{DBLP:conf/iclr/VelickovicCCRLB18}, to prune noisy interactions and disentangle multiple user intents~\citep{DBLP:conf/icml/Ma0KW019, DBLP:conf/sigir/ZhangL0WSLSZDZ22, DBLP:conf/sigir/WangJZ0XC20, DBLP:conf/www/WangHWYL0C21}.
More recent lines of research have extended graph-based recommender systems in several directions. Semi-supervised approaches~\citep{DBLP:conf/cikm/HuangXW0Y22} and contrastive learning frameworks~\citep{DBLP:conf/nips/KhoslaTWSTIMLK20} mitigate the sparsity of user–item graphs through carefully designed graph augmentations~\citep{DBLP:conf/www/LinTHZ22, DBLP:conf/iclr/Cai0XR23, DBLP:conf/sigir/0001OM22, DBLP:conf/sigir/YuY00CN22, DBLP:conf/sigir/WuWF0CLX21, DBLP:journals/tkde/YuXCCHY24}. 
Parallel efforts addressed fundamental limitations of message passing by tackling over-smoothing~\citep{DBLP:conf/aaai/ChenLLLZS20}, over-squashing~\citep{DBLP:conf/iclr/ToppingGC0B22}, or both~\citep{DBLP:conf/cikm/GiraldoSBM23}, inspiring architectures that go beyond standard neighborhood aggregation~\citep{DBLP:conf/cikm/MaoZXLWH21, DBLP:conf/cikm/ShenWZSZLL21, DBLP:conf/cikm/PengSM22, DBLP:conf/sigir/WuWS0CX24}. 
More recently, some works have proposed to promote uniformity and alignment in graph CF \citep{DBLP:conf/cikm/YangLWYLMY23}, while graph diffusion models~\citep{DBLP:conf/iclr/VignacKSWCF23} have also emerged as a promising direction for denoising interaction graphs and improving robustness in recommendation tasks~\citep{DBLP:conf/wsdm/JiangYXH24}, and large language models (LLMs) have been exploited to enhance the node representations and edges within the historical user-item graph \citep{DBLP:conf/wsdm/WeiRTWSCWYH24}. 
To illustrate the evolution of this research area, Figure~\ref{fig:timeline} presents a non-exhaustive timeline of representative GNN-based recommendation models.

Despite extensive research on GNN-based recommender systems, a fundamental question remains insufficiently answered: “Why do these models perform so well?” 
Existing work has explored both theoretical and experimental perspectives in an attempt to explain their success. Theoretically, several studies suggest that graph convolution operations act as smoothing filters on node features, enabling GNNs to capture latent structural patterns in recommendation graphs~\citep{DBLP:conf/cikm/ShenWZSZLL21}. More recent research has highlighted issues such as over-smoothing and over-correlation, motivating adaptive techniques designed to mitigate these effects~\citep{DBLP:conf/sigir/WuWS0CX24}. Experimentally, efforts have focused on modularizing GNN pipelines and generalizing architectures across recommendation scenarios~\citep{DBLP:conf/wsdm/WangZS22}; then, some works have compared both GNN-based \citep{DBLP:conf/recsys/AnelliMPBSN23} and GNN session-based \citep{DBLP:conf/recsys/ShehzadJ25} approaches with other carefully-tuned recommender systems, outlining performance pitfalls which are not commonly outlined in the original works; finally other authors identified node representation learning and neighborhood exploration as critical components, studying the performance of graph-based CF beyond accuracy metrics~\citep{ DBLP:conf/ecir/AnelliDNMPP23, DBLP:conf/recsys/AnelliDNSFMP22}.

However, these investigations largely overlook a crucial factor: the role of graph topology in shaping GNN performance. Ultimately, answering the question “Why do GNN-based recommender systems work so well?” requires addressing a deeper one:
“Which topological characteristics of user–item graphs do these models capture, and do those properties influence their performance?”

Our intuition is supported by recent developments in graph representation learning. A growing line of work has begun to interrogate how structural properties (such as topology and node features) affect GNN performance~\citep{DBLP:conf/icml/WuSZFYW19, DBLP:conf/www/WeiZH22, DBLP:conf/www/0002ZPNG0CH22, DBLP:conf/iclr/KlicperaBG19, DBLP:conf/iclr/ZhaoA20, DBLP:conf/icml/Abu-El-HaijaPKA19, DBLP:conf/nips/LuanZCP19, DBLP:conf/icml/XuLTSKJ18, DBLP:conf/iclr/RongHXH20, DBLP:journals/corr/abs-2308-09570}. For instance, it is now well established that GNNs excel in node classification tasks on graphs with high homophily~\citep{DBLP:conf/nips/ZhuYZHAK20}. Likewise, recent studies show that node degree can significantly influence GNN performance in both node classification and link prediction tasks~\citep{DBLP:conf/iclr/01600ZLCSD24}. Additionally, in recommender systems, existing studies have hinted that dataset characteristics (such as sparsity) can affect performance~\citep{DBLP:journals/tmis/AdomaviciusZ12, DBLP:conf/sigir/DeldjooNSM20, DBLP:conf/recsys/MalitestaPAMNS24}; more recently, the works by \citet{DBLP:conf/ecir/GioiaPBN26} and \citet{DBLP:conf/sigir/GioiaPBN26} have investigated how spectral graph models can differently shape the topology of the user-item graph, while the study conducted by \citet{DBLP:conf/recsys/FiorettiLPMPNN25} has assessed the impact of LLM augmentations on the user-item graph from a topological perspective. Yet, a systematic investigation of topological factors specifically in recommendation is still missing, leaving a key aspect of GNN-based CF unexplored.

In light of the above, our work offers a topology-centered perspective on GNN-based recommendation, systematically relating the performance of representative models to the structural properties of the user–item graphs on which they operate. Building on this analysis, we derive a set of take-home messages on when and why specific strategies are effective, and we identify outstanding challenges and future research directions for topology-aware recommender systems. In the following, we outline the structure of the monograph (Section~\ref{sec:index}), the main notation and terminology (Section~\ref{sec:notation}), and selected datasets and GNN models to exemplify our analysis (Sections~\ref{sec:data} and~\ref{sec:models}).

\section{Structure of this monograph}\label{sec:index}
This work begins by introducing basic notions about graph neural networks (Section~\ref{ch:background}) to provide the \textbf{foundations} to understand and categorize graph-based recommendation algorithms (Section~\ref{ch:pipeline}). 
This allows us to present a clear \textbf{taxonomy} of the methods and a \textbf{pipeline} of the recommendation process. 
This includes an overview of how GNN-based recommender systems have evolved, from early approaches that adapted shallow graph representation learning techniques to more recent trends involving semi-supervised and contrastive learning, to approaches translating graph-based recommendation to the spectral domain to address problems such as over-smoothing. 

Sections~\ref{ch:char_class} and~\ref{ch:char_topo} define several \textbf{dataset characteristics}, either classical or topological. How data is structured, shaped, or distributed across various dimensions impacts differently recommendation algorithms depending on their nature, as observed in the literature. Hence, we here divide the presentation of these dataset characteristics according to their topological interpretation.

Then, we examine how various \textbf{(topological) dataset characteristics} impact the recommendation performance of GNN algorithms (Section~\ref{ch:topo_gnn}). This is achieved by re-interpreting popular topological graph properties under the lens of recommendation and highlighting their (explicit) presence within the models' formulations. 
Afterwards, in Section~\ref{ch:framework}, we present an \textbf{explanatory framework} that involves generating a large set of small graph recommendation datasets encompassing a wide range of topological structures and testing on them the recommendation performance of selected GNN-based models. 
Hence, \textbf{performance explanations} are derived by jointly considering the models' results and the dataset properties. Building upon the topological re-interpretations provided on classical GNNs-based recommendation, in Section \ref{ch:recent_trends}, we offer an overview of recent trends in graph-based recommendation, more specifically those involving \textbf{spectral} methodologies and \textbf{LLMs}-based approaches, underlining topological implications of such novel strategies.

Based on these conclusions, Section~\ref{ch:message} presents the most important \textbf{take-home messages} to be considered when dealing with GNN-based recommendation algorithms.
This, together with the \textbf{challenges} and \textbf{future directions} included in Section~\ref{ch:future}, conclude the monograph and paves the way for further research, emphasizing issues and problems with more potential to be considered interesting for the community and industry practitioners.

\section{Notations and terminology}\label{sec:notation}

We introduce the useful notation and terminology for this monograph. While this section provides most of the notations adopted in the following pages, we highlight that each section in this monograph might sometimes contain more specific notations tailored to a particular formulation that is not covered here.

We use boldface capital letters to represent multi-dimensional vectors (e.g., $\mathbf{A}$), lowercase letters for constants and coefficients (e.g., $t$), and Greek uppercase/lowercase letters for classical and topological dataset properties, respectively (we summarize them in Table \ref{tab:shorthand}). Note that, for the degree distribution only, we do not introduce a symbolic notation, as its behavior is more effectively analyzed through its empirical shape. For functions, we indicate them in a textual, human-readable format, such as $\text{LeakyReLU}(\cdot)$ or $\text{exp}(\cdot)$ for the LeakyReLU activation function \citep{DBLP:journals/corr/XuWCL15} and the exponential function, respectively. Moreover, we use the calligraphic notation for loss functions and sets, for instance $\mathcal{L}_{\text{BPR}}$ and $\mathcal{U}, \mathcal{I}$ to refer to the Bayesian-personalized ranking loss function (BPR, introduced by \citet{DBLP:conf/uai/RendleFGS09}) and the sets of users and items, respectively.

Regarding vectors, we indicate specific entries of these vectors with subscripts, such as $A_{ij}$ to indicate the $i$-th row and $j$-th column (the boldface notation is removed here as we refer to a single scalar element); we adopt the ``*'' notation to refer to all entries for one dimension (e.g., $\mathbf{A}_{i*}$ refers to the $i$-th row and all columns for matrix $\mathbf{A}$). Conversely, we use superscripts mainly to refer to updated versions of vectors during the message passing algorithm, such as $\mathbf{E}_u^{(l)}$ to refer to the updated version of the vector $\mathbf{E}_u$ after $l$ message passing layers. Additionally, we adopt the hat and tilde notations (e.g., $\hat{\mathbf{E}}_u$ and $\tilde{\mathbf{A}}$) to indicate vectors that have been completely updated to compute a prediction, and a normalized vector (as in the case of the normalized adjacency matrix), respectively. Finally, we leverage the operator $[\cdot]$ to refer to a specific view of a vector, as in the case of graph contrastive learning (e.g., $\mathbf{E}_u[t]$ to indicate the $t$-th view of vector $\mathbf{E}_u$).

Terminology-wise, we highlight that we refer to the whole family of models covered in this monograph in different variants, such as ``graph-based recommender systems'', ``GNNs-based recommender systems'', or ``graph CF recommender systems''.

\begin{table}[!t]
\caption{The table summarizes the data characteristics analyzed in this work, the symbols used to refer to them, and their shorthand notation.}
\label{tab:shorthand}
\begin{adjustbox}{max width=\columnwidth}
\begin{tabular}{l|l|l|l}
\toprule
\textbf{Type} & \textbf{Characteristics} & \textbf{Symbol} & \textbf{Shorthand} \\
\cmidrule{1-4}
\multirow{5}{*}{\textit{Classical}} 
    & Space size & $\Psi$ & \textsc{SpaceSize} \\
    & Shape & $\Phi$ & \textsc{Shape} \\
    & Density & $\Delta$ & \textsc{Density} \\
    & Sparsity & $\Lambda$ & \textsc{Sparsity} \\
    & Gini user & $\text{K}_{\mathcal{U}}$ & \textsc{Gini-U} \\
    & Gini item & $\text{K}_{\mathcal{I}}$ & \textsc{Gini-I} \\
\hline
\multirow{7}{*}{\textit{Topological}} 
    & Average degree user & $\sigma_{\mathcal{U}}$ & \textsc{AvgDegree-U} \\
    & Average degree item & $\sigma_{\mathcal{I}}$ & \textsc{AvgDegree-I} \\
    & Avg. clustering coefficient user & $\gamma_{\mathcal{U}}$ & \textsc{AvgClustC-U} \\
    & Avg. clustering coefficient item & $\gamma_{\mathcal{I}}$ & \textsc{AvgClustC-I} \\
    & Degree assortativity user & $\rho_{\mathcal{U}}$ & \textsc{Assort-U} \\
    & Degree assortativity item & $\rho_{\mathcal{I}}$ & \textsc{Assort-I} \\
    & Degree distribution & -- & \textsc{DegreeDistr} \\
\bottomrule
\end{tabular}
\end{adjustbox}
\end{table}

\section{Datasets referenced in this monograph}\label{sec:data}

As one of the main rationales behind this monograph is to offer a dataset-centric perspective on graph-based recommendation (with specific focus on topology), the next sections introduce both classical and topological properties of the recommendation dataset. To complement the theoretical description of such characteristics, we select popular datasets from the recommendation literature, and compute values for these properties, thus providing the reader with also a practical illustration. For the selection, we follow the dataset list collected in \citep{DBLP:conf/sigir/MancinoBF0MPN25} and limit ourselves to the six most popular datasets in the recommendation literature. 

\begin{itemize}
\item \textbf{MovieLens 1M}\footnote{\url{https://grouplens.org/datasets/movielens/}}~\citep{DBLP:journals/tiis/HarperK16}. A widely used benchmark dataset containing 
movie ratings collected from the MovieLens movie recommendation platform.
\item \textbf{Yelp}\footnote{\url{https://business.yelp.com/data/resources/open-dataset/}}. The Yelp Open Dataset contains interactions between users and businesses, where user feedback is provided in the form of written reviews.
\item \textbf{Amazon Beauty}\footnote{\url{https://amazon-reviews-2023.github.io/}}~\citep{DBLP:journals/corr/abs-2403-03952}. A large-scale dataset collected in 2023 from the Amazon website, containing user reviews for products in the Beauty category.
\item \textbf{Amazon Books}\footnote{\url{https://amazon-reviews-2023.github.io/}}~\citep{DBLP:journals/corr/abs-2403-03952}. A large-scale dataset collected in 2023 from the Amazon website, containing user reviews for products in the Books category.
\item \textbf{Gowalla}\footnote{\url{https://snap.stanford.edu/data/loc-gowalla.html}}~\citep{DBLP:conf/kdd/ChoML11__rep}. A dataset derived from the location-based social network Gowalla, containing 
user check-ins recorded between Feb. 2009 and Oct. 2010.
\item \textbf{Last.fm}\footnote{\url{https://grouplens.org/datasets/hetrec-2011/}}~\citep{DBLP:conf/recsys/2011hetrec}. A dataset derived from the Last.fm music platform, containing users’ listening preferences and interactions with musical artists.
\end{itemize}

To simplify the analysis and reduce noise, we filtered the datasets by applying an iterative user–item k-core with different depths: 15 for Amazon Beauty and LastFM, 25 for Amazon Books, MovieLens, Gowalla, and Yelp. The statistics obtained after filtering are reported in~\Cref{tab:referenced_datasets}, where the datasets are ordered by their total number of interactions, from the highest to the lowest.

\begin{table}[!t]
\centering
\caption{Statistics of the recommendation datasets referenced in this monograph after filtering, ordered by their total number of interactions (the number of edges in the corresponding bipartite graph).}
\label{tab:referenced_datasets}
\resizebox{0.95\linewidth}{!}{
\begin{tabular}{lrrr}
\toprule
\textbf{Dataset} & \textbf{\# Users} & \textbf{\# Items} & \textbf{\# Interactions} \\
\midrule
\href{https://grouplens.org/datasets/movielens/}{MovieLens}     & 5,611                              & 2,932                              & 983,705                                   \\
\href{https://business.yelp.com/data/resources/open-dataset/}{Yelp}          & 14,691                             & 12,482                             & 819,986                                   \\
\href{https://amazon-reviews-2023.github.io/}{Amazon Beauty} & 8,425                              & 11,509                             & 304,396                                   \\
\href{https://amazon-reviews-2023.github.io/}{Amazon Books}  & 3,250                              & 2,450                              & 141,331                                   \\
\href{https://snap.stanford.edu/data/loc-gowalla.html}{Gowalla}       & 2,426                              & 2,195                              & 131,715                                   \\
\href{https://grouplens.org/datasets/hetrec-2011/}{LastFM}        & 1,601                              & 951                                & 53,491  \\
\bottomrule
\end{tabular}
} 
\end{table}

\section{Models referenced in this monograph}\label{sec:models}

Alongside the selected dataset characteristics, we also consider a list of eleven graph-based recommendation approaches from the literature, to re-interpret their formulations under the topological lens. The motivation behind this selection is twofold. On the one side, these models constitute a comprehensive (but non-exhaustive) group of approaches spanning well-established (i.e., counting several citations on Google Scholar) and more recent solutions, encompassing diversified strategies from shallow GNNs architectures to more refined techniques involving graph contrastive learning. On the other side, prior research (such as the work by \citet{DBLP:conf/recsys/AnelliMPBSN23}) has evidenced that most of these approaches are constantly tested as strong graph-based baselines for novel proposed methodologies. In the following, we briefly summarize their architecture and rationales, displaying their publication year and venue in Table \ref{tab:referenced_models}, where we sort models according to the chronological order of their publication.

\begin{itemize}
    \item \textbf{NGCF.} Neural graph collaborative filtering \citep{DBLP:conf/sigir/Wang0WFC19} is one of the pioneer approaches in graph-based recommendation. The model implements a GCN-like message passing, where each propagation layer takes also into account the inter-dependencies between the ego and neighborhood nodes. 
    \item \textbf{DGCF.} Disengangled graph collaborative filtering \citep{DBLP:conf/sigir/WangJZ0XC20} proposes to divide user and item vectors into intents to represent the reasons behind each recorded user-item interaction and weight their importance.
    \item \textbf{LightGCN.} Light graph convolutional network \citep{DBLP:conf/sigir/0001DWLZ020} simplifies the GCN message passing from previous approaches by removing the features transformation and non-linearities from the message passing procedure, while still reaching improved recommendation performance.
    \item \textbf{SGL.} Self-supervised graph learning \citep{DBLP:conf/sigir/WuWF0CLX21} builds different augmented views of the user-item graph and applies a contrastive procedure between them to limit the sparsity problem of the recommendation dataset. 
    \item \textbf{UltraGCN.} Ultra simplification of graph convolutional network \citep{DBLP:conf/cikm/MaoZXLWH21} tailors crucial issues in graph learning, such as over-smoothing and usually-unexplored node relationships (such as item-item). To this end, it proposes a mathematical simplification of the GCN message passing layer that (in one step) approximates an ideal infinite-layer message passing, and specific loss components to further address over-smoothing and consider item-item interactions.
    \item \textbf{GFCF.} Graph filter-based collaborative filtering \citep{DBLP:conf/cikm/ShenWZSZLL21} re-interprets graph convolution in recommendation under the lens of graph signal processing, by formulating a unified graph filtering framework for most approaches in CF. On such a basis, it proposes a closed-form solution that performs better than other trainable approaches in the literature.
    \item \textbf{SVD-GCN.} The authors in \citep{DBLP:conf/cikm/PengSM22} applies singular value decomposition (SVD) to the user-item interaction matrix to calculate, through a closed-form solution, vector representations for users and items. On such a basis, they design additional loss function components (besides BPR) that take into account user-user and item-item similarities.
    \item \textbf{SimGCL.} Simplification of graph contrastive learning \citep{DBLP:conf/sigir/YuY00CN22} questions the role of traditional graph augmentations (such as node and edge dropout), proposing a new augmentation strategy applied on node vectors during the message passing procedure. The approach further implements a uniformity loss on the learned representations.
    \item \textbf{LightGCL.} The authors in \citep{DBLP:conf/iclr/Cai0XR23} design a double message passing procedure in graph contrastive learning, one acting locally on the perturbed graph structure, and the other acting globally by truncating the SVD representation of users and items. The two views are contrasted in the loss function, which also implements a hinge loss unlike many other solutions leveraging BPR.
    \item \textbf{GraphAU.} The model proposed in \citep{DBLP:conf/cikm/YangLWYLMY23} re-adapts the principles of uniformity and alignment to graph-based recommendation by setting up a trade-off loss between uniformity and alignment on the user-item graph structure.
    \item \textbf{XSimGCL.} Extremely simple graph contrastive learning \citep{DBLP:journals/tkde/YuXCCHY24} surpasses the traditional concept of graph contrastive learning in recommendation by leveraging the same augmented node view for the final recommendation task. Moreover, the contrastive loss accounts for two versions of the same node vectors computed at different propagation layers in the message passing.
\end{itemize}

\begin{table}[!t]
\centering
\caption{GNN-based recommendation models referenced in this monograph, sorted by publication year.}\label{tab:referenced_models}
\resizebox{\textwidth}{!}{
\begin{tabular}{llll}
\toprule
\textbf{Model} & \textbf{Paper} & \textbf{Venue} \\ \midrule
NGCF & \parbox[t]{9cm}{Neural Graph Collaborative Filtering \citep{DBLP:conf/sigir/Wang0WFC19}} & SIGIR'19 \\ \\
DGCF & \parbox[t]{9cm}{Disentangled Graph Collaborative Filtering \citep{DBLP:conf/sigir/WangJZ0XC20}} & SIGIR'20 \\ \\
LightGCN & \parbox[t]{9cm}{LightGCN: Simplifying and Powering Graph Convolution Network for Recommendation \citep{DBLP:conf/sigir/0001DWLZ020}} & SIGIR'20 \\ \\
SGL & \parbox[t]{9cm}{Self-supervised Graph Learning for Recommendation \citep{DBLP:conf/sigir/WuWF0CLX21}} & SIGIR'21 \\ \\
UltraGCN & \parbox[t]{9cm}{UltraGCN: Ultra Simplification of Graph Convolutional Networks for Recommendation \citep{DBLP:conf/cikm/MaoZXLWH21}} & CIKM'21 \\ \\
GFCF & \parbox[t]{9cm}{How Powerful is Graph Convolution for Recommendation? \citep{DBLP:conf/cikm/ShenWZSZLL21}} & CIKM'21 \\ \\ 
SVD-GCN & \parbox[t]{9cm}{SVD-GCN: A Simplified Graph Convolution Paradigm for Recommendation \citep{DBLP:conf/cikm/PengSM22}} & CIKM'22 \\ \\ 
SimGCL & \parbox[t]{9cm}{Are Graph Augmentations Necessary? Simple Graph Contrastive Learning for Recommendation \citep{DBLP:conf/sigir/YuY00CN22}} & SIGIR'22 \\ \\
LightGCL & \parbox[t]{9cm}{LightGCL: Simple Yet Effective Graph Contrastive Learning for Recommendation \citep{DBLP:conf/iclr/Cai0XR23}} & ICLR'23 \\ \\ 
GraphAU & \parbox[t]{9cm}{Graph-based Alignment and Uniformity for Recommendation \citep{DBLP:conf/cikm/YangLWYLMY23}} & CIKM'23 \\ \\
XSimGCL & \parbox[t]{9cm}{XSimGCL: Towards Extremely Simple Graph Contrastive Learning for Recommendation \citep{DBLP:journals/tkde/YuXCCHY24}} & TKDE'24 \\
\bottomrule
\end{tabular}
} 
\end{table}
\chapter{Background on graph neural networks}\label{ch:background}

This section provides the indispensable background notions regarding graph neural networks (GNNs). After formally defining graphs, the section presents the message passing algorithm, which is the basic procedure behind most of the GNN architectures from the literature. Then, it introduces two of the pioneer GNN models, namely, graph convolutional network (GCN) and graph attention network (GAT), whose procedures inspired many graph-based recommendation approaches. Finally, it describes the task of link prediction in graph representation learning, for whom recommendation represents one of the possible instantiations. 

\section{Basic notions about graphs}
This section begins with some basic notions about graphs as data structures. After a formal definition of what a graph is, along with its multiple variations, we formalize the adjacency matrix and node features, which represent the main components to define a graph.

\subsection{Definition of graph}
We define a graph through the set of nodes $\mathcal{V}$ and edges $\mathcal{E}$ connecting such nodes, namely, $\mathcal{G} = (\mathcal{V}, \mathcal{E})$. Specifically, an edge between node $v \in \mathcal{V}$ and $w \in \mathcal{V}$ exists if the two nodes are connected in the graph. When edges in a graph come with a direction (e.g., node $v$ is connected to node $w$ but not vice versa), the graph is \textit{directed}; otherwise, if the edge direction is not defined (e.g., $v$ is connected to $w$ and also the vice versa holds) the graph is \textit{undirected}. 

Moreover, nodes in a graph may belong to disjoint partitions, namely, $\mathcal{V} = \mathcal{V}_1 \; \cup \; \mathcal{V}_2, \; \dots, \; \cup \; \mathcal{V}_p$, having $\mathcal{V}_i \cap \mathcal{V}_j = \emptyset \; , \forall i \neq j$. This type of graph, named \textit{heterogeneous}, has edges that can connect nodes from different partitions following a specific rationale. A special case of heterogeneous graphs is the \textit{multipartite} one, a graph where edges only connect nodes from different partitions. Among multipartite graphs, \textit{bipartite} graphs contain nodes that belong to two different partitions, and nodes from one partition can only be connected to nodes from the other partition. 

\subsection{Adjacency matrix and node features}

An \textit{adjacency matrix} represents the graph $\mathcal{G}$ into matrix format. By assigning each node in the graph an ordering (i.e., nodes are represented by specific rows and columns in the matrix), the adjacency matrix $\mathbf{A} \in \mathbb{R}^{|\mathcal{V}| \times |\mathcal{V}|}$ is designed such that $A_{vw} = a$ if there exists an edge connecting $v$ and $w$, 0 otherwise. In a directed graph, $A_{vw} \neq A_{wv}$ in general, while in an undirected graph, $\mathbf{A}$ is symmetric and $A_{vw} = A_{wv}$. 

While the entry $A_{vw} = a$ is a real-valued number that represents the weight of the edge connecting the nodes $v$ and $w$, it is common practice (for the sake of simplicity) to consider binary adjacency matrices where $A_{vw} = \{0, 1\}$ and $A_{vw} = 1$ if there exists an edge between the two nodes, 0 otherwise. In multipartite graphs there exist entire portions of the adjacency matrix whose entries are zero, as in such a family of graphs some node partitions are not connected to one another.  

Moreover, for the training of graph neural networks (see later), the adjacency matrix is usually transformed to obtain the symmetric normalized adjacency matrix (i.e., $\tilde{\mathbf{A}}$):
\begin{equation}
    \tilde{\mathbf{A}} = \mathbf{D}^{-\frac{1}{2}} \mathbf{A} \mathbf{D}^{-\frac{1}{2}},
\end{equation}
where the purpose of this normalization is to flatten the weight importance of all nodes in the graph based upon their degree to tackle possible instabilities during the optimization of the graph neural network.

In addition, another important concept is the \textit{neighborhood} of a node $v$, defined as the set of nodes which are connected to $v$ through one edge. Let $\mathcal{N}_v = \{w \in \mathcal{V} \; | \; A_{vw} = 1\}$ be the neighborhood of $v$.

Finally, nodes in a graph may be associated with attributes or features describing them. Generally, such features are formally represented through a real-valued matrix $\mathbf{X} \in \mathbb{R}^{|\mathcal{V}| \times F}$ where $F$ is the dimensionality of the feature matrix.

\section{The message passing algorithm}
Independently on the graph learning strategy adopted, the most atomic building block of any graph neural network architecture lies in the \textit{message passing} algorithm \citep{DBLP:conf/icml/GilmerSRVD17}. 

In its most generalized version, such a procedure works by \textit{updating} the latent representation of each node (defined as \textit{ego} node in this context) through the information conveyed by the nodes directly connected to the ego one (defined as \textit{neighbor} nodes). The information from the neighborhood is usually referred to as \textit{messages}, which are \textit{aggregated} and used to update the representation of the ego node. Finally, the message passing schema is iteratively applied for $L$ layers, and the various node representations are eventually \textit{combined} to obtain a unique representation for each node. In the following, a formal definition for each step, namely, \textit{message aggregation}, \textit{node embedding update}, and \textit{layer combination}, is presented \citep{DBLP:series/synthesis/2020Hamilton}.

\subsection{Message aggregation}
Let $\mathbf{X}_{v}^{(l-1)} \in \mathbb{R}^{F_{l-1}}$ be the representation of node $v \in \mathcal{V}$ at layer $l-1$, with $1 \leq l \leq L$, and $F_{l-1}$ the embedding dimension at layer $l-1$. Note that we have $\mathbf{X}_v^{(0)} = \mathbf{X}_v$. We obtain the aggregated messages from the neighbor nodes of $v$ through the following:
\begin{equation}
    \mathbf{M}_v^{(l-1)} = \text{Aggregate}^{(l)}(\{\mathbf{X}_w^{(l-1)} \; | \; w \in \mathcal{N}_v\}),
\end{equation}
where $\mathbf{M}_v^{(l-1)} \in \mathbb{R}^{F_{l-1}}$ is the aggregated message from the neighbor nodes of $v$ at layer $l-1$ and $\text{Aggregate}^{(l)}(\cdot)$ is the aggregation function over the neighbor nodes of $v$. For instance, a popular choice for the $\text{Aggregate}^{(l)}(\cdot)$ function is the element-wise addition:
\begin{equation}
    \mathbf{M}_v^{(l-1)} = \sum_{w \in \mathcal{N}_v} A_{vw} \mathbf{X}_w^{(l-1)},
\end{equation}
where $A_{vw}$ corresponds to the weight of the edge between $v$ and $w$ (if the nodes are connected). As stated above, the adjacency matrix may be normalized into the symmetric normalized adjacency matrix:
\begin{equation}
    \mathbf{M}_v^{(l-1)} = \sum_{w \in \mathcal{N}_v} \frac{A_{vw}}{\sqrt{|\mathcal{N}_v||\mathcal{N}_w|}} \mathbf{X}_w^{(l-1)},
\end{equation}
where the denominator involves the node degrees of the ego node and each neighbor node.

\subsection{Node embedding update}
Once the aggregated message from the neighbor nodes has been calculated, the node embedding can be updated to obtain the representation at layer $l$:
\begin{equation}
    \mathbf{X}_v^{(l)} = \text{Update}^{(l)}(\mathbf{X}_v^{(l-1)}, \mathbf{M}_v^{(l-1)}),
\end{equation}
where $\mathbf{X}_v^{(l)} \in \mathbb{R}^{F_{l}}$ is the updated embedding representation of node $v$ at layer $l$, $\text{Update}^{(l)}(\cdot)$ is the update function for the ego node at layer $l$, and $\mathbf{X}_v^{(l-1)}$ is usually included to leverage the current representation of the ego node into the message passing formulation. For instance, a popular choice to design the $\text{Update}^{(l)}(\cdot)$ function is the feature transformation of the two inputs:
\begin{equation}
    \mathbf{X}_v^{(l)} = \text{Activate}(\mathbf{W}_{\text{ego}}^{(l)}\mathbf{X}_v^{(l-1)} + \mathbf{W}_{\text{neigh}}^{(l)}\mathbf{M}_v^{(l-1)}),
\end{equation}
where $\text{Activate}(\cdot)$ is the activation, such as the Sigmoid or ReLU, while $\mathbf{W}_{\text{ego}}^{(l)} \in \mathbb{R}^{F_{l-1} \times F_{l}}$ and $\mathbf{W}_{\text{neigh}}^{(l)} \in \mathbb{R}^{F_{l-1} \times F_{l}}$ are the weights referring to the ego node and the message aggregated through the $\text{Aggregate}^{(l)}(\cdot)$ function, respectively. 

\subsection{Layer combination}
The final stage is the combination of all node embedding representations obtained in each layer $l$. The layer combination is obtained through:
\begin{equation}
    \hat{\mathbf{X}}_v = \text{Combine}(\{\mathbf{X}_v^{(0)}, \; \mathbf{X}_v^{(1)}, \; \dots, \; \mathbf{X}_v^{(l)}, \; \dots, \; \mathbf{X}_v^{(L)}\}),
\end{equation}
where $\text{Combine}(\cdot)$ is defined through different operations, such as the element-wise addition, the average, the concatenation, or it is simply obtained as the last node update from layer $L$.

\subsection{Matrix format and self-loops}
The formulations provided above for the message passing algorithm are expressed at node level. However, in most of the cases, the message passing algorithm can be expressed into a more compact matrix format which comprises all graph nodes at once:
\begin{equation}
    \mathbf{X}^{(l)} = \text{Activate}(\mathbf{X}^{(l-1)}\mathbf{W}_{\text{ego}}^{(l)} + \mathbf{A}\mathbf{X}^{(l-1)}\mathbf{W}_{\text{neigh}}^{(l)}),
\end{equation}
where $\mathbf{X}^{(l-1)} \in \mathbb{R}^{|\mathcal{V}| \times F_{l-1}}$ includes all node embeddings. Generally, the $\text{Update}^{(l)}(\cdot)$ function can be simplified by adding self-loops in the adjacency matrix, namely, the ego node is added to the set of neighbor nodes during the message aggregation:
\begin{equation}
     \mathbf{X}^{(l)} = \text{Activate}((\mathbf{A} + \mathbf{I})\mathbf{X}^{(l-1)}\mathbf{W}^{(l)}),
\end{equation}
where $\mathbf{A} + \mathbf{I}$ is used to add self-loops to the adjacency matrix, and the presence of a single $\mathbf{W}^{(l)}$ matrix suggests $\mathbf{W}_{\text{ego}}^{(l)}$ and $\mathbf{W}_{\text{neigh}}^{(l)}$ are shared into one matrix. 

\section{Popular graph neural network architectures}
The literature recognizes different graph neural network architectures based on the specific strategies and operations adopted for message passing. In the following, we present and formalize the most popular ones, namely, graph convolutional network (GCN) and graph attention network (GAT), whose architectures inspired many graph-based recommender systems.

\subsection{Graph convolutional network}
The graph convolutional network architecture (GCN) proposed by~\citet{DBLP:conf/iclr/KipfW17} is one of the pioneer works in graph neural networks. The layer is defined as:
\begin{equation}
    \mathbf{X}^{(l)} = \text{ReLU}(\tilde{\mathbf{A}}\mathbf{X}^{(l-1)}\mathbf{W}^{(l)}),
\end{equation}
where $\tilde{\mathbf{A}} = (\mathbf{D} + \mathbf{I})^{-\frac{1}{2}}(\mathbf{A} + \mathbf{I})(\mathbf{D} + \mathbf{I})^{-\frac{1}{2}}$ is the symmetric normalized adjacency matrix with self-loops obtained through the identity matrix $\mathbf{I} \in \mathbb{R}^{|\mathcal{V}| \times |\mathcal{V}|}$. 

\subsection{Graph attention network}
The graph attention network architecture (GAT) was first introduced by the work of~\citet{DBLP:conf/iclr/VelickovicCCRLB18} and it leverages attention mechanisms to weight the relative importance of each neighbor node to its corresponding ego node. Specifically, the message passing schema of GAT is defined as:
\begin{equation}
    \mathbf{X}^{(l)} = \text{ReLU}(\hat{\mathbf{A}}\mathbf{X}^{(l-1)}\mathbf{W}^{(l)}),
\end{equation}
where $\hat{\mathbf{A}} \in \mathbb{R}^{|\mathcal{V}| \times |\mathcal{V}|}$ is the weight matrix obtained via attention mechanisms. Specifically, by considering the ego node $v \in \mathcal{V}$ along with its neighbor nodes $\forall w \in \mathcal{N}_v$, the attention mechanism is formulated as:
\begin{equation}
    \hat{A}_{vw} = \frac{\text{exp}(\text{LeakyReLU}(\mathbf{T}[\mathbf{W}\mathbf{X}_v^{(l-1)} || \mathbf{W}\mathbf{X}_w^{(l-1)}]))}{\sum_{w \in \mathcal{N}_v} \text{exp}(\text{LeakyReLU}(\mathbf{T}[\mathbf{W}\mathbf{X}_v^{(l-1)} || \mathbf{W}\mathbf{X}_w^{(l-1)}]))},
\end{equation}
where $\text{exp}(\cdot)$ is the exponential function, $\mathbf{T} \in \mathbb{R}^{1 \times 2F_{l-1}}$ is the projection matrix, and $||$ is the concatenation operation.

\section{The link prediction task}
This last part is devoted to presenting a popular task in graph representation learning, namely, link prediction, since recommendation represents one of its possible instantiations. 

As for other tasks in graph representation learning (e.g., node and graph classification), it is possible to formalize link prediction through the \textit{encoder}-\textit{decoder} framework \citep{DBLP:journals/jmlr/ChamiAPR022}. In its most generic version, an \textit{encoder} function maps the adjacency matrix and initial node features of a graph to an embedded representation of the nodes:
\begin{equation}
    \textsc{Encoder: } \mathbb{R}^{|\mathcal{V}| \times |\mathcal{V}|} \times \mathbb{R}^{|\mathcal{V}| \times F_0} \rightarrow \mathbb{R}^{|\mathcal{V}| \times F_L}.
\end{equation}
From the above, the \textit{encoder} function is formulated as any message passing procedure repeated for $L$ propagation layers. Regarding the \textit{decoder} function, it takes the encoded representation of the nodes as input, and produces some outputs which depend on the considered task. 

Link prediction is the task of predicting if an edge exists between any two nodes of a graph. Specifically, we consider a subset of graph edges on which the graph learning model will be trained (i.e., the training set), and mask the remaining ones (i.e., the test set). The graph learning model will be trained to correctly predict the presence and absence of an edge in the training set, to eventually be evaluated on the predictions for the unseen links in the test set. Thus, the decoder function in link prediction is defined as:
\begin{equation}
    \textsc{Decoder: } \mathbb{R}^{|\mathcal{V}| \times F_L} \rightarrow \mathbb{R}^{|\mathcal{V}| \times |\mathcal{V}|}, 
\end{equation}
where we aim to reconstruct the whole adjacency matrix and make predictions on the existence of edges in the graph. Practically speaking, the decoder is generally designed as a function computing all similarities between any pairs of learned node embeddings, for instance through their dot product:
\begin{equation}
    \hat{\mathbf{A}} = \hat{\mathbf{X}}^{\top}\hat{\mathbf{X}}.
\end{equation}
\chapter{The graph CF recommendation pipeline}\label{ch:pipeline}

\begin{sidewaysfigure*}
\centering
\includegraphics[width=\textwidth]{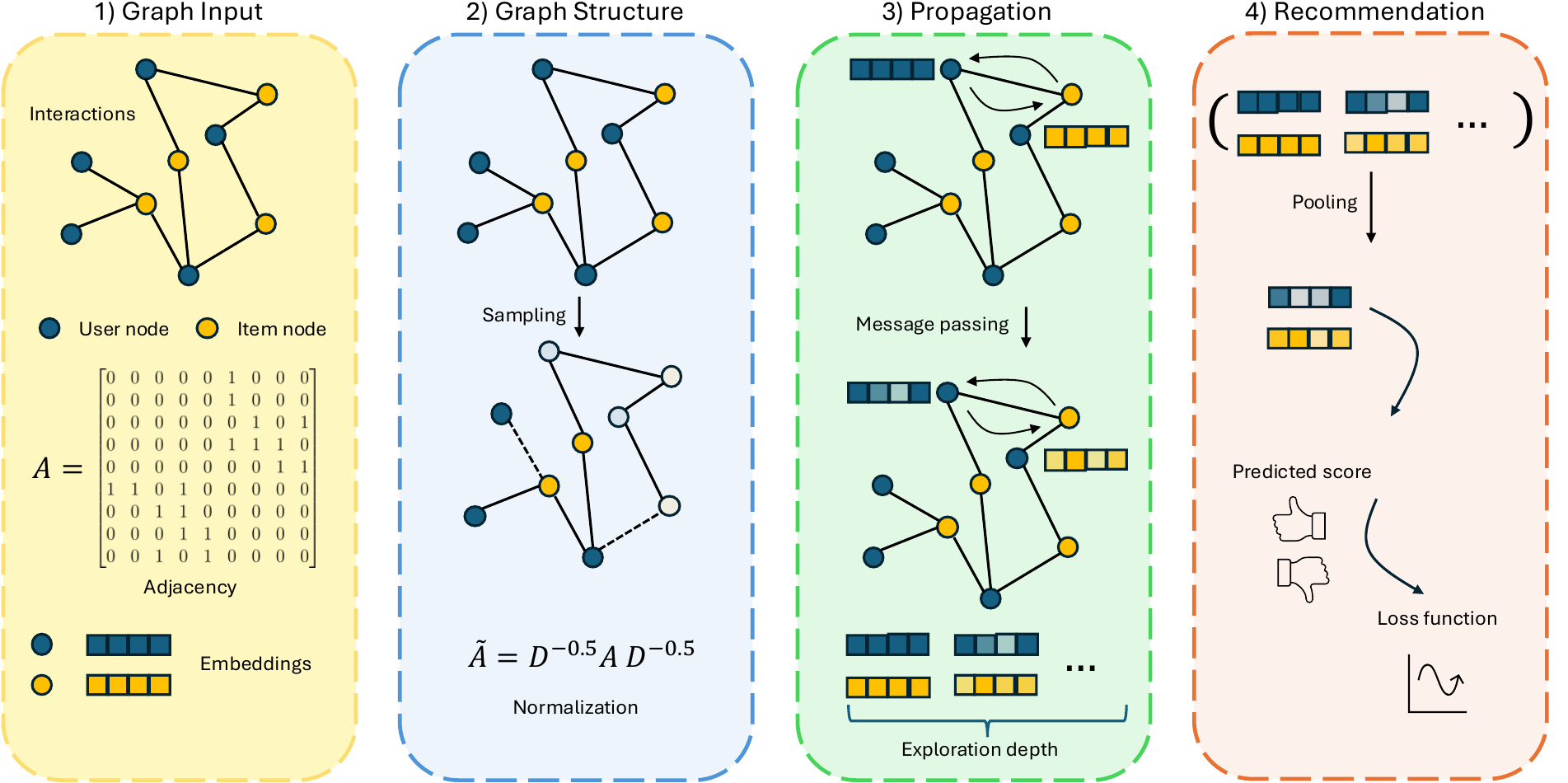}
\caption{Overall graph CF pipeline, consisting of four main steps: 1) graph input, where we consider the user-item graph structure with optional node embeddings; 2) graph structure, where we perform optional modifications to the graph structure to improve its informativeness; 3) propagation, where we optionally propagate messages at multiple hops through the message passing algorithm; 4) recommendation, where we get a compact representation of users' and items' embeddings and exploit them for the final task.
}
\label{fig:pipeline}
\end{sidewaysfigure*}

This section provides the required technical background regarding graph-based collaborative filtering (CF). By considering the main solutions depicted in the literature, we recognize some recurrent strategy patterns that help us designing a general graph CF recommendation pipeline. In the following, we focus on each stage of the pipeline, namely, the graph input data, the processing of the graph structure, the propagation procedures, and the final recommendation task (Figure \ref{fig:pipeline}). Finally, we formally define a model taxonomy over the selected eleven recommendation models. 

\section{Graph input}

The graph input data, in the minimum settings, requires one piece of information: the bipartite user-item interaction graph, formally represented by an adjacency matrix. Optionally, users' and items' nodes may come with (learnable) features, represented as node embeddings. In the following, we formalize both components.

\subsection{Interactions}

Let $\mathcal{U}$ and $\mathcal{I}$ be the sets of users and items, respectively, with $|\mathcal{U}| = U$ and $|\mathcal{I}| = I$ as the total number of users and items in the catalog. Then, let $\mathbf{R} \in \mathbb{R}^{U \times I}$ be the user-item interaction matrix where, assuming an implicit feedback setting, $R_{ui} = 1$ if user $u \in \mathcal{U}$ interacted with item $i \in \mathcal{I}$, 0 otherwise. We build the adjacency matrix $\mathbf{A} \in \mathbb{R}^{(U + I) \times (U + I)}$ which represents the bi-directional connections among users and items according to $\mathbf{R}$:
\begin{equation}
    \mathbf{A} = \begin{bmatrix}
    0 & \mathbf{R} \\
    \mathbf{R}^\top & 0 
    \end{bmatrix}.
\end{equation}
On the above premises, we formally define the user-item bipartite and undirected graph $\mathcal{G} = \{\mathcal{U} \cup \mathcal{I}, \mathbf{A}\}$. We recall that, in a bipartite graph, we see connections existing only among nodes from the two partitions (i.e., users with items and vice versa). However, there exist other types of connections which are not explicitly encoded, such as the user-user connections and the item-item ones. To obtain those, let us first partition the adjacency matrix as follows:
\begin{equation}
    \label{eq:partition}
    \mathbf{A} = \begin{bmatrix}
    \mathbf{A}^{\mathcal{U}\mathcal{U}} & \mathbf{A}^{\mathcal{U}\mathcal{I}} \\
    \mathbf{A}^{\mathcal{I}\mathcal{U}} & \mathbf{A}^{\mathcal{I}\mathcal{I}}
    \end{bmatrix},
\end{equation}
where, in the original user-item graph, $\mathbf{A}^{\mathcal{U}\mathcal{U}} = 0$ and $\mathbf{A}^{\mathcal{I}\mathcal{I}} = 0$. Nevertheless, if we consider the user- and item-\textit{projected} graphs as $\mathcal{G}_\mathcal{U} = \{\mathcal{U}, \mathbf{A}^{\mathcal{UU}}\}$ and $\mathcal{G}_\mathcal{I} = \{\mathcal{I}, \mathbf{A}^{\mathcal{II}}\}$, we can explicitly derive those connections. In this respect, let $\mathbf{R}^{\mathcal{UU}}$ and $\mathbf{R}^{\mathcal{II}}$ be the user-user and item-item interaction matrices:
\begin{equation}
    \mathbf{R}^{\mathcal{UU}} = \mathbf{R} \cdot \mathbf{R}^\top, \qquad \mathbf{R}^{\mathcal{II}} = \mathbf{R}^\top \cdot \mathbf{R},
\end{equation}
which store the co-occurrences among users and items, respectively. Trivially, the corresponding adjacency matrices $\mathbf{A}^{\mathcal{UU}}$ and $\mathbf{A}^{\mathcal{II}}$ are obtained as:
\begin{equation}
    \mathbf{A}^{\mathcal{UU}} = \mathbf{R}^{\mathcal{UU}}, \qquad \mathbf{A}^{\mathcal{II}} = \mathbf{R}^{\mathcal{II}}.
\end{equation}

\subsection{Node representation}

We optionally represent each user and item node with its own features. In this respect, the tendency is to leverage latent factor-based representations, which are carefully initialized and trained end-to-end in the downstream task. Specifically, let $\mathbf{E}_u \in \mathbb{R}^F$ and $\mathbf{E}_i \in \mathbb{R}^F$ be the embedding representations for user $u$ and item $i$, respectively, where $F$ is their vector dimension (i.e., $F << U, I$). Given the iterative nature of several approaches (i.e., message passing, see later), we may need to update these node embeddings for multiple hops. Thus, we will use the notation $\mathbf{E}^{(l)}$ to denote the updated representation of the node latent factor $\mathbf{E}$ at hop $l$ (i.e., $0 \le l \le L$), with $\mathbf{E}^{(0)} = \mathbf{E}$ as the initialization. 

While the most recurrent strategy for node representation is the one described above, it is worth underlining other important solutions from the literature. Indeed, approaches such as~\citep{DBLP:conf/sigir/WangJZ0XC20} disentangle node embeddings into intents; that is, each embedding portion refers to a specific aspect (intent) underlying the user-item interaction. Then, other solutions~\citep{DBLP:conf/cikm/ShenWZSZLL21, DBLP:conf/cikm/PengSM22, DBLP:conf/iclr/Cai0XR23} decompose the interaction matrix $\mathbf{R}$ through, for example, singular value decomposition, to access the intrinsic properties of the user-item interaction structure and embed it into the node representation.

\section{Graph structure}

Noteworthy, we may use some tailored procedures to modify the graph input structure so that the information conveyed by the recorded interactions is more useful to the downstream recommendation task. Specifically, the literature indicates two main strategies, namely, adjacency normalization and graph sampling. In the following, we deepen into both techniques. 

\subsection{Normalization}
\label{sec:normalization}

To limit the over-penalization of nodes with very few interactions in the user-item graph, a common approach is to normalize the adjacency matrix through the symmetric adjacency normalization. Let $\mathcal{N}_u = \{i \in \mathcal{I} \;|\; R_{ui} = 1\}$ and $\mathcal{N}_i = \{u \in \mathcal{U} \;|\; R_{ui} = 1\}$ be the set of one-hop neighbor nodes of $u$ and $i$, respectively, with $|\mathcal{N}_u|$ and $|\mathcal{N}_i|$ their cardinalities (i.e., the number of interactions for user $u$ and item $i$). Then, let $\mathbf{D} \in \mathbb{R}^{(U + I) \times (U + I)}$ be the diagonal degree matrix, where $D_{v,v} = \sum R_{v, *} = |\mathcal{N}_v|$ is the degree of the generic graph node $v$. Similarly to what we did for the adjacency matrix (Equation \ref{eq:partition}), we can partition it as:
\begin{equation}
    \label{eq:partition_degree}
    \mathbf{D} = \begin{bmatrix}
        \mathbf{D}^{\mathcal{UU}} & \mathbf{D}^{\mathcal{UI}} \\
        \mathbf{D}^{\mathcal{IU}} & \mathbf{D}^{\mathcal{II}}
    \end{bmatrix}.
\end{equation}
Then, the symmetric normalization of the adjacency matrix is calculated as:
\begin{equation}
    \tilde{\mathbf{A}} = \mathbf{D}^{-\frac{1}{2}} \mathbf{A} \mathbf{D}^{-\frac{1}{2}},
\end{equation}
where the generic entry is as follows:
\begin{equation}
    \tilde{A}_{u, i} = \frac{1}{\sqrt{|\mathcal{N}_u|}\sqrt{|\mathcal{N}_i|}}.
\end{equation}
This type of normalization prevents nodes with low degree (i.e., few interactions) from being penalized over nodes with high degree (i.e., numerous interactions). Various approaches in the literature~\citep{DBLP:conf/sigir/Wang0WFC19, DBLP:conf/sigir/0001DWLZ020, DBLP:conf/cikm/PengSM22, DBLP:conf/cikm/YangLWYLMY23, DBLP:journals/tkde/YuXCCHY24} adopt the symmetric normalization, with some variants that apply it during the learning of the adjacency matrix~\citep{DBLP:conf/sigir/WangJZ0XC20} and that reformulate it to tackle some issues such as the over-smoothing effect~\citep{DBLP:conf/cikm/MaoZXLWH21}.

\subsection{Sampling}

Graph structure sampling aims at exploring specific paths in the graph structure and/or regularizing the training of the models by encouraging the agreement between different views of the same user-item graph. 

Random walk (i.e., RW) is a stochastic process on graphs that simulates the path that, starting from seed nodes, traverses the neighbors with a uniform probability up to a given depth. As for training regularization~\citep{DBLP:conf/sigir/Wang0WFC19, DBLP:conf/sigir/WuWF0CLX21, DBLP:conf/iclr/Cai0XR23}, there are approaches~\citep{DBLP:conf/www/ShuXLWKM22} which follow a similar rationale to dropout in machine learning, such as node, edge, and message dropout. Node and edge dropout randomly mask nodes and edges in the user-item graph, respectively, with all connected edges and nodes. Concretely, node dropout (i.e., ND) masks rows and/or columns of the adjacency matrix, while edge dropout (i.e., ED) masks specific entries in the adjacency matrix. On the other hand, message dropout (i.e., MD) deals with the application of dropout to propagated messages during the message passing schema. The concept will be clearer in the next section, where we formally introduce the message passing procedure. 

\section{Propagation}
To effectively exploit the user-item graph structure, a popular strategy is to propagate users' and items' node information across the graph topology. While this procedure is commonly known as message passing, each recommendation model may perform it in a different manner, or even replace it with ad-hoc proxies. Additionally, when the message passing occurs, it is interesting to understand up to which depth we want to make the message passing run, as different strategies may heavily impact the final performance. In the next paragraphs, we will define these outlined aspects in the graph CF pipeline. 

\subsection{Message passing}
The message passing algorithm is re-proposed in recommendation to propagate messages coming from the neighborhood of each node at multiple iterative hops, thus effectively distilling the collaborative signal~\citep{DBLP:conf/sigir/Wang0WFC19}. Let $v \in \mathcal{U} \cup \mathcal{I}$ be either a user or item node. The most general formulation for the message passing on node $v$ is:
\begin{equation}
    \label{eq:hop-l}
    \mathbf{E}^{(l)}_v = {\text{Update}^{(l)}}(\mathbf{E}_v^{(l-1)}, \text{Aggregate}^{(l)}(\{\mathbf{E}_w^{(l-1)} \; | \; w \in \mathcal{N}_v\})),
\end{equation}
where $\text{Update}(\cdot)$ and $\text{Aggregate}(\cdot, \cdot)$ are the update and the aggregate functions, respectively. The aggregate function collects the last representations of nodes from the ego neighbors to obtain the messages; when the ego node is a \textit{user}, the messages are aggregated from the directly-connected \textit{items} (those which have been interacted by the user). After that, the update function generates the new version of the ego node representation through the collected messages, optionally combining it with the ego node itself.

While message passing is the standard solution observed in graph CF, we highlight that more recent models provide proxy versions of it~\citep{DBLP:conf/cikm/MaoZXLWH21}, or avoid using it (e.g., singular value decomposition as in~\citep{DBLP:conf/cikm/ShenWZSZLL21, DBLP:conf/cikm/PengSM22}). The reasons for these choices are to be found in the propagation depth, an aspect that we focus on in the following section.

\subsection{Exploration depth}\label{sec:exploration_depth}

Another important aspect is the \textit{depth} of exploration when we perform the message passing algorithm. Indeed, each chosen depth may have a different meaning when it comes to a bipartite graph (as in our case). For instance, a reworking of Equation~\ref{eq:hop-l} on the user side for $l \in \{2, 3\}$ allows \textit{same-} and \textit{different-}type node representation emerge~\citep{DBLP:conf/recsys/AnelliDNSFMP22, DBLP:conf/ecir/AnelliDNMPP23}. For simplicity, we omit the $\text{Update}(\cdot)$ function and simply aggregate embeddings from neighbor nodes:

\begin{align}
\footnotesize
\label{eq:hop-rewrite}
    \begin{aligned}
    \text{\textbf{\textit{Same}}} & 
    \begin{cases} \underbrace{\mathbf{E}_u^{(2)}}_{(\text{\textbf{user}})} &= \text{Aggregate}^{(2)}(\{\text{Aggregate}^{(1)}(\{\underbrace{\mathbf{E}^{(0)}_{u'}}_{\text{\textbf{user}}} \; | \; u' \in \mathcal{N}_i\}) \; | \; i \in \mathcal{N}_u \}),
    \\
    \end{cases}\\
    \text{\textbf{\textit{Diff.}}} & \begin{cases} \underbrace{\mathbf{E}_u^{(3)}}_{(\text{\textbf{user}})} &= \text{Aggregate}^{(3)}(\{\cdots\text{Aggregate}^{(1)}(\{\underbrace{\mathbf{E}_{i'}^{(0)}}_{\text{\textbf{item}}} \; | \; i' \in \mathcal{N}_{u'}\})\cdots \; | \; i \in \mathcal{N}_u \}).
    \\
    \end{cases}
    \end{aligned}
\end{align}

To better clarify the extent of Equation~\ref{eq:hop-rewrite}, after an \textbf{even} and an \textbf{odd} number of explored hops in the user-item bipartite graph, the message passing works by leveraging \textit{same-} and \textit{different-}type node connections (i.e., U-U/I-I and U-I/I-U as evident from Equation~\ref{eq:hop-rewrite}). While the existing literature does not always consider the two scenarios as distinct, it becomes fundamental to assess the influence of different node-node connections explored during the message passing. Formally, we may count the explored hops as follows: $\mathbf{E}^{(2l)}, \forall l \in \{1, 2, \dots, \frac{L}{2}\}$ as obtained through $l$ \textbf{same-}type node connections (denoted as \textit{same-l}), and $\mathbf{E}^{(2l - 1)}, \forall l \in \{1, 2, \dots, \frac{L}{2}\}$ as obtained through $l$ \textbf{different-}type node connections (denoted as \textit{different-l}).

On another level, exploration depth has been shown to negatively affect performance for GNNs due to the over-smoothing effect~\citep{DBLP:conf/iclr/ZhaoA20, DBLP:conf/aaai/ChenLLLZS20}. Particularly, it is commonly observed that performing message passing for a high number of hops may drastically decrease the performance of the GNN, as node embeddings tend to become increasingly similar in the latent space, thus losing informativeness. The same effect occurs in recommendation, where it has been empirically demonstrated that after $l = 3$ explorations, 
performance generally starts to worsen, e.g., \citep{DBLP:conf/sigir/0001DWLZ020, DBLP:conf/sigir/Wang0WFC19}. As this appears to be counteractive to the idea of exploring larger portions of the user-item interaction graph to effectively distill the collaborative signal, some approaches have begun proposing novel techniques that provide proxies to the traditional message passing~\citep{DBLP:conf/cikm/MaoZXLWH21}; the idea is to modify the usual procedure to propagate messages at higher hops and still avoid over-smoothing. 

\section{Recommendation}

For the final stage of the pipeline, we use the updated representations of the users' and items' nodes to predict the user-item interaction score and (optionally) calculate the loss function of the overall framework. Before the score prediction, in the message passing settings, we also provide a summarized version of the node embeddings coming from each propagation step. In the following, we explore these two aspects.

\subsection{Layer combination}

The final stage in the message passing algorithm is the combination of all node embedding representations obtained in each hop $l$. Let $\mathbf{E}^{(0)} = \mathbf{E}$ be the node representation for the hop 0, which corresponds to the initial embedding representation of the node. Then, the layer combination is obtained as:
\begin{equation}
    \hat{\mathbf{E}} = \text{Combine}(\{\mathbf{E}^{(0)}, \; \mathbf{E}^{(1)}, \; \dots, \; \mathbf{E}^{(l)}, \; \dots, \; \mathbf{E}^{(L)}\}),
\end{equation}
where $\text{Combine}(\cdot)$ can be different operations. Possible solutions include: 
\begin{itemize}
    \item element-wise addition ($\sum_{0 \leq l \leq L} \mathbf{E}^{(l)}$)
    \item weighted element-wise addition ($\sum_{0 \leq l \leq L} w_l\mathbf{E}^{(l)}$)
    \item mean ($\frac{1}{L} \sum_{0 \leq l \leq L} \mathbf{E}^{(l)}$)
    \item concatenation ($\{\mathbf{E}^{(0)} \; || \; \mathbf{E}^{(1)} \; || \; \dots \; || \; \mathbf{E}^{(l)} \; \dots \; || \; \mathbf{E}^{(L)}\}$)
    \item taking the last obtained representation ($\mathbf{E}^{(L)}$).
\end{itemize}

\subsection{Loss}

In the case of trainable recommendation models, the final node embeddings are commonly combined through the dot product to predict the interaction score, as observed in traditional approaches for collaborative filtering. Thus, most of the considered models optimize a loss function which, in many cases, adopts the Bayesian personalized ranking (BPR)~\citep{DBLP:conf/uai/RendleFGS09} loss with additional and ad-hoc components. 

Let $\mathcal{T} = \{(u, i, j) \; | \; i \in \mathcal{N}_u \land j \in \mathcal{I} \setminus \mathcal{N}_u\}$ be a set of triples, where $\mathcal{N}_u$ is the set of all positive items for user $u$ (i.e., its one-hop neighbor nodes) and $\mathcal{I} \setminus \mathcal{N}_u$ is the set of all negative items for user $u$. BPR seeks to maximize the posterior probability that each user $u$ prefers a positive item $i$ over a negative item $j$:
\begin{equation}
\begin{split}
    \mathcal{L}_{\text{BPR}} = & - \sum_{(u, i, j) \sim \mathcal{T}} \text{ln } \text{Sigmoid}(\hat{R}_{ui} - \hat{R}_{uj}) = \\
    &- \sum_{(u, i, j) \sim \mathcal{T}} \text{ln } \text{Sigmoid}(\hat{\mathbf{E}}_u^{\top}\hat{\mathbf{E}}_i - \hat{\mathbf{E}}_u^{\top}\hat{\mathbf{E}}_j),
\end{split}
\end{equation}
where $\text{Sigmoid}(\cdot)$ is the sigmoid function, and $\hat{R}_{ui}, \hat{R}_{uj}$ are the predicted scores for the pairs of user $u$ with the positive item $i$, and user $u$ with the negative item $j$. Starting from the BPR basis, some works propose to include additional loss components to reach certain constraints and objectives. We will deepen into them in the last part of this section, where we provide a formal model taxonomy on graph CF based upon the depicted pipeline. 

\section{Model taxonomy}
\label{sec:models_formulation}

Based upon the graph CF recommendation pipeline described above, in conclusion, we report the main technical aspects regarding eleven state-of-the-art graph CF models that we selected for this monograph, namely, NGCF \citep{DBLP:conf/sigir/Wang0WFC19}, DGCF~\citep{DBLP:conf/sigir/WangJZ0XC20}, LightGCN~\citep{DBLP:conf/sigir/0001DWLZ020}, SGL~\citep{DBLP:conf/sigir/WuWF0CLX21}, UltraGCN~\citep{DBLP:conf/cikm/MaoZXLWH21}, GFCF~\citep{DBLP:conf/cikm/ShenWZSZLL21}, SVD-GCN~\citep{DBLP:conf/cikm/PengSM22}, SimGCL \citep{DBLP:conf/sigir/YuY00CN22}, LightGCL \citep{DBLP:conf/iclr/Cai0XR23}, GraphAU \citep{DBLP:conf/cikm/YangLWYLMY23}, and XSimGCL \citep{DBLP:journals/tkde/YuXCCHY24}. Indeed, it allows us to derive a formal model taxonomy representing graph CF, and we depict it in Table~\ref{tab:taxonomy}.

\clearpage
\begin{sidewaystable*}[!t]
\centering
\caption{Model taxonomy as observed on the selected seven representative graph CF approaches from the recent literature. As described in the graph CF pipeline, the taxonomy accounts for four main aspects: 1) graph input, 2) graph structure, 3) propagation, and 4) recommendation. Models follow the chronological order.}
\label{tab:taxonomy}
\begin{adjustbox}{max width=\columnwidth}
\begin{tabular}{l|cc|cc|cc|cc}
\toprule
\multicolumn{1}{c}{\multirow{3}{*}{\textbf{Models}}} & \multicolumn{2}{c}{\textbf{(1) Graph Input}} & \multicolumn{2}{c}{\textbf{(2) Graph Structure}} & \multicolumn{2}{c}{\textbf{(3) Propagation}} & \multicolumn{2}{c}{\textbf{(4) Recommendation}} \\ \cmidrule{2-9}
\multicolumn{1}{c}{} & \multicolumn{1}{c}{\textit{Interactions*}} & \multicolumn{1}{c}{\begin{tabular}{c}\textit{Node}\\\textit{Repr.}\end{tabular}} & \multicolumn{1}{c}{\textit{Norm.}} & \multicolumn{1}{c}{\textit{Sampling**}} & \multicolumn{1}{c}{\begin{tabular}{c}\textit{Message}\\\textit{Passing}\end{tabular}} & \multicolumn{1}{c}{\begin{tabular}{c}\textit{Exploration}\\\textit{Depth}\end{tabular}} & \multicolumn{1}{c}{\textit{Pooling}} & \multicolumn{1}{c}{\textit{Loss***}} \\ \cmidrule{1-9}
\multicolumn{1}{l|}{NGCF} & U-I & latent fact. & \cmark & MD/ND & \cmark & 3 & concat & BPR \\
\multicolumn{1}{l|}{DGCF} & U-I & intents & \cmark & \xmark & \cmark & 1 & sum & BPR/IDL \\
\multicolumn{1}{l|}{LightGCN} & U-I & latent fact. & \cmark & \xmark & \cmark & 3 & w-sum & BPR \\
\multicolumn{1}{l|}{SGL} & U-I & latent fact. & \xmark & ND/ED/RW & \cmark & 3 & w-sum & BPR/SSL \\
\multicolumn{1}{l|}{UltraGCN} & U-I/I-I & latent fact. & \cmark & \xmark & \cmark & $+\infty$ & \xmark & BPR/CL/IL \\
\multicolumn{1}{l|}{GFCF} & U-I/I-I & decomp. & \xmark & \xmark & \xmark & 1 & \xmark & \xmark \\
\multicolumn{1}{l|}{SVD-GCN} & U-I/U-U/I-I & decomp. & \cmark & \xmark & \xmark & 1 & \xmark & BPR/UL/IL\\
SimGCL & U-I & latent fact. & \cmark & \xmark & \cmark & 3 & avg & BPR/CL/UNL\\
LightGCL & U-I & latent fact./decomp. & \cmark & ND/ED & \cmark & 2 & sum & HL/CL \\
GraphAU & U-I & latent fact. & \cmark & \xmark & \cmark & 4 & w-sum & AL/UNL \\
XSimGCL & U-I & latent fact. & \cmark & \xmark & \cmark & 5 & w-sum & BPR/CL \\
\bottomrule
\multicolumn{9}{l}{\footnotesize\textit{*U-I: user-item, I-I: item-item, U-U: user-user}}\\
\multicolumn{9}{l}{\footnotesize\textit{**MD: message dropout, ND: node dropout, ED: edge dropout, RW: random walk}}\\
\multicolumn{9}{l}{\footnotesize\textit{***BPR: Bayesian personalized ranking, IDL: independence loss, SSL: self-supervised loss}}\\
\multicolumn{9}{l}{\textit{\footnotesize\phantom{***}CL: constraint loss, UL: user loss, IL: item loss, UNL: uniform loss, HL: hinge loss, AL: alignment loss}}
\end{tabular}
\end{adjustbox}
\end{sidewaystable*}
\clearpage

\subsection{NGCF} The authors from~\citep{DBLP:conf/sigir/Wang0WFC19} propose neural graph collaborative filtering (NGCF). At each hop, the model aggregates neighborhood messages and the inter-dependencies among the ego and the neighborhood nodes. The message passing is formalized as:
\begin{equation}
\footnotesize
    \label{eq:hop-l-ngcf}
    \mathbf{E}_u^{(l)} = \text{LeakyReLU}\left(\sum_{i \in \mathcal{N}_u}\left(\mathbf{W}_{\text{neigh}}^{(l)}\tilde{A}_{ui}\mathbf{E}_i^{(l - 1)} + \mathbf{W}_{\text{inter}}^{(l)}\left(\mathbf{E}^{(l - 1)}_i\odot\mathbf{E}^{(l - 1)}_u\right) + \mathbf{b}^{(l)}\right)\right),
\end{equation}
where LeakyReLU is the activation function, $\mathbf{W}^{(l)}_{\text{neigh}} \in \mathbb{R}^{F_{l - 1} \times F_l}$ and $\mathbf{W}^{(l)}_{\text{inter}} \in \mathbb{R}^{F_{l - 1} \times F_l}$ are the neighborhood and inter-dependencies weight matrices, respectively, $\mathbf{b}^{(l)} \in \mathbb{R}^{F_l}$ is the bias term, while $\odot$ is the Hadamard product. 

\subsection{DGCF} \cite{DBLP:conf/sigir/WangJZ0XC20} design a message passing that calculates the importance of neighborhood nodes for ego nodes by disentangling the intents underlying each user-item interaction. Therfore, they propose the following formulation:
\begin{equation}
    \mathbf{E}^{(l)}_{u}[t] = \sum\limits_{i \in \mathcal{N}_u[t]} \frac{A_{ui}[t]}{\sqrt{|\mathcal{N}_u[t]|}\sqrt{|\mathcal{N}_i[t]|}} \mathbf{E}_{i}^{(l - 1)}[t],
\end{equation}
where the operator $[t]$ indicates we are considering embeddings, matrices, and sets with reference to the t-\textit{th} intent, having $1 \leq t \leq T$. Then, assuming the chosen intents should be independent, the authors also introduce an independence loss (i.e., IDL) formulated as:
\begin{equation}
    \mathcal{L}_{\text{IDL}} = \sum_{t = 1}^{T} \sum_{t'=t+1}^{T} \text{dCor}(\hat{\mathbf{E}}[t], \hat{\mathbf{E}}[t']),
\end{equation}
where $\text{dCor}()$ is a distance correlation function calculated among all pairs of intents in the setting. 

\subsection{LightGCN} \cite{DBLP:conf/sigir/0001DWLZ020} propose a light graph convolutional network, dubbed as LightGCN, to simplify the message passing algorithm from NGCF. The idea is to drop feature transformations (i.e., the weight matrices and biases) and the non-linearity applied after the message aggregation. The final formulation is:
\begin{equation}
    \label{eq:hop-l-lightgcn}
    \mathbf{E}_u^{(l)} = \sum_{i \in \mathcal{N}_u}\tilde{A}_{ui}\mathbf{E}_{i}^{(l - 1)}.
\end{equation}

\subsection{SGL} Self-supervised graph learning~\citep{DBLP:conf/sigir/WuWF0CLX21} (SGL) brings self-supervised~\citep{DBLP:conf/cikm/HuangXW0Y22} and contrastive~\citep{DBLP:conf/nips/KhoslaTWSTIMLK20} learning to graph CF. It uses the same message passing formulation as LightGCN, but learning different views of nodes; these are obtained through sampling strategies on the adjacency matrix (i.e., ND, ED, and RW). Concretely, we have:
\begin{equation}
    \mathbf{E}_u^{(l)}[t_1] = \sum_{i \in \mathcal{N}_u} \tilde{A}_{ui}[t_1] \mathbf{E}^{(l -1)}_i[t_2], \quad \mathbf{E}_u^{(l)}[t_2] = \sum_{i \in \mathcal{N}_u} \tilde{A}_{ui}[t_2] \mathbf{E}^{(l-1)}_i[t_2],   
\end{equation}
where, with a notation abuse, we re-use the same operator [t] from DGCF to indicate the two different views for the ego node, obtained through two sampling strategies. Then, the idea is to encourage the similarity between the two views of the same node, and the dissimilarity between the two views of different nodes:
\begin{equation}
    \mathcal{L}_{\text{SSL}} = \sum_{u \in \mathcal{U}} - \text{ln} \frac{\text{exp}(\text{Cosine}(\hat{\mathbf{E}}_u[t_1], \hat{\mathbf{E}}_u[t_2]))}{\sum_{v \in \mathcal{U} \setminus \{u\}} \text{exp}(\text{Cosine}(\hat{\mathbf{E}}_u[t_1], \hat{\mathbf{E}}_v[t_2]))},
\end{equation}
where the self-supervised loss function (i.e., SSL) is only shown on the user side, but the symmetric holds for the item side, while $\text{Cosine}(\cdot)$ is the cosine similarity. This formulation of the $\mathcal{L}_{\text{SSL}}$ is better known as the InfoNCE loss \citep{DBLP:journals/jmlr/GutmannH10}.

\subsection{UltraGCN} Ultra simplification of graph convolutional network \citep{DBLP:conf/cikm/MaoZXLWH21} (UltraGCN) addresses some crucial issues in graph CF. Specifically, the authors propose a novel message passing algorithm that works as a proxy to the infinite-layer propagation through a single (simplified) hop. Moreover, the adjacency matrix is normalized through a novel version of the symmetric normalization accounting for the asymmetric weighting of connected nodes in U-U and I-I connections. Thus, the single-hop message passing formulation is:
\begin{equation}
    \hat{\mathbf{E}}_u = \sum_{i \in \mathcal{N}_u} \tilde{A}'_{ui} \mathbf{E}_{i}, \quad
    \hat{\mathbf{E}}_i = \sum\limits_{u \in \mathcal{N}_i^{(1)}} \tilde{A}'_{iu} \; \mathbf{E}_{u},
\end{equation}
where we do not use any superscript as the message passing occurs in one iteration only, and $\tilde{\mathbf{A}}'$ is the novel version of the symmetric normalization as described above. Then, two loss components are introduced to tackle the over-smoothing effect and learn from the usually-unexplored type of node relationships such as item-item. The former is called constraint loss (i.e., CL), while the latter is called item loss (i.e., IL). In detail, we have:
\begin{equation}
\begin{split}
    \mathcal{L}_{\text{CL}} = & - \sum\limits_{(u,i) \sim \mathcal{T}_+} t_+ \text{ln}(\text{Sigmoid}(\hat{\mathbf{E}}_u^{\top} \hat{\mathbf{E}}_i)) \; + \\
    & - \sum\limits_{(u, j) \sim \mathcal{T}_-} t_- \text{ln}(\text{Sigmoid}(-\hat{\mathbf{E}}_u^\top \hat{\mathbf{E}}_j)),
\end{split}
\end{equation}
where $\mathcal{T}_+ = \{(u, i) \; | \; R_{ui} = 1\}$ and $\mathcal{T}_- = \{(u, j)\; | \; R_{uj} = 0\}$, $t_+, t_-$ (again, a notation abuse) are node degree-based coefficients calculated on the user-item graph, and $\text{Sigmoid}(\cdot)$ is the sigmoid function. Regarding the item loss, it is calculated by considering the item-projected graph, thus it is formalized as follows:
\begin{equation}
    \mathcal{L}_{\text{IL}} = - \sum\limits_{(u, i) \sim \mathcal{T}_+} \text{ } \sum\limits_{j \in \text{topK}(\mathcal{T}_+^{\mathcal{II}})} t_-^{\mathcal{II}} \text{ln}(\text{Sigmoid}(\hat{\mathbf{E}}_u^\top \hat{\mathbf{E}}_j)),
\end{equation}
where topK is the function returning the top $K$ values of a set, $\mathcal{T}_+^{\mathcal{II}} = \{(i, j) \; | \; A^{\mathcal{II}}_{ij} \neq 0\}$, and $t_-^{\mathcal{II}}$ is a node degree-based coefficient calculated on the item-item graph. 

\subsection{GFCF} Graph filter-based collaborative filtering~\citep{DBLP:conf/cikm/ShenWZSZLL21} (GFCF) questions the role of graph convolutional network into recommendation by leveraging graph signal processing theory. Indeed, the authors show that several existing approaches in CF may fall into one unified framework based upon graph convolution, to eventually propose a closed-form algorithm that proves to be a strong baseline against other trainable and computationally-expensive (graph-based) approaches in CF.
\begin{equation}
\footnotesize
    \mathbf{\hat{R}}_{u*} = \mathbf{R}_{u*}\left(\left(\tilde{\mathbf{A}}^{\mathcal{UI}}\right)^{\top}\tilde{\mathbf{A}}^{\mathcal{IU}} + t \left(\mathbf{D}^{\mathcal{II}}\right)^{-\frac{1}{2}}\text{topK}(\mathbf{U})\text{topK}(\mathbf{U})^{\top}\left(\mathbf{D}^{\mathcal{II}}\right)^{+\frac{1}{2}}\right),
\end{equation}
where we are re-using the notation we introduced in Equation~\ref{eq:partition} and Equation~\ref{eq:partition_degree} to indicate different partitions of the (normalized) adjacency matrix and degree matrix, while $t$ is a tunable coefficient, and $\mathbf{U}$ is the vector collecting the singular vectors of $\tilde{\mathbf{A}}^{\mathcal{UI}}$.

\subsection{SVD-GCN} Similarly to what was observed with GFCF, the model proposed by~\citep{DBLP:conf/cikm/PengSM22} leverages the conceptual similarities between graph convolution and singular value decomposition (SVD) to design a novel approach which also explores user-user and item-item co-occurrences. Concretely, the authors propose a light-weight trainable recommender model, where user and item embeddings are obtained as:
\begin{equation}
    \hat{\mathbf{E}}_u = (\mathbf{U}) \text{exp}(t\mathbf{V})\mathbf{W},
\end{equation}
where $\mathbf{U}$ is the vector collecting the singular vectors of $\tilde{\mathbf{A}}^{\mathcal{UI}}$ (as in GFCF), $t$ is a tunable parameter, $\mathbf{V}$ is vector of the largest singular values on $\tilde{\mathbf{A}}^{\mathcal{UI}}$, and $\mathbf{W}$ is a trainable matrix. Finally, the model optimizes the following user (item) loss function (i.e., UL/IL) on the user-user (item-item) co-occurrences graph (here, we only report the user side):
\begin{equation}
    \mathcal{L}_{\text{UL}} = \sum_{u \in \mathcal{U}} \text{ } \sum_{(u, v) \sim \mathcal{T}^{\mathcal{UU}}_+, \; (u, w) \sim \mathcal{T}^{\mathcal{UU}}_-} \text{ln} \; \text{Sigmoid}(\hat{\mathbf{E}}_u^{\top} \hat{\mathbf{E}}_v - \hat{\mathbf{E}}_u^{\top} \hat{\mathbf{E}}_w),
\end{equation}
where $\mathcal{T}_+^{\mathcal{UU}} = \{(u, v) \; | \; A_{uv}^{\mathcal{UU}} \neq 0\}$ and $\mathcal{T}_-^{\mathcal{UU}} = \{(u, w) \; | \; A_{uw}^{\mathcal{UU}} = 0\}$. 

\subsection{SimGCL}
The authors in \citep{DBLP:conf/sigir/YuY00CN22} empirically observe that graph augmentations, such as those proposed in \citep{DBLP:conf/sigir/WuWF0CLX21}, are not the actual reason behind the improved recommendation performance, but rather uniformity of the node embeddings' distribution. Thus, they propose SimGCL, short for Simplification of Graph Contrastive Learning, a novel approach that simplifies graph contrastive learning by regulating the uniformity. Concretely, they propose to perturb the node embeddings during the message passing procedure in this way:
\begin{equation}
\begin{aligned}
    & \mathbf{E}_u^{(l)}[t_1] = \sum_{i \in \mathcal{N}_u} \tilde{A}_{ui}\mathbf{E}^{(l -1)}_i[t_1] + \Delta^{(l)}[t_1] \\
    & \mathbf{E}_u^{(l)}[t_2] = \sum_{i \in \mathcal{N}_u} \tilde{A}_{ui} \mathbf{E}^{(l-1)}_i[t_2] + \Delta^{(l)}[t_2], 
\end{aligned}
\end{equation}
where $||\Delta^{(l)}[t_*]||_2 = \epsilon$, $\Delta^{(l)}[t_*] = \bar{\Delta}^{(l)}[t_*] \odot\text{sign}(\mathbf{E}_i^{(l-1)})$, and $\bar{\Delta}^{(l)}[t_*] \sim \text{uniform}(0,1)$. Finally, the model optimizes a composite loss with BPR and a CL (as in SGL), along with a uniform loss (UNL) to ensure the uniformity of the learned representations:
\begin{equation}
    \mathcal{L}_{\text{UNL}} = \text{ln}\left(\underset{u,v}{\mathbb{E}}\left[\text{exp}\left(-2\left\lVert\frac{\mathbf{E}_u}{||{\mathbf{E}_u}||_2} - \frac{\mathbf{E}_v}{||{\mathbf{E}_v}||_2}\right\rVert_2^2\right)\right]\right),
\end{equation}
where $u,v$ are iid from the set of most popular items in the catalog. 

\subsection{LightGCL} 

The authors in \citep{DBLP:conf/iclr/Cai0XR23} propose a novel approach named LightGCL, a simple yet effective graph contrastive learning solution. The framework consists of a double message passing procedure acting in the local and global view of the user-item interaction graph. As for the local view, the model adopts the same propagation layer proposed in \citep{DBLP:conf/sigir/XiaHXZYH22}, namely:
\begin{equation}
    \mathbf{E}_u^{(l)} = \text{LeakyReLU}\left( \text{drop}(\tilde{\mathbf{A}}_{u*})\mathbf{E}^{(l -1)}_i\right) + \mathbf{E}^{(l -1)}_i,
\end{equation}
where drop() corresponds to the edge dropout function to avoid overfitting, and a residual connection with the previous representation of the node embedding is adopted. Conversely, at global view, the authors propose to approximate the adjacency matrix through its SVD, and in particular the approximated SVD where top \textit{K} singular values are retained:
\begin{equation}
    \mathbf{G}_u^{(l)} = \text{LeakyReLU}\left( \hat{\mathbf{A}}_{u*}\mathbf{E}^{(l -1)}_i\right) = \text{LeakyReLU}\left(\hat{\mathbf{U}}_{K}\hat{\mathbf{S}}_{K}\hat{\mathbf{V}}_{K}^{\top}\mathbf{E}^{(l -1)}_i\right),
\end{equation}
where $\hat{\mathbf{U}}_K \in \mathbb{R}^{U \times K}$ and $\hat{\mathbf{V}}_K \in \mathbb{R}^{I \times K}$ are the eigenvectors of the adjacency matrix when considering the top \textit{K} largest columns of both vectors, and $\hat{\mathbf{S}}_K \in \mathbb{R}^{K \times K}$ is the diagonal matrix of the top \textit{K} largest eigenvalues. On such a basis, a contrastive loss is defined between the local and global views of the same node embeddings, obtaining:
\begin{equation}
    \mathcal{L}_{\text{CL}} = \sum_{u \in \mathcal{U}} \; \sum_{l \in \{0, \dots, L\}}- \text{ln} \frac{\text{exp}(\text{Sim}(\mathbf{E}^{(l)}_u, \mathbf{G}^{(l)}_u) / t)}{\sum_{v \in \mathcal{U} \setminus \{u\}} \text{exp}(\text{Sim}(\mathbf{E}^{(l)}_u, \mathbf{G}^{(l)}_v) / t)},
\end{equation}
where $\mathbf{E}_*^{(l)}$ and $\mathbf{G}_*^{(l)}$ refer to the local and global views computed at layer $l$, respectively, while $t$ is the common temperature coefficient to rescale the similarity function $\text{Sim}(\cdot)$. Note that, in the above equation, the model performs node dropout to select specific nodes for the calculation to prevent overfitting. Alongside the contrastive loss, the model pursues the recommendation task by optimizing the hinge pairwise loss:  
\begin{equation}
    \mathcal{L}_{\text{HL}} = \sum_{u \in \mathcal{U}} \; \sum_{(u, i, j) \in \mathcal{T}} \text{max}\left(0, 1 - \hat{\mathbf{E}}_u^\top \hat{\mathbf{E}}_i\right).
\end{equation}

\subsection{GraphAU}

The authors in \citep{DBLP:conf/cikm/YangLWYLMY23} leverage the rationale from \citep{DBLP:conf/kdd/WangYM000M22} to adapt the uniformity and alignment paradigm for graph-based recommendation within their GraphAU framework. First, they refine the node embeddings for users and items through the LightGCN message passing layer. On such a basis, the overall loss is calculated as a trade-off between the alignment (AL) and uniformity (UNL) losses, where the former is specifically tailored in this work for graph learning:
\begin{equation}
    \mathcal{L} = \frac{1}{|\mathcal{T}|} \sum_{(u,i) \in \mathcal{T}} \mathcal{L}_{\text{AL}} + \frac{t_1}{2} \mathcal{L}_{\text{UNL}},
\end{equation}
where the AL is calculated as:
\begin{equation}
    \mathcal{L}_{\text{AL}} = \sum_{l \in \{0, \dots, L\}}\frac{(t_2)^{l}}{2}\left(||{\mathbf{E}_u^{(0)}} - \mathbf{E}_i^{(l)}||^2 + ||{\mathbf{E}_i^{(0)}} - \mathbf{E}_u^{(l)}||^2 \right).
\end{equation}
Note that, in the above equations, $t_1$ and $t_2$ correspond to the trade-off coefficient between the AL and the UNL and the importance weight of the AL at each propagation layer, respectively.

\subsection{XSimGCL}

Unlike previous contrastive learning approaches (such as SimGCL), the model proposed in \citep{DBLP:journals/tkde/YuXCCHY24}, named XSimGCL (short for extremely simple graph contrastive learning), suggests using the same contrasted view of the node embeddings also for the recommendation task. In addition to this, the authors adopt a cross-layer contrastive module which calculates the InfoNCE between the same node view calculated at different propagation layers. First, the model computes the augmented version of the embedding at layer $l$ as in SimGCL, but only for one view $t$:
\begin{equation}
    \mathbf{E}_u^{(l)}[t] = \sum_{i \in \mathcal{N}_u} \tilde{A}_{ui}\mathbf{E}^{(l -1)}_i[t] + \Delta^{(l)}[t].
\end{equation}
Then, after $L$ propagation layers, the loss is calculated as the combination of BPR and the InfoNCE where the contrasting views are obtained from the same augmented node embedding but at layers $L$ and at a specific layer $l^* < L$:
\begin{equation}
\begin{aligned}
    \mathcal{L}_{\text{BPR}} & = - \sum_{(u, i, j) \in \mathcal{T}} \text{ln } \text{Sigmoid}(\hat{\mathbf{E}}_u[t]^{\top} \hat{\mathbf{E}}_i[t] - \hat{\mathbf{E}}_u[t]^{\top} \hat{\mathbf{E}}_j[t]), \\
    \mathcal{L}_{\text{CL}} & =  \sum_{i \in \mathcal{I}}- \text{ln} \frac{\text{exp}(\text{Sim}(\hat{\mathbf{E}}_i, \mathbf{E}^{(l^*)}_i))}{\sum_{j \in \mathcal{I} \setminus \{i\}} \text{exp}(\text{Sim}(\hat{\mathbf{E}}_i, \mathbf{E}^{(l^*)}_j))}.
\end{aligned}
\end{equation}
\chapter{Classical dataset characteristics}\label{ch:char_class}

The pioneering work by \citet{DBLP:journals/tmis/AdomaviciusZ12} first demonstrated the existence of a correlation between the intrinsic characteristics of recommender system datasets and the final performance achieved by learning algorithms. 
This finding opened the possibility of anticipating algorithmic behavior based on \textbf{dataset properties}, even before model training takes place. 
Before delving into the topological characteristics, which are the focus of the following section, we first introduce the so-called \textbf{classical dataset characteristics}, that is, those already well established in the literature and whose effectiveness has been consistently demonstrated across a variety of recommendation settings~\citep{DBLP:conf/sigir/DeldjooNSM20, DBLP:journals/ipm/DeldjooBN21, DBLP:conf/recsys/MalitestaPAMNS24}.

\section{Space size}
The space size quantifies the total number of possible user–item interactions in a dataset.
When comparing two datasets with the same number of observed interactions, the one with a larger space size exhibits a lower probability of observing a link between a given user and item.
Consequently, the space size provides an estimate of the potential concentration of interactions among users and items.
It is formally defined as:

\begin{equation}
\label{eq:spacesize}
\Psi = U * I,
\end{equation}
where U and I denote the total number of users and items, respectively.
The metric ranges from 0 (no possible interactions) to $+\infty$ as the dataset size grows.

\section{Shape}
While the space size captures the magnitude of the interaction space, it does not provide any information about the balance between the number of users and items.
Indeed, two datasets with the same space size may differ substantially in their composition; for instance, one may have many users and few items, while the other exhibits the opposite pattern.
This relationship is expressed by the shape, defined as:

\begin{equation}
\Phi = \frac{U}{I}.
\end{equation}

A user–item ratio of 1 indicates a perfect balance between the number of users and items.
Values close to 0 correspond to datasets with relatively few users and many items, whereas very high (potentially unbounded) values indicate the opposite configuration.

\section{Density and sparsity}\label{sec:density}
Data sparsity is one of the defining characteristics of recommendation datasets, as the recommendation task typically involves suggesting items from a large catalog with which each user has interacted only to a limited extent.
A highly sparse dataset conveys less information than a denser one, given the same space size.
In such sparse settings, the probability that two users (or items) share interactions with the same items (or users) decreases, which in turn weakens the effectiveness of collaborative filtering approaches.

To define data sparsity, it is first needed to introduce data density, which measures the proportion of existing interactions out of all possible ones:

\begin{equation}
\Delta = \frac{E}{U * I},
\end{equation}
where $E = |\{(u, i)\;|\;\mathbf{R}_{u, i} = 1\}|$ is the number of interactions existing among users and items in the recommendation data.

Since the denominator represents the total number of possible interactions, the density can equivalently be expressed as:

\begin{equation}
\Delta = \frac{R}{\Psi},
\end{equation}
where $\Psi$ denotes the space size introduced in Equation \ref{eq:spacesize}.
The density ranges from 0 (maximum sparsity) to 1 (maximum density).
Accordingly, sparsity is defined as the complement of density:

\begin{equation}
\Lambda = 1 - \Delta.
\end{equation}

\section{Gini coefficient}
\label{sec:gini}

The Gini coefficient measures the inequality in the frequency distribution of users or items. In practice, it quantifies how unevenly users and items are exposed within a dataset. This metric is particularly relevant in recommendation scenarios, where exposure is typically far from uniform: a small set of highly popular items often dominates the catalog, while the majority of users interact with only a limited subset of items.

The user- and item-level Gini coefficients are defined as follows:
\begin{equation}
    \text{K}_{\mathcal{U}} = 
    \frac{\sum\limits_{u = 1}^{U-1}\sum\limits_{v = u + 1}^{U} |\sigma_u - \sigma_v|}
         {U \sum\limits_{u = 1}^{U} \sigma_u},
    \qquad
    \text{K}_{\mathcal{I}} = 
    \frac{\sum\limits_{i = 1}^{I-1}\sum\limits_{j = i + 1}^{I} |\sigma_i - \sigma_j|}
         {I \sum\limits_{i = 1}^{I} \sigma_i},
\end{equation}
where $U$ and $I$ denote the number of users and items, respectively, and $\sigma_k$ represents the interaction frequency of user $u$ or item $i$. The operator $|\cdot|$ denotes the absolute value.
A Gini coefficient close to 0 corresponds to a perfectly balanced distribution, whereas a value close to 1 indicates a highly unbalanced one.

\section{Empirical evaluation of classical characteristics}
\label{sec:empirical_classical}

To provide an indication of the typical ranges of these metrics and to highlight their variability across datasets, we compute all characteristics for the examples introduced in~\Cref{tab:classic_char}. The datasets are ordered by their number of interactions, from the largest (\textit{MovieLens 1M}) to the smallest (\textit{LastFM}).

The \textbf{space size} ($\Psi$) exhibits substantial variation, spanning more than two orders of magnitude, reflecting the different scales of the underlying user–item interaction spaces. As expected from its definition, this metric is positively related to density: larger interaction spaces tend to be associated with lower densities, particularly in highly sparse, real-world settings such as \textit{Yelp} and Amazon-based datasets. Notably, large values of $\Psi$ can also have practical implications for the size and complexity of recommendation models, especially for embedding-based architectures in which the number of users and items directly determines the dimensionality, and therefore the memory footprint, of the embedding tables.

The \textbf{shape metric} ($\Phi$) also reveals domain-dependent structural differences. Values greater than one indicate that the number of users exceeds the number of items, a condition observed in most datasets, sometimes by a factor of two. \textit{Amazon Books} constitutes an exception, with a shape below one, suggesting a catalog-heavy configuration. Such contrasts align with the intrinsic nature of each domain: e-commerce collections typically feature extensive item catalogs, whereas media- or location-oriented platforms tend to exhibit a more balanced or user-heavy structure.

The dataset \textbf{density} ($\Delta$) is uniformly low, ranging from 0.003 to 0.060, and reflects the well-known sparsity characteristic of recommendation data. \textit{MovieLens 1M} appears as an outlier with a substantially higher density, which contributes to its reputation as a comparatively “easy” benchmark and partly explains the strong performance typically reported on this dataset relative to more sparse alternatives \citep{DBLP:journals/tois/FanJZS24, DBLP:conf/um/AnelliBNJP22}.

The \textbf{Gini coefficients} ($\text{K}_{\mathcal{U}}$, $\text{K}_{\mathcal{I}}$) further illuminate differences in interaction distributions. Inequality levels are generally higher for items than for users, consistent with longstanding observations regarding popularity concentration. A notable case is \textit{LastFM}, where the item-side Gini is substantially higher than the user-side value. This indicates that artists are far more unevenly exposed than users, with a small number of very popular artists attracting a disproportionate share of interactions. This behavior is characteristic of music consumption, where a limited set of artists typically dominates listening patterns across the platform.

Finally, several domain-level patterns emerge. In this sense, location-based (\textit{Gowalla}) and music-oriented (\textit{LastFM}) datasets tend to exhibit moderate densities and shape values above one. E-commerce datasets (\textit{Amazon Beauty} and \textit{Amazon Books}) are extremely sparse and display moderate inequality patterns. \textit{Yelp} combines one of the largest interaction spaces with one of the lowest densities, underscoring the complexity of review-based platforms. In contrast, \textit{MovieLens 1M} stands out as structurally more compact, denser, and more skewed, reinforcing its role as a convenient yet less representative benchmark for contemporary recommendation research.

\begin{table*}[t]
\centering
\footnotesize
\caption{Calculated classical characteristics for our selected recommendation datasets.}
\label{tab:classic_char}
\begin{tabular}{lccccc}
\toprule
\textbf{Dataset} & 
\textbf{$\Psi$} &
\textbf{$\Phi$} &
\textbf{$\Delta$} &
\textbf{$\text{K}_{\mathcal{U}}$ } &
\textbf{$\text{K}_{\mathcal{I}}$ } \\
\midrule
MovieLens 1M & 1.65$\; \cdot\;10^7$  & 1.914  & 0.060  & 0.507  & 0.548   \\
Yelp      & 1.83$\; \cdot\;10^8$ & 1.177  & 0.004  & 0.341  & 0.376   \\
Amazon Beauty & 9.70$\; \cdot\;10^7$  & 1.327  & 0.018  & 0.254  & 0.332   \\
Amazon Books & 7.95$\; \cdot\;10^6$  & 0.732  & 0.003  & 0.408  & 0.304   \\
Gowalla & 1.52$\; \cdot\;10^6$  & 1.105  & 0.025  & 0.327  & 0.369   \\
LastFM & 5.33$\; \cdot\;10^6$  & 1.683  & 0.035  & 0.155  & 0.484   \\
\bottomrule
\end{tabular}
\end{table*}
\chapter{Topological dataset characteristics}\label{ch:char_topo}

Graph neural networks (GNNs) are designed to represent recommendation data as a bipartite, undirected graph. These graphs possess inherent topological properties that can and often are leveraged by GNN-based recommenders to enhance their representation learning capabilities.

This hypothesis is further supported by recent literature on representation learning that examines the role of graph structure (encompassing both topology and node features) in the performance of GNN models~\citep{DBLP:conf/icml/WuSZFYW19, DBLP:conf/www/WeiZH22, DBLP:conf/www/0002ZPNG0CH22, DBLP:conf/iclr/KlicperaBG19, DBLP:conf/iclr/ZhaoA20, DBLP:conf/icml/Abu-El-HaijaPKA19,  DBLP:conf/nips/LuanZCP19, DBLP:conf/icml/XuLTSKJ18, DBLP:conf/iclr/RongHXH20, DBLP:journals/corr/abs-2308-09570}. Additionally, other studies have established connections between dataset characteristics and recommender system performance, reinforcing the plausibility of this relationship~\citep{DBLP:journals/tmis/AdomaviciusZ12, DBLP:conf/sigir/DeldjooNSM20, DBLP:conf/recsys/MalitestaPAMNS24}.
 
This section provides a formal definition of four topological properties in graphs~\citep{DBLP:journals/socnet/LatapyMV08, PhysRevE.67.026126, PhysRevE.67.026112}: \textit{node degree}, \textit{clustering coefficient}, \textit{degree assortativity}, and \textit{degree distribution}. Then, it reinterprets these properties under the lens of recommender systems.

\section{Node degree} 
Given a generic user $u$ and item $i$, let 
$\mathcal{N}_u = \{i \in \mathcal{I} \;|\; R_{ui} = 1\}$ and $\mathcal{N}_i = \{u \in \mathcal{U} \;|\; R_{ui} = 1\}$ denote their  respective neighborhood sets, as previously defined in Section~\ref{sec:normalization}. This definition naturally extends to $l$-hops, where $\mathcal{N}^{(l)}_u$ and $\mathcal{N}^{(l)}_i$ are the neighboring nodes of the user $u$ and the item $i$, respectively, at $l$-hops distance. Then, the user and item \textbf{average node degrees} are defined as follows:
\begin{equation}
    \sigma_\mathcal{U} = \frac{1}{U} \sum_{u \in \mathcal{U}} \sigma_u, \qquad \sigma_\mathcal{I} = \frac{1}{I} \sum_{i \in \mathcal{I}} \sigma_i,
\end{equation}
where $\sigma_u = |\mathcal{N}_u^{(1)}|$ and $\sigma_i = |\mathcal{N}_i^{(1)}|$ denote the node degrees of user $u$ and item $i$, respectively.

\begin{reinterpretation}[Node degree]
In the user-item graph, the node degree represents the number of items a user has interacted with or the number of users who have interacted with an item. 
It serves as a measure of user activity and item popularity within the dataset. 
A low node degree indicates a potential cold-start issue in recommendation, where cold users exhibit low engagement on the platform, and cold items correspond to niche products.
\end{reinterpretation}

\begin{tcolorbox}[colback=black!5!white,colframe=black!75!black,title=\textsc{\bfseries Node Degree in RecSys}]
  \textit{A graphical representation of the node degree interpretation in the user-item graph. On the left, the user node degree is a measure of user activity. On the right, the item node degree distinguishes popular from niche items.}
  
\begin{minipage}{0.45\textwidth}
\vspace{1.5em}
    \centering
    \includegraphics[width=\linewidth]{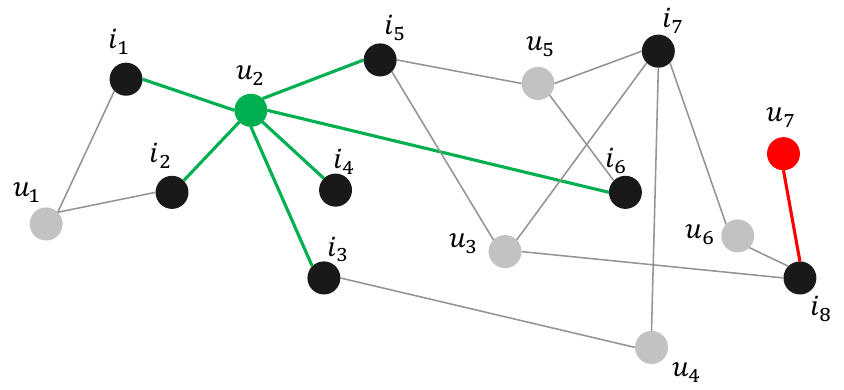} 
    {\footnotesize (a) \textcolor{forestgreen(web)}{\textbf{active}} user \textit{vs.} \textcolor{red}{\textbf{inactive}} user}
\end{minipage}
\hfill
\begin{minipage}{0.45\textwidth}
\vspace{1.5em}
    \centering
    \includegraphics[width=\linewidth]{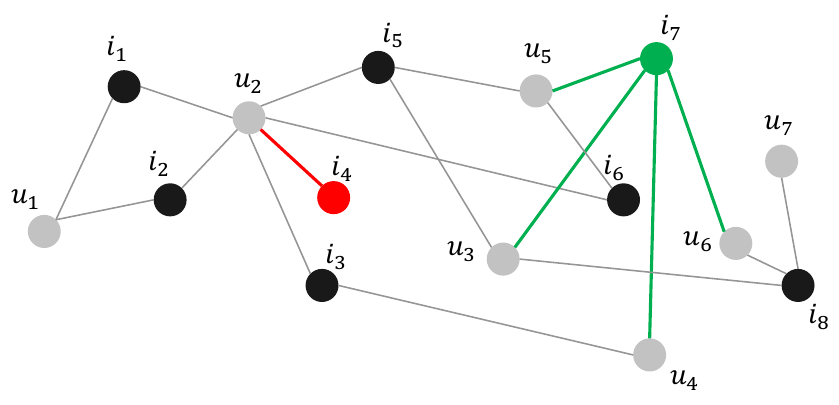}
    {\footnotesize (b) \textcolor{forestgreen(web)}{\textbf{popular}} item \textit{vs.} \textcolor{red}{\textbf{niche}} item}
\end{minipage}
  
\end{tcolorbox}

\section{Clustering coefficient} 
For each partition in a bipartite graph, it is important to recognize nodes with overlapping neighborhoods. Regardless of their size, these overlaps reveal clusters of interconnected nodes.
Let $v$ and $w$ be two nodes belonging to the same partition (e.g., user nodes). Their similarity can be measured using the intersection-over-union ratio of their neighborhoods~\citep{DBLP:journals/socnet/LatapyMV08}.
Formally, the clustering coefficient is defined as follows:
\begin{equation}
    \gamma_{v} = \frac{\sum_{w \in \mathcal{N}^{(2)}_v}\gamma_{v, w}}{|\mathcal{N}_v^{(2)}|}, \qquad \text{with} \quad  \gamma_{v, w} = \frac{|\mathcal{N}^{(1)}_{v} \cap \mathcal{N}^{(1)}_{w}|}{|\mathcal{N}^{(1)}_{v} \cup \mathcal{N}^{(1)}_{w}|},
\end{equation}
where, if $v, w\in \mathcal{U}$ are user nodes, we have $\mathcal{N}_v^{(2)} = \{w \in \mathcal{U} \; | \; \exists i \in \mathcal{I}, i \in \mathcal{N}_v \; \land \; i \in \mathcal{N}_w\}$ as the second-order neighborhood set of $v$. Considering second-order neighborhoods involves measuring similarities between nodes that are connected at a distance of (a multiple of) 2 hops. In a bipartite graph, this corresponds to identifying clusters of nodes within the same partition.
Then, the \textbf{average clustering coefficient} on the node sets $\mathcal{U}$ and $\mathcal{I}$ is defined as follows:
\begin{equation}
    \gamma_{\mathcal{U}} = \frac{1}{U} \sum_{u \in \mathcal{U}} \gamma_u, \qquad \gamma_{\mathcal{I}} = \frac{1}{I} \sum_{i \in \mathcal{I}} \gamma_i.
\end{equation}

\begin{reinterpretation}[Clustering coefficient]
The clustering coefficient quantifies the extent to which a user (or item) node shares interactions with neighboring users (or items). High clustering coefficient values indicate a significant number of co-occurrences among nodes within the same partition. For instance, when considering the user-side formulation, the average clustering coefficient increases when multiple users interact with the same set of items. This intuition aligns with the core principle of collaborative filtering: users with overlapping interactions are likely to exhibit similar preferences. 
\end{reinterpretation}

\begin{tcolorbox}[colback=black!5!white,colframe=black!75!black,title=\textsc{\bfseries Clustering Coefficient in RecSys}]
\textit{Graphical representation of the clustering coefficient interpretation in the user-item graph. On the left, an example illustrates two similar users sharing the $67\%$ of the interacted items. On the right, two neighboring users exhibit different preferences.}

\begin{minipage}{0.45\textwidth}
\vspace{1.5em}
    \centering
    \includegraphics[width=\linewidth]{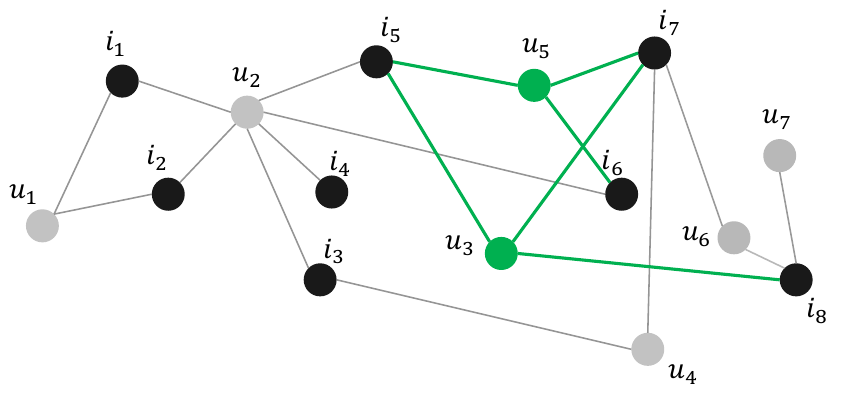} 
    {\footnotesize (a) \textcolor{forestgreen(web)}{\textbf{similar}} user activity ($\gamma_{u_3, u_5} = 0.667$)}
\end{minipage}
\hfill
\begin{minipage}{0.45\textwidth}
\vspace{1.5em}
    \centering
    \includegraphics[width=\linewidth]{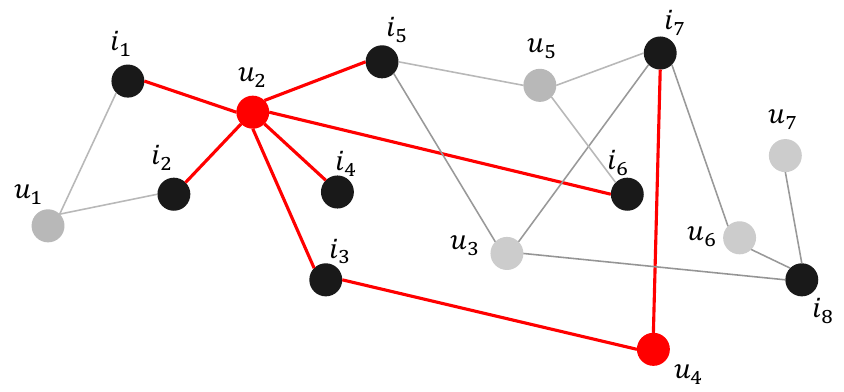}
    {\footnotesize (b) \textcolor{red}{\textbf{different}} user activity ($\gamma_{u_2, u_4} = 0.167$)}
\end{minipage}

\end{tcolorbox}

The clustering coefficient accounts for nodes at 2 distance hops. Nevertheless, we may want to capture properties for even more extended regions of the graph. For this reason, we introduce another topological characteristic that goes beyond the 2-hop distance.   

\section{Degree assortativity}
\label{sec:assortativity} 
In graph structures, nodes sharing similar properties show a tendency to group, and the assortativity quantifies this tendency. 
The definition of assortativity depends on the definition of similarity between nodes one decides to adopt. Therefore, diverse formulations of assortativity exist~\citep{PhysRevE.67.026126}. 
In this work, we focus on the assortativity coefficient based on node degree similarities.

Let $\mathcal{D} = \{d_1, d_2, \dots\}$ represent the set of distinct node degree values in the graph, and let $e_{d_h, d_k}$ denote the proportion of edges linking nodes with degrees $d_h$ and $d_k$. We define $q_{d_h}$ ($q_{d_k}$) as the probability of reaching a node with degree $d_h$ ($d_k$) when traversing from another node with the same degree (i.e., the excess degree distribution). On such a basis, the \textbf{degree assortativity} is given by:
\begin{equation}
    \rho = \frac{\sum\limits_{d_h, d_k}d_h d_k(e_{d_h, d_k} - q_{d_h} q_{d_k})}{std^2_q},
\end{equation}
where $std_q$ represents the standard deviation of the distribution $q$. Due to its formulation, the degree assortativity resembles a correlation measure, such as the Pearson correlation. 
Degree assortativity ranges from $-1$ to $1$, where $1$ indicates the highest similarity and $-1$ the lowest. When the value is below zero, we refer to it as \textit{disassortativity}.

Similarly to the clustering coefficient, we aim to identify similarity patterns among nodes within the same partition. To achieve this, we first project the user-item bipartite graph into the user-user $\mathcal{G}_{\mathcal{U}}$ and the item-item graph $\mathcal{G}_{\mathcal{I}}$. We then compute the degree assortativity for $\mathcal{G}_{\mathcal{U}}$ and $\mathcal{G}_{\mathcal{I}}$, denoted as $\rho_{\mathcal{U}}$ and $\rho_{\mathcal{I}}$, respectively.

\begin{reinterpretation}[Degree assortativity]
The degree assortativity calculated user- and item-wise represents the tendency of users with the same activity level on the platform and items with the same popularity to gather, respectively. Since we calculate the degree assortativity on the complete user-user and item-item co-occurrence graphs, we deem this characteristic to provide a broader view of the dataset than the clustering coefficient. For this reason, we borrow the concept of search space traversal depth in search algorithms theory to give an intuition of degree assortativity. That is, we re-interpret degree assortativity in recommendation as a topological characteristic showing a strong look-ahead nature.
\end{reinterpretation}

\begin{tcolorbox}[colback=black!5!white,colframe=black!75!black,title=\textsc{\bfseries Degree Assortativity in RecSys}]
\textit{Graphical representation of the degree assortativity interpretation in the user-user graph. In this example, user nodes with the same degrees do not gather close in the user-user graph, thus resulting in a low value of degree assoratitvity}.

\begin{minipage}{0.45\textwidth}
\vspace{1.5em}
    \centering
    \includegraphics[width=\linewidth]{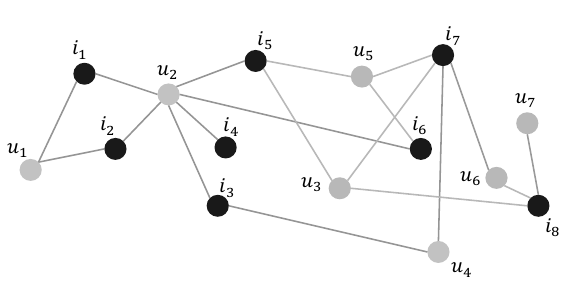} 
    {\footnotesize (a) original user-item graph}
\end{minipage}
\hfill
\begin{minipage}{0.45\textwidth}
\vspace{1.5em}
    \centering
    \includegraphics[width=\linewidth]{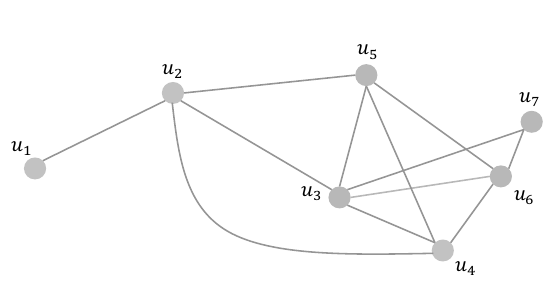}
    {\footnotesize (b) (\textcolor{red}{\textbf{dis}})assortativity ($\rho$ = -0.191)}
\end{minipage}

\end{tcolorbox}

\section{Degree distribution}

A network can be characterized through its degree distribution, defined as the probability distribution of all observed node degrees in the graph. Once again, let $\mathcal{D} = \{d_1, d_2, \dots\}$ be the set of distinct node degree values in the recommendation system, for both users and items. Then, let $n_d = |\{v \in \mathcal{U} \; \cup \; \mathcal{I} \;|\; \sigma_v = d \}|$ be the number of nodes (either user or item) that has degree $d$. The \textbf{degree distribution} of degree $d$ is given by the following formula:
\begin{equation}
    P(d) = \frac{n_d}{U + I}.
\end{equation}

By considering the degree distribution of many real-world networks, it is commonly observed that they resemble the tendency of \textit{scale-free} networks~\citep{PhysRevE.67.026112}: many nodes with very low degrees and few nodes with high degrees (also called \textit{hubs}). The degree distribution of scale-free networks asymptotically follows the power law:
\begin{equation}
    P(d) \sim d^{-\theta}, \text{ with } 2 < \theta < 3.
\end{equation}

As observable in many recommendation datasets, the user-item graph follows this same tendency and can also be considered a \textit{scale-free} network. This trend suggests some important properties of the user-item graph that can be re-interpreted through the lens of recommendation. 

\begin{reinterpretation}[Degree distribution]
The \textit{scale-free} tendency of the recommendation datasets implies some important properties. Many users and items exhibit very few interactions, while a few nodes (\textit{hubs}) have numerous interactions, indicating they are highly active (in the case of users) and highly popular (in the case of items). As observed in other scale-free networks, this means that hubs will generally increase in their degree as the user-item graph evolves over time, leading to growing disparity between users and items (and sparsity of the dataset). All such properties are linked to the tendency of recommendation models to favor highly active users and highly popular items. 
\end{reinterpretation}

\section{Empirical evaluation of topological characteristics}

As previously done for the classical dataset characteristics (Section \ref{sec:empirical_classical}), here we analyze and discuss possible value trends in the topological dataset characteristics for our selected datasets. The computed values are reported in~\Cref{tab:topo_char}, where datasets are sorted by their number of interactions, from the largest (\textit{MovieLens 1M}) to the smallest (\textit{LastFM}) one.

In terms of \textbf{node degree} ($\sigma$), we observe almost aligned trends across all datasets, where the number of recorded interactions on the user and item side generally settles around 30-60. A slight difference is evident if considering the user and item levels separately, since the number of interactions on the item side are usually higher than on the user side. This is a clear indication of how items' popularity in the recommendation catalogs tend to be a prevalent phenomenon, representing a strong factor towards recommending most popular items (quite common for many recommendation models). The only deviation from the trend is represented by \textit{MovieLens 1M} dataset, where the node degree is substantially higher than that of the other datasets. Indeed, as already evidenced for the classical dataset characteristics, this user-item graph appears to be extremely dense.

Regarding the \textbf{clustering coefficient} ($\gamma$), values are (again) quite aligned across datasets. Moreover, and unlike the node degree case, here we observe alignment also between the user and the item side, suggesting similar trends on the two levels. The highest values are measured for \textit{MovieLens 1M} and \textit{LastFM}, indicating that users (items) might have overlapping interactions in these two datasets. Put into other words: a certain group of users have interacted with the same items, and vice versa. We explain this trend by considering, once again, the density of these two datasets: indeed, the two user-item graphs show the most dense topological structures. Thus, it is reasonable to assume that providing high-quality recommendations in these two cases is easier than for the other datasets, actually and effectively exploiting the collaborative signal during the models' training.

Some interesting insights are derived for the \textbf{degree assortativity} property ($\rho$). As observed, many computed values are negative, suggesting a dis-assortativity behavior. That is, users with the same activity levels do not tend to gather close in the user-user graph; the same can be affirmed for items and their popularity levels. A possible explanation for this phenomenon is as follows. When a new user enters the platform, she will likely tend to follow the preference patterns expressed by other highly-active users, which represent (with a certain degree of approximation) the average preferences in the whole community. Likewise, new items that are introduced into the platform will likely follow the interaction trends of already popular items in the system. The only exception to the rule is observed for \textit{Yelp}, where the degree assortativity on both user and item sides is positive and strongly (if compared to the other values) assortative. 

\begin{figure}[!t]
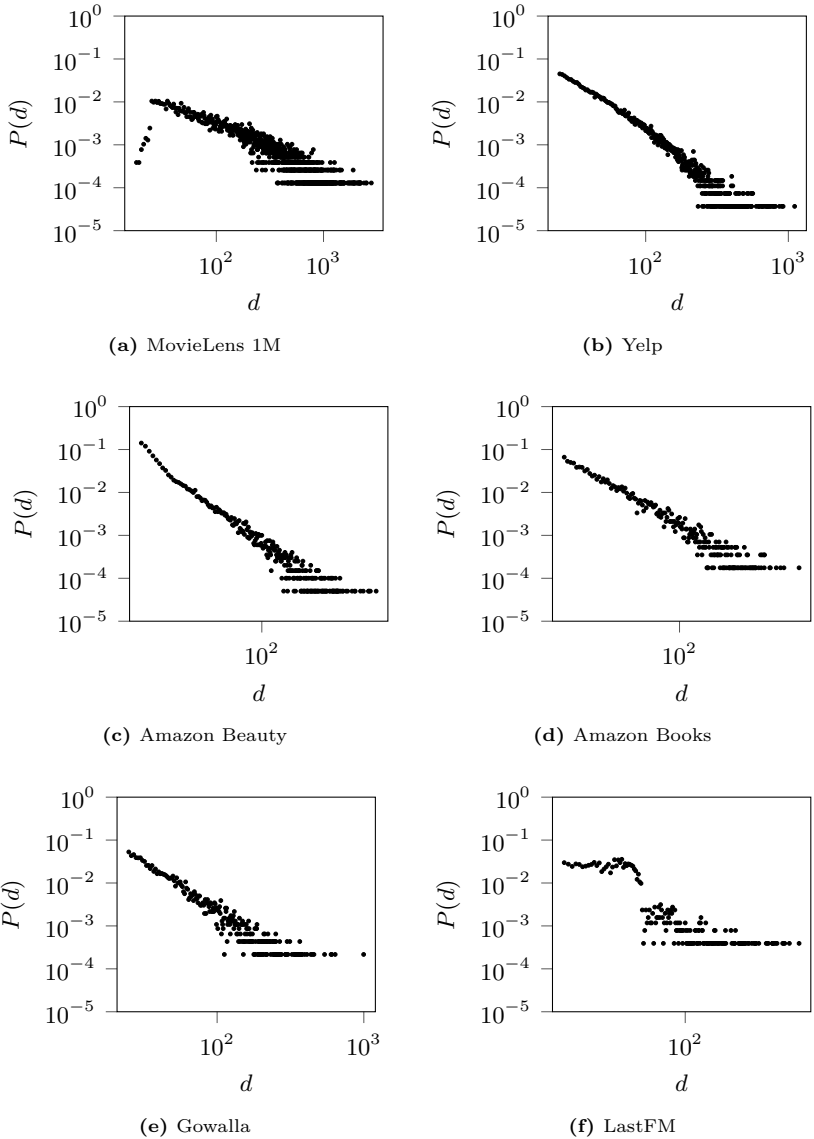

\centering

\begin{subfigure}{0.47\textwidth}
    \centering
    \input{figures/scale_free_movielens}
    \caption{MovieLens 1M}
\end{subfigure}
\begin{subfigure}{0.47\textwidth}
    \centering
    \input{figures/scale_free_yelp}
    \caption{Yelp}
\end{subfigure}

\vspace{0.35cm}

\begin{subfigure}{0.47\textwidth}
   \centering
    \begin{tikzpicture}

\definecolor{black}{RGB}{0,0,0}
\definecolor{dimgray85}{RGB}{85,85,85}
\definecolor{gainsboro220}{RGB}{220,220,220}
\definecolor{green01270}{RGB}{0,127,0}

\begin{axis}[
tick label style = {font=\footnotesize}, 
label style      = {font=\footnotesize},
axis background/.style={fill=white},
scale=0.5,
axis line style={black},
log basis x={10},
log basis y={10},
tick align=outside,
tick pos=left,
xmin=12.4683232539308, xmax=727.844459529795,
xmode=log,
xtick style={color=dimgray85},
xlabel={$d$},
xtick={0.1,1,10,100,1000,10000,100000},
xticklabels={
  \(\displaystyle {10^{-1}}\),
  \(\displaystyle {10^{0}}\),
  \(\displaystyle {10^{1}}\),
  \(\displaystyle {10^{2}}\),
  \(\displaystyle {10^{3}}\),
  \(\displaystyle {10^{4}}\),
  \(\displaystyle {10^{5}}\)
},
ymin=1e-05, ymax=1,
ymode=log,
ylabel={$P(d)$},
ytick style={color=dimgray85},
ytick={1e-06,1e-05,0.0001,0.001,0.01,0.1,1,10},
yticklabels={
  \(\displaystyle {10^{-6}}\),
  \(\displaystyle {10^{-5}}\),
  \(\displaystyle {10^{-4}}\),
  \(\displaystyle {10^{-3}}\),
  \(\displaystyle {10^{-2}}\),
  \(\displaystyle {10^{-1}}\),
  \(\displaystyle {10^{0}}\),
  \(\displaystyle {10^{1}}\)
}
]
\addplot [draw=black, fill=black, mark size=0.7pt, mark=*, only marks]
table{%
x  y
605 5.01655463027992e-05
552 5.01655463027992e-05
540 5.01655463027992e-05
536 5.01655463027992e-05
494 5.01655463027992e-05
473 5.01655463027992e-05
435 5.01655463027992e-05
431 5.01655463027992e-05
414 5.01655463027992e-05
386 5.01655463027992e-05
378 5.01655463027992e-05
359 5.01655463027992e-05
357 5.01655463027992e-05
347 5.01655463027992e-05
342 0.000100331092605598
340 5.01655463027992e-05
333 5.01655463027992e-05
331 5.01655463027992e-05
329 5.01655463027992e-05
328 5.01655463027992e-05
327 5.01655463027992e-05
324 5.01655463027992e-05
323 5.01655463027992e-05
320 5.01655463027992e-05
319 0.000100331092605598
314 5.01655463027992e-05
313 5.01655463027992e-05
309 5.01655463027992e-05
307 5.01655463027992e-05
306 5.01655463027992e-05
305 0.000100331092605598
304 5.01655463027992e-05
302 5.01655463027992e-05
300 0.000100331092605598
296 0.000100331092605598
292 5.01655463027992e-05
291 5.01655463027992e-05
286 0.000100331092605598
278 5.01655463027992e-05
277 5.01655463027992e-05
274 5.01655463027992e-05
272 5.01655463027992e-05
270 0.000100331092605598
269 5.01655463027992e-05
266 5.01655463027992e-05
265 0.000100331092605598
264 5.01655463027992e-05
260 5.01655463027992e-05
259 5.01655463027992e-05
258 5.01655463027992e-05
257 0.000100331092605598
256 0.000100331092605598
252 5.01655463027992e-05
251 0.000100331092605598
246 5.01655463027992e-05
244 0.000150496638908398
243 0.000150496638908398
242 5.01655463027992e-05
241 5.01655463027992e-05
239 5.01655463027992e-05
237 0.000150496638908398
236 0.000150496638908398
235 5.01655463027992e-05
234 0.000100331092605598
233 5.01655463027992e-05
231 5.01655463027992e-05
230 5.01655463027992e-05
229 0.000100331092605598
227 5.01655463027992e-05
224 5.01655463027992e-05
223 0.000100331092605598
222 0.000100331092605598
220 5.01655463027992e-05
219 5.01655463027992e-05
218 0.000100331092605598
217 5.01655463027992e-05
216 0.000200662185211197
215 5.01655463027992e-05
214 0.000150496638908398
213 5.01655463027992e-05
209 0.000100331092605598
207 5.01655463027992e-05
206 5.01655463027992e-05
205 5.01655463027992e-05
204 5.01655463027992e-05
203 5.01655463027992e-05
202 0.000100331092605598
201 0.000150496638908398
200 0.000100331092605598
198 0.000150496638908398
197 5.01655463027992e-05
196 5.01655463027992e-05
194 0.000100331092605598
193 5.01655463027992e-05
192 5.01655463027992e-05
191 0.000250827731513996
190 0.000150496638908398
189 0.000200662185211197
188 0.000200662185211197
187 5.01655463027992e-05
186 5.01655463027992e-05
185 5.01655463027992e-05
184 0.000100331092605598
183 0.000100331092605598
182 0.000100331092605598
181 0.000150496638908398
179 0.000150496638908398
178 0.000150496638908398
177 0.000250827731513996
176 0.000100331092605598
175 0.000100331092605598
174 0.000150496638908398
173 0.000300993277816795
172 0.000250827731513996
171 0.000150496638908398
170 5.01655463027992e-05
169 0.000150496638908398
168 0.000250827731513996
167 0.000100331092605598
166 5.01655463027992e-05
165 0.000100331092605598
164 0.000100331092605598
163 0.000150496638908398
161 0.000401324370422394
160 0.000150496638908398
159 0.000300993277816795
158 0.000351158824119595
157 0.000150496638908398
156 0.000200662185211197
155 0.000150496638908398
154 0.000100331092605598
153 0.000300993277816795
152 0.000250827731513996
151 0.000150496638908398
150 5.01655463027992e-05
149 0.000200662185211197
148 0.000100331092605598
147 0.000351158824119595
146 0.000100331092605598
145 0.000300993277816795
144 0.000401324370422394
143 0.000200662185211197
142 0.000300993277816795
141 5.01655463027992e-05
140 0.000300993277816795
139 0.000351158824119595
138 0.000300993277816795
137 0.000100331092605598
136 0.000250827731513996
135 0.000250827731513996
134 0.000300993277816795
133 0.000300993277816795
132 0.000451489916725193
131 0.000351158824119595
130 0.000401324370422394
129 0.000300993277816795
128 0.000300993277816795
127 0.000551821009330792
126 0.000351158824119595
125 0.000250827731513996
124 0.000250827731513996
123 0.000451489916725193
122 0.000451489916725193
121 0.000501655463027992
120 0.000601986555633591
119 0.000702317648239189
118 0.000601986555633591
117 0.000501655463027992
116 0.000150496638908398
115 0.000351158824119595
114 0.000401324370422394
113 0.000752483194541989
112 0.000501655463027992
111 0.000451489916725193
110 0.000300993277816795
109 0.000752483194541989
108 0.000351158824119595
107 0.000551821009330792
106 0.000702317648239189
105 0.000551821009330792
104 0.000551821009330792
103 0.000551821009330792
102 0.000501655463027992
101 0.000551821009330792
100 0.00100331092605598
99 0.00065215210193639
98 0.00100331092605598
97 0.000601986555633591
96 0.000601986555633591
95 0.00105347647235878
94 0.000953145379753185
93 0.000601986555633591
92 0.000852814287147587
91 0.00110364201866158
90 0.000752483194541989
89 0.000601986555633591
88 0.00065215210193639
87 0.00105347647235878
86 0.00115380756496438
85 0.000953145379753185
84 0.00110364201866158
83 0.00120397311126718
82 0.000852814287147587
81 0.00155513193538678
80 0.000902979833450386
79 0.00135446975017558
78 0.00205678739841477
77 0.00185612521320357
76 0.00105347647235878
75 0.00190629075950637
74 0.00125413865756998
73 0.00120397311126718
72 0.00135446975017558
71 0.00145480084278118
70 0.00220728403732317
69 0.00115380756496438
68 0.00205678739841477
67 0.00185612521320357
66 0.00195645630580917
65 0.00245811176883716
64 0.00170562857429517
63 0.00175579412059797
62 0.00260860840774556
61 0.00225744958362597
60 0.00235778067623156
59 0.00240794622253436
58 0.00270893950035116
57 0.00311026387077355
56 0.00295976723186515
55 0.00336109160228755
54 0.00331092605598475
53 0.00331092605598475
52 0.00366208488010434
51 0.00396307815792114
50 0.00356175378749875
49 0.00451489916725193
48 0.00356175378749875
47 0.00436440252834353
46 0.00436440252834353
45 0.00486605799137153
44 0.00461523025985753
43 0.00511688572288552
42 0.00591953446373031
41 0.0065716865656667
40 0.0066720176582723
39 0.0064211899267583
38 0.00742450085281429
37 0.00802648740844788
36 0.00797632186214508
35 0.011237082371827
34 0.00998294371425705
33 0.0108357580014046
32 0.0116885722885522
31 0.0120397311126718
30 0.013996187418481
29 0.0145480084278118
28 0.0158021470853818
27 0.0171064512892545
26 0.0178087689374937
25 0.0193137353265777
24 0.0227249924751681
23 0.0255844286144276
22 0.0326076050968195
21 0.0372730009029798
20 0.0468044547005117
19 0.0570382261462827
18 0.0712350757499749
17 0.091200963178489
16 0.119594662385873
15 0.141416675027591
};
\end{axis}

\end{tikzpicture}
    \caption{Amazon Beauty}
\end{subfigure}
\begin{subfigure}{0.47\textwidth}
   \centering
    \begin{tikzpicture}

\definecolor{black}{RGB}{0,0,0}
\definecolor{dimgray85}{RGB}{85,85,85}
\definecolor{gainsboro220}{RGB}{220,220,220}
\definecolor{green01270}{RGB}{0,127,0}

\begin{axis}[
tick label style = {font=\footnotesize}, 
label style      = {font=\footnotesize},
axis background/.style={fill=white},
scale=0.5,
axis line style={black},
log basis x={10},
log basis y={10},
tick align=outside,
tick pos=left,
xmin=21.7081545230072, xmax=484.840845814192,
xmode=log,
xtick style={color=dimgray85},
xlabel={$d$},
xtick={0.1,1,10,100,1000,10000,100000},
xticklabels={
  \(\displaystyle {10^{-1}}\),
  \(\displaystyle {10^{0}}\),
  \(\displaystyle {10^{1}}\),
  \(\displaystyle {10^{2}}\),
  \(\displaystyle {10^{3}}\),
  \(\displaystyle {10^{4}}\),
  \(\displaystyle {10^{5}}\)
},
ymin=1e-05, ymax=1,
ymode=log,
ylabel={$P(d)$},
ytick style={color=dimgray85},
ytick={1e-06,1e-05,0.0001,0.001,0.01,0.1,1,10},
yticklabels={
  \(\displaystyle {10^{-6}}\),
  \(\displaystyle {10^{-5}}\),
  \(\displaystyle {10^{-4}}\),
  \(\displaystyle {10^{-3}}\),
  \(\displaystyle {10^{-2}}\),
  \(\displaystyle {10^{-1}}\),
  \(\displaystyle {10^{0}}\),
  \(\displaystyle {10^{1}}\)
}
]
\addplot [draw=black, fill=black, mark size=0.7pt, mark=*, only marks]
table{%
x  y
421 0.000175438596491228
351 0.000175438596491228
337 0.000175438596491228
320 0.000175438596491228
298 0.000175438596491228
296 0.000175438596491228
278 0.000350877192982456
277 0.000175438596491228
276 0.000350877192982456
270 0.000350877192982456
267 0.000175438596491228
265 0.000350877192982456
259 0.000175438596491228
250 0.000175438596491228
249 0.000175438596491228
247 0.000175438596491228
243 0.000175438596491228
242 0.000175438596491228
239 0.000175438596491228
236 0.000175438596491228
234 0.000175438596491228
231 0.000175438596491228
227 0.000350877192982456
226 0.000175438596491228
225 0.000350877192982456
221 0.000175438596491228
220 0.000175438596491228
218 0.000526315789473684
215 0.000175438596491228
213 0.000175438596491228
210 0.000175438596491228
208 0.000175438596491228
206 0.000350877192982456
203 0.000175438596491228
202 0.000350877192982456
201 0.000175438596491228
200 0.000350877192982456
199 0.000526315789473684
197 0.000175438596491228
196 0.000526315789473684
195 0.000526315789473684
194 0.000175438596491228
193 0.000175438596491228
192 0.000175438596491228
191 0.000175438596491228
189 0.000175438596491228
188 0.000175438596491228
187 0.000526315789473684
186 0.000350877192982456
184 0.000175438596491228
183 0.000175438596491228
182 0.000175438596491228
181 0.000175438596491228
180 0.000526315789473684
178 0.000350877192982456
177 0.000526315789473684
176 0.000175438596491228
175 0.000701754385964912
174 0.000350877192982456
172 0.000175438596491228
171 0.000526315789473684
170 0.000175438596491228
169 0.000526315789473684
168 0.000175438596491228
167 0.000526315789473684
166 0.000526315789473684
165 0.000526315789473684
164 0.000350877192982456
163 0.000175438596491228
162 0.000350877192982456
161 0.000701754385964912
160 0.000526315789473684
159 0.000350877192982456
158 0.000350877192982456
157 0.000350877192982456
156 0.000526315789473684
155 0.000701754385964912
154 0.000526315789473684
153 0.000350877192982456
152 0.000350877192982456
151 0.000175438596491228
150 0.000526315789473684
149 0.000350877192982456
148 0.000175438596491228
147 0.00105263157894737
146 0.000350877192982456
144 0.00105263157894737
143 0.000526315789473684
142 0.000701754385964912
141 0.000175438596491228
140 0.000175438596491228
139 0.000175438596491228
138 0.000701754385964912
137 0.000526315789473684
136 0.00087719298245614
135 0.000526315789473684
134 0.000701754385964912
133 0.00105263157894737
132 0.000526315789473684
131 0.000350877192982456
130 0.0012280701754386
129 0.00157894736842105
128 0.000701754385964912
127 0.000526315789473684
126 0.00175438596491228
125 0.000701754385964912
124 0.000350877192982456
123 0.00105263157894737
122 0.00087719298245614
121 0.000701754385964912
120 0.00140350877192982
119 0.0012280701754386
118 0.00087719298245614
117 0.00105263157894737
116 0.00105263157894737
115 0.0012280701754386
114 0.00140350877192982
113 0.00105263157894737
112 0.00087719298245614
111 0.000701754385964912
110 0.000701754385964912
109 0.00140350877192982
108 0.00140350877192982
107 0.00245614035087719
106 0.00087719298245614
105 0.00087719298245614
104 0.00192982456140351
103 0.00140350877192982
102 0.00157894736842105
101 0.00157894736842105
100 0.00263157894736842
99 0.00192982456140351
98 0.00245614035087719
97 0.00192982456140351
96 0.00175438596491228
95 0.0012280701754386
94 0.00210526315789474
93 0.00157894736842105
92 0.00210526315789474
91 0.00315789473684211
90 0.00210526315789474
89 0.00210526315789474
88 0.00385964912280702
87 0.00421052631578947
86 0.00263157894736842
85 0.00210526315789474
84 0.00333333333333333
83 0.00157894736842105
82 0.00333333333333333
81 0.00315789473684211
80 0.00263157894736842
79 0.00385964912280702
78 0.00280701754385965
77 0.00333333333333333
76 0.00315789473684211
75 0.00508771929824561
74 0.00473684210526316
73 0.00456140350877193
72 0.00508771929824561
71 0.00578947368421053
70 0.00526315789473684
69 0.00543859649122807
68 0.00596491228070175
67 0.0043859649122807
66 0.00649122807017544
65 0.00368421052631579
64 0.00508771929824561
63 0.00631578947368421
62 0.00649122807017544
61 0.0056140350877193
60 0.00333333333333333
59 0.00701754385964912
58 0.00736842105263158
57 0.00947368421052632
56 0.00894736842105263
55 0.00719298245614035
54 0.00771929824561404
53 0.00964912280701754
52 0.0126315789473684
51 0.0107017543859649
50 0.0135087719298246
49 0.0121052631578947
48 0.0129824561403509
47 0.0110526315789474
46 0.0135087719298246
45 0.0143859649122807
44 0.0124561403508772
43 0.016140350877193
42 0.0196491228070175
41 0.0173684210526316
40 0.0208771929824561
39 0.0175438596491228
38 0.0233333333333333
37 0.0201754385964912
36 0.0252631578947368
35 0.0240350877192982
34 0.0275438596491228
33 0.0342105263157895
32 0.0315789473684211
31 0.0384210526315789
30 0.0391228070175439
29 0.0384210526315789
28 0.0482456140350877
27 0.05
26 0.0531578947368421
25 0.0663157894736842
};
\end{axis}

\end{tikzpicture}
    \caption{Amazon Books}
\end{subfigure}

\vspace{0.35cm}

\begin{subfigure}{0.47\textwidth}
    \centering
    \begin{tikzpicture}

\definecolor{black}{RGB}{0,0,0}
\definecolor{dimgray85}{RGB}{85,85,85}
\definecolor{gainsboro220}{RGB}{220,220,220}
\definecolor{green01270}{RGB}{0,127,0}

\begin{axis}[
tick label style = {font=\footnotesize}, 
label style      = {font=\footnotesize},
axis background/.style={fill=white},
scale=0.5,
axis line style={black},
log basis x={10},
log basis y={10},
tick align=outside,
tick pos=left,
xmin=20.792286521438, xmax=1198.76185691751,
xmode=log,
xtick style={color=dimgray85},
xlabel={$d$},
xtick={0.1,1,10,100,1000,10000,100000},
xticklabels={
  \(\displaystyle {10^{-1}}\),
  \(\displaystyle {10^{0}}\),
  \(\displaystyle {10^{1}}\),
  \(\displaystyle {10^{2}}\),
  \(\displaystyle {10^{3}}\),
  \(\displaystyle {10^{4}}\),
  \(\displaystyle {10^{5}}\)
},
ymin=1e-05, ymax=1,
ymode=log,
ylabel={$P(d)$},
ytick style={color=dimgray85},
ytick={1e-06,1e-05,0.0001,0.001,0.01,0.1,1,10},
yticklabels={
  \(\displaystyle {10^{-6}}\),
  \(\displaystyle {10^{-5}}\),
  \(\displaystyle {10^{-4}}\),
  \(\displaystyle {10^{-3}}\),
  \(\displaystyle {10^{-2}}\),
  \(\displaystyle {10^{-1}}\),
  \(\displaystyle {10^{0}}\),
  \(\displaystyle {10^{1}}\)
}
]
\addplot [draw=black, fill=black, mark size=0.7pt, mark=*, only marks]
table{%
x  y
997 0.000216403375892664
638 0.000216403375892664
605 0.000216403375892664
598 0.000216403375892664
540 0.000216403375892664
457 0.000216403375892664
446 0.000216403375892664
435 0.000216403375892664
410 0.000216403375892664
395 0.000216403375892664
380 0.000216403375892664
378 0.000216403375892664
371 0.000216403375892664
367 0.000216403375892664
366 0.000432806751785328
365 0.000216403375892664
361 0.000216403375892664
357 0.000216403375892664
355 0.000216403375892664
345 0.000216403375892664
344 0.000216403375892664
339 0.000216403375892664
337 0.000216403375892664
336 0.000216403375892664
324 0.000216403375892664
313 0.000432806751785328
304 0.000216403375892664
300 0.000432806751785328
297 0.000216403375892664
296 0.000216403375892664
293 0.000216403375892664
291 0.000216403375892664
274 0.000216403375892664
272 0.000216403375892664
271 0.000216403375892664
267 0.000432806751785328
264 0.000216403375892664
263 0.000216403375892664
258 0.000216403375892664
255 0.000216403375892664
254 0.000216403375892664
252 0.000432806751785328
251 0.000649210127677992
250 0.000216403375892664
249 0.000216403375892664
247 0.000432806751785328
246 0.000432806751785328
245 0.000216403375892664
244 0.000216403375892664
242 0.000432806751785328
241 0.000432806751785328
240 0.000216403375892664
237 0.000432806751785328
232 0.000216403375892664
230 0.000432806751785328
228 0.000432806751785328
227 0.000432806751785328
222 0.000649210127677992
220 0.000432806751785328
219 0.000216403375892664
218 0.000432806751785328
217 0.000432806751785328
216 0.000216403375892664
214 0.000216403375892664
212 0.000432806751785328
211 0.000649210127677992
209 0.000432806751785328
208 0.000649210127677992
207 0.000216403375892664
206 0.000432806751785328
205 0.000216403375892664
202 0.000432806751785328
201 0.000432806751785328
200 0.000216403375892664
199 0.000216403375892664
197 0.000432806751785328
196 0.000216403375892664
194 0.000216403375892664
193 0.000216403375892664
192 0.000432806751785328
191 0.000432806751785328
190 0.000216403375892664
189 0.000216403375892664
187 0.00108201687946332
186 0.000865613503570656
184 0.000865613503570656
183 0.000432806751785328
182 0.000216403375892664
181 0.000865613503570656
179 0.000216403375892664
178 0.000216403375892664
176 0.000432806751785328
174 0.000649210127677992
173 0.000432806751785328
172 0.000432806751785328
171 0.000432806751785328
170 0.000432806751785328
169 0.000649210127677992
168 0.000432806751785328
167 0.000649210127677992
166 0.000432806751785328
165 0.000865613503570656
164 0.00129842025535598
162 0.000432806751785328
161 0.000865613503570656
160 0.000649210127677992
159 0.000865613503570656
158 0.00108201687946332
157 0.000432806751785328
155 0.000649210127677992
154 0.00108201687946332
153 0.000432806751785328
152 0.000432806751785328
151 0.000216403375892664
150 0.00108201687946332
149 0.00129842025535598
148 0.000432806751785328
147 0.00108201687946332
146 0.00108201687946332
144 0.000649210127677992
143 0.000432806751785328
142 0.00151482363124865
141 0.000865613503570656
140 0.000432806751785328
139 0.00108201687946332
138 0.000649210127677992
137 0.00108201687946332
136 0.000649210127677992
135 0.00108201687946332
134 0.00108201687946332
133 0.00129842025535598
132 0.00108201687946332
131 0.00216403375892664
130 0.000865613503570656
129 0.00173122700714131
128 0.00129842025535598
127 0.00151482363124865
126 0.00129842025535598
125 0.00129842025535598
124 0.00173122700714131
123 0.000649210127677992
122 0.00108201687946332
121 0.00151482363124865
120 0.00129842025535598
119 0.00129842025535598
118 0.00108201687946332
117 0.000432806751785328
116 0.00259684051071197
115 0.00216403375892664
114 0.00259684051071197
113 0.00151482363124865
112 0.000216403375892664
111 0.00151482363124865
110 0.000865613503570656
109 0.00108201687946332
108 0.00129842025535598
107 0.000649210127677992
106 0.00389526076606795
105 0.00194763038303398
104 0.00281324388660463
103 0.00216403375892664
102 0.00259684051071197
101 0.000865613503570656
100 0.0023804371348193
99 0.00108201687946332
98 0.00259684051071197
97 0.00324605063838996
96 0.00346245401428262
95 0.00216403375892664
94 0.00259684051071197
93 0.00216403375892664
92 0.00216403375892664
91 0.00324605063838996
90 0.00346245401428262
89 0.00216403375892664
88 0.00281324388660463
87 0.00324605063838996
86 0.00411166414196061
85 0.00346245401428262
84 0.00194763038303398
83 0.00411166414196061
82 0.00194763038303398
81 0.0030296472624973
80 0.00346245401428262
79 0.00324605063838996
78 0.0030296472624973
77 0.0030296472624973
76 0.00497727764553127
75 0.00497727764553127
74 0.00346245401428262
73 0.00432806751785328
72 0.00432806751785328
71 0.0054100843973166
70 0.00627569790088725
69 0.00389526076606795
68 0.0054100843973166
67 0.00497727764553127
66 0.00562648777320926
65 0.00454447089374594
64 0.00411166414196061
63 0.00714131140445791
62 0.00757411815624324
61 0.0106037654187405
60 0.00562648777320926
59 0.00627569790088725
58 0.00822332828392123
57 0.00887253841159922
56 0.0106037654187405
55 0.0101709586669552
54 0.00995455529106254
53 0.0132006059294525
52 0.00843973165981389
51 0.0112529755464185
50 0.0144990261848085
49 0.0129842025535598
48 0.0129842025535598
47 0.0127677991776672
46 0.0136334126812378
45 0.0147154295607011
44 0.0155810430642718
43 0.0160138498160571
42 0.0166630599437351
41 0.0164466565678425
40 0.0181778835749838
39 0.0164466565678425
38 0.0212075308374811
37 0.0186106903267691
36 0.0255355983553343
35 0.0216403375892664
34 0.025752001731227
33 0.0255355983553343
32 0.0318112962562216
31 0.0335425232633629
30 0.0387362042847868
29 0.0385198009088942
28 0.0389526076606795
27 0.0458775156892448
26 0.0432806751785328
25 0.05280242371781
};
\end{axis}

\end{tikzpicture}
    \caption{Gowalla}
\end{subfigure}
\begin{subfigure}{0.47\textwidth}
    \centering
    \begin{tikzpicture}

\definecolor{black}{RGB}{0,0,0}
\definecolor{dimgray85}{RGB}{85,85,85}
\definecolor{gainsboro220}{RGB}{220,220,220}
\definecolor{green01270}{RGB}{0,127,0}

\begin{axis}[
tick label style = {font=\footnotesize}, 
label style      = {font=\footnotesize},
axis background/.style={fill=white},
scale=0.5,
axis line style={black},
log basis x={10},
log basis y={10},
tick align=outside,
tick pos=left,
xmin=12.4776703869462, xmax=716.479897509777,
xmode=log,
xtick style={color=dimgray85},
xlabel={$d$},
xtick={0.1,1,10,100,1000,10000,100000},
xticklabels={
  \(\displaystyle {10^{-1}}\),
  \(\displaystyle {10^{0}}\),
  \(\displaystyle {10^{1}}\),
  \(\displaystyle {10^{2}}\),
  \(\displaystyle {10^{3}}\),
  \(\displaystyle {10^{4}}\),
  \(\displaystyle {10^{5}}\)
},
ymin=1e-05, ymax=1,
ymode=log,
ylabel={$P(d)$},
ytick style={color=dimgray85},
ytick={1e-06,1e-05,0.0001,0.001,0.01,0.1,1,10},
yticklabels={
  \(\displaystyle {10^{-6}}\),
  \(\displaystyle {10^{-5}}\),
  \(\displaystyle {10^{-4}}\),
  \(\displaystyle {10^{-3}}\),
  \(\displaystyle {10^{-2}}\),
  \(\displaystyle {10^{-1}}\),
  \(\displaystyle {10^{0}}\),
  \(\displaystyle {10^{1}}\)
}
]
\addplot [draw=black, fill=black, mark size=0.7pt, mark=*, only marks]
table{%
x  y
596 0.000391849529780564
512 0.000391849529780564
476 0.000391849529780564
467 0.000391849529780564
461 0.000391849529780564
422 0.000391849529780564
412 0.000391849529780564
402 0.000391849529780564
395 0.000391849529780564
394 0.000391849529780564
391 0.000391849529780564
368 0.000391849529780564
362 0.000391849529780564
356 0.000391849529780564
317 0.000391849529780564
304 0.000391849529780564
301 0.000391849529780564
298 0.000391849529780564
294 0.000391849529780564
284 0.000391849529780564
272 0.000391849529780564
267 0.000391849529780564
257 0.000391849529780564
256 0.000391849529780564
250 0.000391849529780564
249 0.000391849529780564
245 0.000391849529780564
243 0.000391849529780564
242 0.000391849529780564
240 0.000391849529780564
239 0.000391849529780564
238 0.000391849529780564
235 0.000391849529780564
234 0.000391849529780564
231 0.000391849529780564
220 0.000783699059561128
218 0.000391849529780564
216 0.000391849529780564
214 0.000391849529780564
208 0.000391849529780564
205 0.000783699059561128
189 0.000391849529780564
188 0.000783699059561128
186 0.000391849529780564
183 0.000391849529780564
182 0.000391849529780564
178 0.000391849529780564
177 0.000391849529780564
172 0.000391849529780564
169 0.000391849529780564
167 0.000391849529780564
166 0.000391849529780564
162 0.000391849529780564
160 0.000783699059561128
159 0.000391849529780564
158 0.000783699059561128
157 0.000391849529780564
154 0.000391849529780564
153 0.000391849529780564
152 0.000391849529780564
151 0.000391849529780564
147 0.000391849529780564
145 0.000391849529780564
143 0.000391849529780564
142 0.000391849529780564
141 0.000391849529780564
140 0.000391849529780564
138 0.00117554858934169
135 0.000391849529780564
134 0.000391849529780564
133 0.000391849529780564
131 0.000783699059561128
130 0.000783699059561128
128 0.00156739811912226
127 0.00117554858934169
126 0.000391849529780564
125 0.00117554858934169
124 0.000783699059561128
123 0.000783699059561128
122 0.000391849529780564
121 0.000391849529780564
120 0.000391849529780564
118 0.000783699059561128
117 0.000783699059561128
115 0.000391849529780564
113 0.000391849529780564
111 0.000391849529780564
110 0.000391849529780564
109 0.000391849529780564
108 0.000783699059561128
107 0.000391849529780564
106 0.000391849529780564
105 0.000391849529780564
104 0.000783699059561128
103 0.000391849529780564
102 0.000391849529780564
101 0.000783699059561128
100 0.00117554858934169
97 0.000783699059561128
96 0.000391849529780564
95 0.000783699059561128
94 0.000783699059561128
93 0.000783699059561128
92 0.00117554858934169
91 0.000783699059561128
90 0.000391849529780564
89 0.00117554858934169
88 0.000783699059561128
87 0.000783699059561128
86 0.00117554858934169
85 0.00235109717868339
84 0.000391849529780564
82 0.00274294670846395
81 0.00195924764890282
80 0.00235109717868339
79 0.00117554858934169
78 0.00195924764890282
77 0.00156739811912226
76 0.00235109717868339
75 0.00117554858934169
74 0.000783699059561128
73 0.00195924764890282
72 0.00117554858934169
71 0.000391849529780564
70 0.00156739811912226
69 0.00235109717868339
68 0.00313479623824451
67 0.00156739811912226
66 0.00274294670846395
65 0.00156739811912226
64 0.00156739811912226
63 0.00117554858934169
62 0.00274294670846395
61 0.000391849529780564
60 0.00195924764890282
59 0.00117554858934169
58 0.00117554858934169
57 0.00235109717868339
56 0.00156739811912226
55 0.00117554858934169
54 0.00235109717868339
53 0.000783699059561128
52 0.000391849529780564
51 0.00235109717868339
50 0.00979623824451411
49 0.0105799373040752
48 0.0160658307210031
47 0.0121473354231975
46 0.0192006269592476
45 0.0211598746081505
44 0.0246865203761755
43 0.0258620689655172
42 0.0270376175548589
41 0.0278213166144201
40 0.0254702194357367
39 0.0282131661442006
38 0.0231191222570533
37 0.0356583072100313
36 0.0297805642633229
35 0.0344827586206897
34 0.0250783699059561
33 0.0348746081504702
32 0.0235109717868339
31 0.0172413793103448
30 0.0274294670846395
29 0.0239028213166144
28 0.0211598746081505
27 0.0184169278996865
26 0.0293887147335423
25 0.0254702194357367
24 0.0301724137931034
23 0.0270376175548589
22 0.0258620689655172
21 0.0246865203761755
20 0.0239028213166144
19 0.0270376175548589
18 0.0278213166144201
17 0.024294670846395
16 0.0274294670846395
15 0.0297805642633229
};
\end{axis}

\end{tikzpicture}
    \caption{LastFM}
\end{subfigure}

\caption{Computed degree distributions for all the selected datasets in this monograph. On the $x$ and $y$ axes, we report the degree values (both user and item side) and the probability distribution (in log scale), respectively.}
\label{fig:deg_dist}
\end{figure}

Finally, we plot the \textbf{degree distribution} of all the selected datasets in Figure \ref{fig:deg_dist}. As already stated above, all degree distributions closely follow a power law tendency, suggesting a scale-free nature, as observed in most real-world social networks. Note that, only in this case, we do not compute and report any numerical value in Table~\ref{tab:topo_char} for this topological property because it is more effective to visualize it through its distributional shape (Figure \ref{fig:deg_dist}). That is also why we do not associate this property with any symbol as previously done in Table~\ref{tab:shorthand}. 
\begin{table*}[t]
\centering
\caption{Calculated topological characteristics for our selected recommendation datasets.}
\label{tab:topo_char}
\footnotesize
\begin{tabular}{lcccccc}
\toprule
\textbf{Dataset} & 
\textbf{$\sigma_{\mathcal{U}}$} &
\textbf{$\sigma_{\mathcal{I}}$} &
\textbf{$\gamma_{\mathcal{U}}$} &
\textbf{$\gamma_{\mathcal{I}}$} &
\textbf{$\rho_{\mathcal{U}}$} &
\textbf{$\rho_{\mathcal{I}}$} \\
\midrule
MovieLens 1M & 175.317        & 335.506  & 0.065  & 0.055 & -0.046    & -0.040 \\
Yelp  & 55.816  & 65.693  & 0.025  & 0.025  & 0.306 & 0.134 \\
Amazon Beauty & 36.130   & 26.449 & 0.021  & 0.025  & -0.039    & 0.021 \\
Amazon Books  & 43.486  & 57.686 & 0.033  & 0.026   & 0.031     & -0.040 \\
Gowalla        & 54.293    & 60.007 & 0.039  & 0.030  & 0.029     & -0.029 \\
LastFM  & 33.411  & 56.247 & 0.064   & 0.031  & -0.029    & -0.120 \\ 
\bottomrule
\end{tabular}
\end{table*}

\chapter{Topological characteristics in GNN-based recommendation} \label{ch:topo_gnn}

In recommender systems, GNN-based models learn user preference by leveraging patterns in the user-item bipartite graph. In this section, we investigate whether, and to what extent, these models implicitly or explicitly incorporate the \textit{topological} properties introduced before. To illustrate this, we consider the selected GNN-based recommenders for this monograph: NGCF \citep{DBLP:conf/sigir/Wang0WFC19}, DGCF~\citep{DBLP:conf/sigir/WangJZ0XC20},  LightGCN~\citep{DBLP:conf/sigir/0001DWLZ020}, SGL \citep{DBLP:conf/sigir/WuWF0CLX21}, UltraGCN \citep{DBLP:conf/cikm/MaoZXLWH21}, GFCF \citep{DBLP:conf/cikm/ShenWZSZLL21}, SVD-GCN~\citep{DBLP:conf/cikm/PengSM22}, SimGCL \citep{DBLP:conf/sigir/YuY00CN22}, LightGCL \citep{DBLP:conf/iclr/Cai0XR23}, GraphAU \citep{DBLP:conf/cikm/YangLWYLMY23}, and XSimGCL \citep{DBLP:journals/tkde/YuXCCHY24}. Specifically, we reinterpret their learning procedures through the lens of the identified \textit{topological} properties, explicitly highlighting their influence.

\section{NGCF~\citep{DBLP:conf/sigir/Wang0WFC19}} 
Similar to GCN, NGCF adopts feature transformation and non-linearities within the message passing algorithm. For each ego node, the neighborhood nodes' contribution is rescaled according to the corresponding value in the normalized adjacency matrix. This property helps reduce disparities between nodes with high and low \textit{node degrees}:
\begin{equation}
\begin{aligned}
    \mathbf{E}_u^{(l)} & = \text{LeakyReLU}\left(\sum_{i \in \mathcal{N}^{(1)}_u}\mathbf{W}_{\text{neigh}}^{(l)}\tilde{A}_{ui}\mathbf{E}_i^{(l - 1)}\right) \\
    & = \text{LeakyReLU}\left(\sum_{i \in \mathcal{N}^{(1)}_u}\mathbf{W}_{\text{neigh}}^{(l)}\frac{A_{ui}}{\sqrt{\sigma_u}\sqrt{\sigma_i}}\mathbf{E}_i^{(l - 1)}\right),
\end{aligned}
\end{equation}
where, for the sake of simplicity, we removed the interdependence factor as it does not encode any topological property. It is worth noticing that $A_{ui} = 1, \text{ } \forall i \in \mathcal{N}^{(1)}_u$, meaning that the denominator solely determines the contribution weighting. 

\section{DGCF~\citep{DBLP:conf/sigir/WangJZ0XC20}} 
This model assumes that user-item interactions are mediated by underlying intents, meaning that user and item representations are learned differently for each specific intent $t$. Consequently, the adjacency matrix $A_{ui}[t]$ and its normalization are intent-dependent.
We reformulate the node embedding as follows:
\begin{equation}
    \mathbf{E}^{(l)}_{u}[t] = \sum\limits_{i \in \mathcal{N}^{(1)}_u[t]} \tilde{A}_{ui}[t] \mathbf{E}_{i}^{(l - 1)}[t] = \sum\limits_{i \in \mathcal{N}_u^{(1)}} \frac{{A}_{ui}[t]}{\sqrt{\sigma_{u}[t]} \sqrt{\sigma_{i}[t]}} \mathbf{E}_{i}^{(l - 1)}[t],
\end{equation}
where the \textit{node degree} serves as a normalization factor that depends on the specific intent $t$.

\section{LightGCN~\citep{DBLP:conf/sigir/0001DWLZ020} }
Here, users and items are represented as the linear aggregation of their neighbors. As in NGCF, each neighbor's contribution is weighted according to the corresponding value in the normalized adjacency matrix. We reformulate it as follows:
\begin{equation}
\label{eq:lightgcn}
    \mathbf{E}_u^{(l)} = \sum_{i \in \mathcal{N}_u^{(1)}}\tilde{A}_{ui}\mathbf{E}_{i}^{(l - 1)} = \sum\limits_{i \in \mathcal{N}_u^{(1)}} \frac{A_{ui}}{\sqrt{\sigma_u} \sqrt{\sigma_{i}}} \mathbf{E}_{i}^{(l - 1)}.
\end{equation}

Again, the \textit{node degrees} of the ego and neighborhood nodes contribute to weight the aggregated message.

\section{SGL~\citep{DBLP:conf/sigir/WuWF0CLX21}}
In SGL, the authors propose to perform graph augmentations (such as node dropout and edge dropout) to obtain two views of the original user-item graph, and ultimately learn two views of each node embedding through the message passing algorithm. In particular, SGL leverages the same message passing as LightGCN, but on the augmented graph views:
\begin{equation}
\begin{aligned}
    \mathbf{E}_u^{(l)}[t_*] & = \sum_{i \in \mathcal{N}^{(1)}_u} \tilde{A}_{ui}[t_*] \mathbf{E}^{(l -1)}_i[t_*] \\ 
    & = \sum_{i \in \mathcal{N}^{(1)}_u}\frac{{A}_{ui}[t_*]}{\sqrt{\sigma_u[t_*]}\sqrt{\sigma_i[t_*]}} \mathbf{E}^{(l -1)}_i[t_*],
\end{aligned}
\end{equation}
where we compute the normalized version of the adjacency matrix in the specific graph view $t_*$. The \textit{node degree} information appears within the augmented views of the graph.

\section{UltraGCN~\citep{DBLP:conf/cikm/MaoZXLWH21}} UltraGCN recognizes three major limitations in GCN-based message-passing for collaborative filtering, namely, (i) the asymmetric weight assignment to connected nodes when considering user-user and item-item relationships; (ii) the impossibility to diversify the importance of each type of relation (i.e., user-item, user-user, item-item) during the message-passing; (iii) the over-smoothing effect when stacking more than 3 layers. To tackle such issues, the authors propose to go beyond the traditional concept of explicit message-passing, and approximate the infinite-layer message-passing through the following:
\begin{equation}
    \hat{\mathbf{E}}_u = \sum\limits_{i \in \mathcal{N}_u^{(1)}} \frac{A_{ui} \sqrt{\sigma_u + 1} \; \mathbf{E}_{i}}{\sigma_u \sqrt{\sigma_{i} + 1}}, \quad
    \hat{\mathbf{E}}_i = \sum\limits_{u \in \mathcal{N}_i^{(1)}} \frac{A_{iu} \sqrt{\sigma_i + 1} \; \mathbf{E}_{u}}{\sigma_{i} \sqrt{\sigma_{u} + 1}}.
\end{equation}

Note that the procedure is not repeated for layers $l > 1$, as the method surpasses the concept of iterative message-passing. During the optimization, the model first minimizes a constraint loss that adopts negative sampling to limit the over-smoothing effect:
\begin{equation}
    \begin{aligned}
    \mathcal{L}_{\text{CL}} = & -  \sum\limits_{(u,i) \sim \mathcal{T}_+ }\frac{1}{\sigma_u} \frac{\sqrt{\sigma_u + 1}}{\sqrt{\sigma_i + 1}} \text{ln}(\text{Sigmoid}(\hat{\mathbf{E}}_u^\top \hat{\mathbf{E}}_i)) \text{ }+ \\
    & - \sum\limits_{(u, j) \sim \mathcal{T}_-} \frac{1}{\sigma_u} \frac{\sqrt{\sigma_u + 1}}{\sqrt{\sigma_j + 1}} \text{ln}(\text{Sigmoid}(-\hat{\mathbf{E}}_u^\top \hat{\mathbf{E}}_j)),
    \end{aligned}
\end{equation}
where $\mathcal{T}_+ = \{(u, i) \; | \; R_{ui} = 1\}$ and $\mathcal{T}_- = \{(u, j)\; | \; R_{uj} = 0\}$, while $\text{ln}(\cdot)$ and $\text{Sigmoid}(\cdot)$ are the logarithm and sigmoid function. Then, they also take into account the item-item projected graph $\mathcal{G}_{\mathcal{I}}$ and minimize:
\begin{equation}
    \mathcal{L}_{\text{IL}} =  - \sum\limits_{(u, i) \sim \mathcal{T}_+} \text{ } \sum\limits_{j \in \text{topK}(\mathbf{R}^{\mathcal{I}\mathcal{I}}_{i*})}\frac{R^{\mathcal{I}\mathcal{I}}_{ij}}{\sigma^{\mathcal{I}\mathcal{I}}_i - R^{\mathcal{I}\mathcal{I}}_{ii}} \sqrt{\frac{\sigma^{\mathcal{I}\mathcal{I}}_i}{\sigma^{\mathcal{I}\mathcal{I}}_j}} \text{ln}(\text{Sigmoid}(\hat{\mathbf{E}}_u^\top \hat{\mathbf{E}}_j)),
\end{equation}
where $\text{topK}(\cdot)$ retrieves the top \textit{K} values of a matrix row-wise, and $\sigma_*^{\mathcal{I}\mathcal{I}}$ is the node degree calculated on $\mathcal{G}_{\mathcal{I}} = \{\mathcal{I, \mathbf{A}^{\mathcal{I}\mathcal{I}}} = \mathbf{R}^{\mathcal{I}\mathcal{I}}\}$. Thus, the model formulation encodes the \textit{node degrees} information in both the user-item and the \textit{item-item} co-occurrence graphs.

\section{GFCF~\citep{DBLP:conf/cikm/ShenWZSZLL21}}
The authors of GFCF re-interpret all main GNN-based recommender systems under the lens of graph signal processing. Eventually, they design a strong closed-form algorithm to predict novel user-item interactions:
\begin{equation}
\footnotesize
    \hat{\mathbf{R}}_{u*} = \mathbf{R}_{u*}\left(\left(\tilde{\mathbf{A}}^{\mathcal{UI}}\right)^{\top}\tilde{\mathbf{A}}^{\mathcal{IU}} + t \left(\mathbf{D}^{\mathcal{II}}\right)^{-\frac{1}{2}}\text{topK}(\mathbf{U})\text{topK}(\mathbf{U})^{\top}\left(\mathbf{D}^{\mathcal{II}}\right)^{+\frac{1}{2}}\right),
\end{equation}
where the superscripts $\mathcal{U}\mathcal{I}$, $\mathcal{I}\mathcal{U}$, and $\mathcal{I}\mathcal{I}$ stand for the user-item, item-user, and item-item partitions of the adjacency matrix, respectively. As $\tilde{\mathbf{A}}^{\mathcal{U}\mathcal{I}} = \left(\mathbf{D}^{\mathcal{U}\mathcal{I}}\right)^{-\frac{1}{2}}\mathbf{A}^{\mathcal{U}\mathcal{I}}\left(\mathbf{D}^{\mathcal{U}\mathcal{I}}\right)^{-\frac{1}{2}}$ and $\tilde{\mathbf{A}}^{\mathcal{I}\mathcal{U}} = \left(\mathbf{D}^{\mathcal{I}\mathcal{U}}\right)^{-\frac{1}{2}}\mathbf{A}^{\mathcal{I}\mathcal{U}}\left(\mathbf{D}^{\mathcal{I}\mathcal{U}}\right)^{-\frac{1}{2}}$, we explicitly highlight the presence of the \textit{node degrees} information of the user-item graph and \textit{item-item} co-occurrence graph within the GFCF formulation. 

\section{SVD-GCN~\citep{DBLP:conf/cikm/PengSM22}}
By reinterpreting the message-passing mechanism introduced in LightGCN, SVD-GCN leverages both even-hop (user-user, item-item) and odd-hop (user-item, item-user) message aggregations, along with the singular vectors and values of the normalized adjacency matrix $\tilde{\mathbf{A}}^{\mathcal{U}\mathcal{I}} = \left(\mathbf{D}^{\mathcal{U}\mathcal{I}}\right)^{-\frac{1}{2}}\mathbf{A}^{\mathcal{U}\mathcal{I}}\left(\mathbf{D}^{\mathcal{U}\mathcal{I}}\right)^{-\frac{1}{2}}$.
The user (item) embeddings are computed as follows:
\begin{equation}
    \hat{\mathbf{E}}_u = (\mathbf{U}) \text{exp}(t_1\mathbf{V})\mathbf{W},
\end{equation}
where $\mathbf{U}$ contains the singular vectors of $\tilde{\mathbf{A}}^{\mathcal{U}\mathcal{I}}$, $t_1$ is a tunable parameter, $\mathbf{V}$ is vector containing the largest singular values on $\tilde{\mathbf{A}}^{\mathcal{U}\mathcal{I}}$, and $\mathbf{W}$ is a trainable weight matrix.
A connection exists between the largest singular value $v_{\max} \in \mathbf{V}$ and the \textit{maximum node degree} $\max(\mathcal{D})$ in the user-item matrix, captured by the following inequality: 
\begin{equation}
    v_{\max} \leq \frac{\max(\mathcal{D})}{\max(\mathcal{D}) + t_2},
\end{equation}
where $t_2$ is an additional tunable hyperparameter controlling the spacing among singular values.
Moreover, to capture \textit{even-hop connections} in the bipartite graph, the loss function incorporates an additional component that depends on user-user (item-item) interactions. Specifically, this component considers \textit{users (items) co-occurrences} in the graph as follows:
\begin{equation}
    \mathcal{L}_{\text{UL}} = \sum_{u \in \mathcal{U}} \text{ } \sum_{(u, v) \sim \mathcal{T}^{\mathcal{U}\mathcal{U}}_+, \; (u, w) \sim \mathcal{T}^{\mathcal{U}\mathcal{U}}_-} \text{ln} \; \text{Sigmoid}(\hat{\mathbf{E}}_u^{\top} \hat{\mathbf{E}}_v - \hat{\mathbf{E}}_u^{\top}\hat{\mathbf{E}}_w),
\end{equation}
where $\mathcal{T}_+^{\mathcal{U}\mathcal{U}} = \{(u, v) \; | \; A_{uv}^{\mathcal{U}\mathcal{U}} \neq 0\}$ and $\mathcal{T}_-^{\mathcal{U}\mathcal{U}} = \{(u, w) \; | \; A_{uw}^{\mathcal{U}\mathcal{U}} = 0\}$, while $\text{Sigmoid}(\cdot)$ is the sigmoid function.  

\section{SimGCL~\citep{DBLP:conf/sigir/YuY00CN22}}
SimGCL surpasses the traditional concept of graph contrastive learning by applying augmentations on the node embeddings rather than on the graph structure. Thus, from a topological perspective, the presence of the \textit{node degree} information is equal to that of other models such as NGCF and LightGCN:
\begin{equation}
\begin{aligned}
    \mathbf{E}_u^{(l)}[t_*] &= \sum_{i \in \mathcal{N}^{(1)}_u} \tilde{A}_{ui}\mathbf{E}^{(l -1)}_i[t_*] + \Delta^{(l)}[t_*] =\\ &= \sum_{i \in \mathcal{N}^{(1)}_u} \frac{A_{ui}}{\sqrt{\sigma_u}\sqrt{\sigma_i}}\mathbf{E}^{(l -1)}_i[t_*] + \Delta^{(l)}[t_*], 
\end{aligned}
\end{equation}
where $t_*$ stands for one of the two computed views.

\section{LightGCL~\citep{DBLP:conf/iclr/Cai0XR23}}
LightGCL proposes a simple yet effective graph contrastive learning solution, where two views of the user and item embeddings (obtained from a global and local perspective on the user-item graph) are contrasted. Regarding the global view, it is obtained through the same message passing formulation proposed in \citep{DBLP:conf/www/XiaLGLLG22}:
\begin{equation}
    \mathbf{E}_u^{(l)} = \text{LeakyReLU}\left( \text{drop}(\tilde{\mathbf{A}}_{u*})\mathbf{E}^{(l -1)}_i\right) + \mathbf{E}^{(l -1)}_i,
\end{equation}
where the model applies the edge dropout function to the normalized adjacency matrix $\tilde{\mathbf{A}} = \mathbf{D}^{-\frac{1}{2}}\mathbf{A}\mathbf{D}^{-{1}{2}}$. Thus, as in other described models, the \textit{node degree} component is explicitly highlighted in the formulation.

\section{GraphAU~\citep{DBLP:conf/cikm/YangLWYLMY23}}
In a similar manner as \citep{DBLP:conf/kdd/WangYM000M22}, the authors redefine the uniformity and alignment paradigm to graph-based recommendation within this model. From a topological viewpoint, GraphAU shows the same contribution as LightGCN, by leveraging the normalized adjacency matrix where the \textit{node degree} is explicitly encoded:
\begin{equation}
    \mathbf{E}_u^{(l)} = \sum_{i \in \mathcal{N}_u^{(1)}}\tilde{A}_{ui}\mathbf{E}_{i}^{(l - 1)} = \sum\limits_{i \in \mathcal{N}_u^{(1)}} \frac{A_{ui}}{\sqrt{\sigma_u} \sqrt{\sigma_{i}}} \mathbf{E}_{i}^{(l - 1)}.
\end{equation}

\section{XSimGCL~\citep{DBLP:journals/tkde/YuXCCHY24}}
Unlike other contrastive models, XSimGCL leverages the same contrasted view of node embeddings for both the recommendation task and the contrastive learning task; it then applies a cross-layer contrastive module, which calculates the InfoNCE between the same node view calculated at different propagation layers. From a topological perspective, XSimGCL follows the same message passing as SimGCL, where the normalized adjacency matrix encodes the \textit{node degree} information:
\begin{equation}
\begin{aligned}
    \mathbf{E}_u^{(l)}[t] &= \sum_{i \in \mathcal{N}^{(1)}_u} \tilde{A}_{ui}\mathbf{E}^{(l -1)}_i[t] + \Delta^{(l)}[t] =\\ &= \sum_{i \in \mathcal{N}^{(1)}_u} \frac{A_{ui}}{\sqrt{\sigma_u}\sqrt{\sigma_i}}\mathbf{E}^{(l -1)}_i[t] + \Delta^{(l)}[t], 
\end{aligned}
\end{equation}
where, this time, only one view is considered and calculated.

\section{Topology and GNN models: an overview}
In Table \ref{tab:topology_models}, we summarize the main topological contributions of each selected GNN model. 

Overall, the most evident trend is the explicit encoding of the \textit{node degree} information within the models' formulations, mainly adopted in the calculation of the normalized adjacency matrix of the user-item and (in one case) the item-item graph. Additionally, some GNN models work on the graph structure of the user-item graph by modifying it or weighting the importance of its different contributions. Finally, a few models do not limit their perspective to the user-item graph, but propose leveraging the user-user and item-item projections to reveal hidden node relationships, which could become fundamental in recommendation. 

\begin{tcolorbox}[colback=black!5!white,colframe=black!75!black,title=\textsc{\bfseries Final remarks}]
\textit{The analysis reveals that GNN-based recommenders explicitly incorporate node degrees in their formulations, while other topological properties, such as clustering coefficient and assortativity, may only be implicitly captured—if at all. Therefore, our explanatory framework, detailed in Section~\ref{ch:framework}, also serves to investigate which topological characteristics GNN-based recommender systems might unintentionally or implicitly learn during their pipeline.}
\end{tcolorbox}

\begin{table}[!t]
\centering
\caption{Selected GNN models for this monograph and how they handle topological properties according to their implemented strategies for the the graph structure, the prediction, and the loss function.}
\label{tab:topology_models}
\footnotesize
\begin{tabular}{lccc}
\toprule
\textbf{Model} & \textbf{Graph structure} & \textbf{Prediction} & \textbf{Loss function} \\ \cmidrule{1-4}
NGCF & \xmark & \parbox[t]{2.8cm}{Node degree in the normalized adjacency matrix} & \xmark \\ \cmidrule{1-4}
DGCF & \parbox[t]{2.8cm}{Separate graph structure for each intent} & \parbox[t]{2.8cm}{Node degree in the normalized adjacency matrix} & \xmark \\ \cmidrule{1-4}
LightGCN & \xmark & \parbox[t]{2.8cm}{Node degree in the normalized adjacency matrix} & \xmark \\ \cmidrule{1-4}
SGL & \parbox[t]{2.8cm}{Node dropout and edge dropout applied to obtain different graph views} & \parbox[t]{2.8cm}{Node degree in the normalized adjacency matrix} & \xmark \\ \cmidrule{1-4}
UltraGCN & \xmark & \parbox[t]{2.8cm}{Node degree for normalizing the adjacency matrix in a custom way} & \parbox[t]{2.8cm}{The item-item co-occurrences are used for the IL component} \\ \cmidrule{1-4}
GFCF & \xmark & \parbox[t]{2.8cm}{Node degree in the normalized adjacency matrix and item-item degree matrix} & \xmark \\ \cmidrule{1-4}
SVD-GCN & \xmark & \parbox[t]{2.8cm}{Node degree in the normalized adjacency matrix and connection between max eigenvalue and max node degree} & \parbox[t]{2.8cm}{The user-user and item-item co-occurrences are used for the UL/IL components} \\ \cmidrule{1-4}
SimGCL & \xmark & \parbox[t]{2.8cm}{Node degree in the normalized adjacency matrix} & \xmark \\ \cmidrule{1-4}
LightGCL & \parbox[t]{2.8cm}{Edge dropout applied to address overfitting} & \parbox[t]{2.8cm}{Node degree in the normalized adjacency matrix} & \xmark \\ \cmidrule{1-4}
GraphAU & \xmark & \parbox[t]{2.8cm}{Node degree in the normalized adjacency matrix} & \xmark \\ \cmidrule{1-4}
XSimGCL & \xmark & \parbox[t]{2.8cm}{Node degree in the normalized adjacency matrix} & \xmark \\
 \bottomrule
 \end{tabular}
\end{table}
\chapter{Understanding the impact: an explanatory framework}\label{ch:framework}

In this section, we present our methodology designed to investigate how both \emph{classical} and \emph{topological} data characteristics influence the performance of the GNN-based recommender. Inspired by related research~\citep{DBLP:journals/tmis/AdomaviciusZ12, DBLP:conf/sigir/DeldjooNSM20,DBLP:conf/recsys/MalitestaPAMNS24}, our objective is to construct an explanatory statistical model capable of discerning how particular dataset attributes relate to recommendation quality. To achieve this, we propose a four-step pipeline: (i) creation of a comprehensive pool of recommendation datasets, (ii) computation of their classical and topological features, (iii) training and evaluation of GNN-based recommenders on these datasets, and (iv) fitting a linear explanatory model to link the observed data characteristics and system performance.

\section{Step 1: Dataset generation through graph sampling} 

A robust statistical investigation requires a sufficiently large sample set to ensure stable inferences. Although one could consider multiple large-scale recommendation datasets from the literature, such an approach would be prohibitively time-consuming and computationally expensive, since it entails training and evaluating complex GNN-based models on numerous large graphs.

To address this limitation, we adopt a strategy akin to previous works~\citep{DBLP:journals/tmis/AdomaviciusZ12, DBLP:conf/sigir/DeldjooNSM20} but with a crucial difference. Instead of relying on traditional sampling procedures from the literature~\citep{DBLP:conf/sigir/WuWF0CLX21, DBLP:conf/kdd/ChenSSH17, DBLP:conf/wsdm/SachdevaWM22}, we employ dedicated \emph{graph sampling} techniques such as node-dropout and edge-dropout~\citep{DBLP:conf/sigir/WuWF0CLX21, DBLP:conf/www/ShuXLWKM22}. By varying dropout rates, we produce multiple smaller recommendation datasets that retain a resemblance to the original sources while exhibiting a broad spectrum of topological variations.

Starting from each bipartite user-item graph, we repeatedly apply the following procedure: select a random dropout rate between $0.7$ and $0.9$, choose a sampling strategy (node-dropout or edge-dropout), and remove nodes or edges accordingly. Each iteration yields a new dataset sample, capturing various topological patterns. Following prior works~\citep{DBLP:journals/tmis/AdomaviciusZ12, DBLP:conf/sigir/DeldjooNSM20}, we generate $600$ samples per original dataset, resulting in a total of $1{,}800$ sampled datasets. The pseudo-code of the algorithms for Step 1 are reported in Algorithms \ref{alg:sampling} and \ref{alg:data-generation}.

\section{Step 2: Calculation of classical and topological features}

After creating the dataset collection, we compute a model-agnostic set of descriptors that capture both \emph{classical} data properties and \emph{topological} graph structure as those described earlier in this monograph. To improve comparability and numerical stability in subsequent modeling, we apply a base-10 logarithmic transformation to highly skewed quantities and standardize all predictors (z-score).

\begin{algorithm}[!t]
\caption{Graph sampling}
\label{alg:sampling}
\SetAlgoLined
\textbf{Input:} Bipartite user-item graph $\mathcal{G}$, dropout rate $\mu$, graph sampling strategy \texttt{\textit{sampling}}.\\
\textbf{Output:} Sampled graph $\mathcal{G}_m$.\\
\uIf{\texttt{sampling} $==$ \texttt{nodeDropout}}{
    $N = (U + I) * (1 - \mu)$ \\
    $\mathcal{V}_m \leftarrow \text{Uniform}_N(\mathcal{U} \cup \mathcal{I})$\\
    $\mathbf{A}_m \leftarrow \text{MaskNode}(\mathbf{A}, \mathcal{V}_m)$
}
\uElseIf{\texttt{sampling} $==$ \texttt{edgeDropout}}{
    $N = E * (1 - \mu)$ \\
    $\mathcal{E}_m \leftarrow \text{Uniform}_N(\mathcal{E}_{u \rightarrow i})$\\
    $\mathbf{A}_m \leftarrow \text{MaskEdge}(\mathbf{A}, \mathcal{E}_m)$\\
    $\mathcal{V}_m \leftarrow \text{Induce}(\mathbf{A}_m)$
}
\phantom{ }\\
$\mathcal{G}_m \leftarrow \{\mathcal{V}_m, \mathbf{A}_m\}$ \\
Return $\mathcal{G}_m$.
\end{algorithm}

\begin{algorithm}[!t]
\SetAlgoLined
\textbf{Input:} Bipartite user-item graph $\mathcal{G}$, number of samples $M$.\\
\textbf{Output:} $M$ sampled graphs.\\
$m \leftarrow 1$\\
$\mathcal{M} = \{\}$\\
\While{$m \leq M$}{
    $\mu \leftarrow \text{Uniform}([0.7, 0.9])$\\
    $\texttt{\textit{sampling}} \leftarrow \text{Uniform}(\{\texttt{nodeDropout}, \texttt{edgeDropout}\})$
    \\
    \texttt{\textcolor{forestgreen(web)}{// We use \Cref{alg:sampling} for graph sampling}}
    \\
    $\mathcal{M} \leftarrow \mathcal{M} \cup \text{GraphSampling}(\mathcal{G}, \mu, \texttt{\textit{sampling}})$\\
    $m \leftarrow m + 1$
}
Return $\mathcal{M}$.
\caption{Sub-dataset generation.}
\label{alg:data-generation}
\end{algorithm}

\section{Step 3: Training and evaluation of the models}

In the third step, we adopt a uniform, model-agnostic evaluation protocol for GNN-based recommender. Each dataset is partitioned into training, validation, and test subsets. When timestamps are available, a chronological hold-out approach is preferred to preserve temporal order; otherwise, a random split is used. Models are trained with early stopping based on the best validation score on a ranking-oriented metric (e.g., Recall@K or NDCG@K), under a fixed compute budget (maximum epochs and patience). Hyperparameters are either set via a lightweight validation-driven search or fixed to canonical defaults to ensure comparability across runs. Final effectiveness is reported on the test set and, when applicable, averaged over multiple random seeds to reduce variance. The resulting scores constitute the dependent variable $\mathbf{y}$ used by the explanatory model in Step~4.

\section{Step 4: Constructing an explanatory statistical model}

Finally, we employ a statistical model to determine how the aforementioned features (independent variables) explain variations in the recommender system’s effectiveness (dependent variable). Consistent with recent literature~\citep{DBLP:journals/tmis/AdomaviciusZ12,DBLP:conf/sigir/DeldjooNSM20}, we adopt a linear regression model to facilitate interpretability. This choice allows us to derive straightforward explanations about the directional relationships and significance of each dataset characteristic.

Formally, let $\mathbf{X}_c \in \mathbb{R}^{M \times C}$ be the matrix of feature values for $M$ samples and $C$ characteristics, and let $\mathbf{y}$ be the vector representing the performance outcomes. The linear model is given by:
\begin{equation}
\mathbf{y} = \boldsymbol{\epsilon} + \theta_0 + \boldsymbol{\theta}_c \mathbf{X}_c,
\label{eq:linear_model}
\end{equation}
where $\theta_0$ is the intercept (representing the expected value of $\mathbf{y}$ for mean-centered data) and $\boldsymbol{\theta}_c$ is the vector of coefficients associated with the features. We estimate these parameters via Ordinary Least Squares (OLS):
\begin{equation}
(\theta_0^*, \boldsymbol{\theta}_c^*) = \min_{\theta_0, \boldsymbol{\theta}_c} \frac{1}{2}\|\mathbf{y}-\theta_0-\boldsymbol{\theta}_c \mathbf{X}_c\|_2^2.
\end{equation}

By maximizing the $R^2$ coefficient and analyzing the signs and magnitudes of the estimated coefficients, we reveal which characteristics most strongly correlate with improvements or declines in GNN-based recommendation performance. This approach complements existing guidelines on regression analysis~\citep{gareth2013introduction} and provides interpretable insights into how data properties influence the effectiveness of next-generation recommender systems.

\section{Impact of characteristics}
\label{sec:impact}

In this section, we discuss how to interpret the output of the explanatory framework introduced in Step~4. Rather than focusing on a specific experimental setting, we provide a general reading guide for understanding how \emph{classical} and \emph{topological} dataset characteristics relate to the performance of GNN-based recommender systems.

Given a linear regression model where each characteristic is associated with a coefficient, the \emph{sign} of a coefficient indicates whether increases in that characteristic tend to be associated with improvements or deteriorations in the chosen performance metric. Positive coefficients suggest a direct correspondence (larger values of the characteristic relate to higher performance), whereas negative coefficients indicate an inverse correspondence.

The \emph{magnitude} of a coefficient, once variables have been properly scaled, offers a notion of relative importance: characteristics with larger absolute coefficients are those for which small changes in the underlying data are more strongly reflected in the model’s performance. However, these magnitudes should be interpreted in conjunction with measures of model fit and statistical significance, rather than in isolation.

A third element concerns \emph{statistical significance}. Even when a coefficient has a non-negligible magnitude, it should only be interpreted as evidence of a systematic effect if standard significance tests support it. Non-significant coefficients are better viewed as inconclusive: the available data do not provide enough evidence to claim a stable association between the corresponding characteristic and performance.

Finally, characteristics rarely act independently. Correlations among them, as well as possible interaction effects, may cause changes in one characteristic to co-occur with changes in others. For this reason, interpreting the impact of individual coefficients should always be done with the broader correlation structure in mind, and with reference to the conceptual grouping of characteristics into \emph{classical} (space size, shape, sparsity-related factors) and \emph{topological} (degree, clustering, assortativity) properties.

\subsection{Classical characteristics}

From an interpretive standpoint, classical characteristics offer a coarse-grained description of the user-item space. Properties such as the overall size of the interaction space, the balance between users and items, the density of observed interactions, and the concentration of activity on a subset of users or items help to characterize how much information is available to the model and how it is distributed.

When coefficients associated with density-related measures are positive, they typically indicate that models benefit from richer interaction data, since additional observations provide more evidence for learning robust user and item representations. Conversely, negative coefficients associated with shape-related measures indicate detrimental imbalances between the number of users and items. For example, scenarios with more users than items (or vice versa) may lead to configurations where the model has a limited chance to observe diverse interaction patterns.

Inequality indices (i.e., user- and item-side Gini coefficients) allow us to reason about how skewed the interaction distributions are. A positive coefficient on the item-side index, for instance, suggests that models tend to perform better when there is a clearer separation between popular and less popular items, which aligns with the tendency of many recommenders to focus on highly interacted items. In contrast, weak or unstable effects on the user-side index indicate that strong heterogeneity in user activity is not necessarily beneficial or harmful per se, and may interact with other characteristics such as sparsity or shape.

Overall, classical characteristics describe the \emph{amount} and \emph{allocation} of available information in the interaction matrix. Their coefficients indicate the extent to which simple aggregate statistics of the data are sufficient to account for variations in model performance.

\subsection{Topological characteristics}

Topological characteristics refine this picture by focusing on the structure of the user--item graph beyond aggregate counts. Average degrees, clustering coefficients, and assortativity capture how interactions are organized in terms of local neighborhoods and connectivity patterns, and thus how well the graph supports message passing.

Coefficients associated with user and item average degree indicate how sensitive a model is to the richness of local neighborhoods. Positive values suggest that having more interactions per node provides the model with more informative neighborhoods from which to aggregate signals, effectively enabling better representation learning. Differences between user- and item-side degrees can reveal whether it is more beneficial to increase the activity of users (e.g., by providing more feedback per user) or to enhance the catalog-side connectivity.

Clustering coefficients describe the tendency of nodes to form tightly connected triadic structures and, in the bipartite setting, the existence of dense local patterns of shared interactions. Positive coefficients on user- or item-side clustering typically signal that models are able to exploit such local redundancy, as it reinforces the collaborative signal by repeatedly exposing similar neighborhoods. When coefficients differ between user and item clustering, this may indicate an asymmetric sensitivity of the model to user- versus item-centric local structures.

Degree assortativity captures whether high-degree nodes tend to connect to other high-degree nodes, or instead to low-degree ones. Positive coefficients on assortativity suggest that the model benefits from graphs where nodes of similar degree are often connected, as this “lookahead” structure enhances the propagation of signals along high-activity regions of the graph. Negative coefficients, in contrast, indicate scenarios where connecting nodes of very different degrees may lead to better information diffusion for the model under consideration.

Taken together, topological characteristics should be read as describing the \emph{shape} and \emph{organization} of the interaction graph. Their coefficients reveal how different GNN architectures leverage specific structural patterns and where their inductive biases align, or conflict, with the topology of the data.

\begin{tcolorbox}[colback=black!5!white,colframe=black!75!black,title=\textsc{\bfseries Final remarks}]
\textit{The framework reveals how characteristics relate to model performance through: (i) coefficient signs, which indicate whether a characteristic leads to performance gains/losses; (ii) its magnitude and significance, which quantify the strength and reliability of this association; (iii) the broader correlation among characteristics, which prevents overly simplistic interpretations. Classical characteristics capture the available information and its distribution, while the topological ones describe the user-item graph.}
\end{tcolorbox}

\subsection{Application to a topology-aware evaluation study}

The methodological principles outlined above can be instantiated in concrete evaluation scenarios. A notable example is the topology-aware study of GNN-based recommender systems presented in the work by~\cite{DBLP:conf/recsys/MalitestaPAMNS24}. In that study, the explanatory framework is applied to real-world recommendation datasets and a set of representative GNN architectures. In particular, it considers LightGCN \citep{DBLP:conf/sigir/0001DWLZ020}, DGCF \citep{DBLP:conf/sigir/WangJZ0XC20}, and SVD-GCN \citep{DBLP:conf/cikm/PengSM22} as exemplary GNN-based models, evaluated on Yelp2018~\citep{DBLP:conf/cikm/PengSM22}, Gowalla~\citep{DBLP:conf/sigir/0001DWLZ020}, and Amazon-Book~\citep{DBLP:conf/www/WangHWYL0C21}.

More specifically, the case study revisits the set of classical and topological characteristics introduced in earlier sections and summarized in Table~\ref{tab:shorthand}. Classical properties summarize space size, shape, density, and inequality of interactions (via Gini coefficients on the user and item sides), following prior work on dataset characterization~\citep{DBLP:journals/tmis/AdomaviciusZ12, DBLP:conf/sigir/DeldjooNSM20}. Then, topological properties capture the structural aspects of the user-item bipartite graph, including the average degree (for users and items), average clustering coefficients (user and item projections), and degree assortativity on both sides.

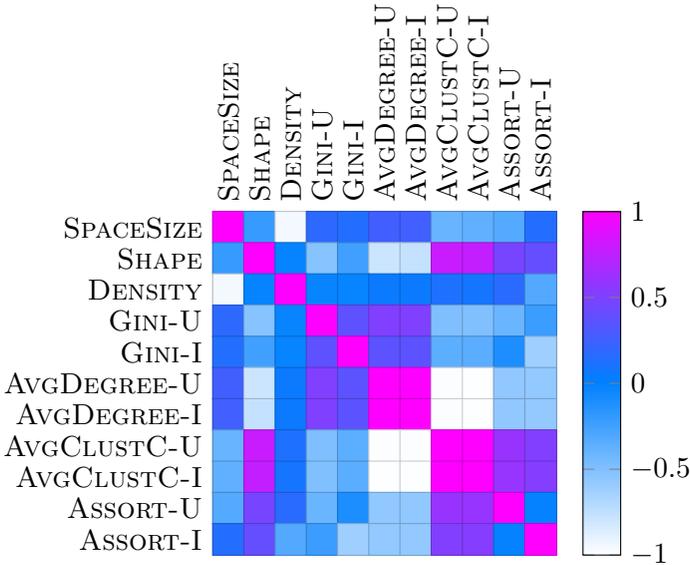
\begin{figure}[!t]
\centering
\begin{tikzpicture}
\begin{axis}[
    axis equal image,
    scatter, 
    colormap/cool, 
    colorbar, 
    point meta min=-1,
    point meta max=1,
    grid=none, 
    minor tick num=1, 
    tickwidth=0pt,
    y dir=reverse, 
    xticklabel pos=right, 
    xticklabel style={rotate=90},
    xtick={0,1,2,3,4,5,6,7,8,9,10},
    xticklabels={\textsc{SpaceSize}, \textsc{Shape}, \textsc{Density}, \textsc{Gini-U}, \textsc{Gini-I}, \textsc{AvgDegree-U}, \textsc{AvgDegree-I}, \textsc{AvgClustC-U}, \textsc{AvgClustC-I}, \textsc{Assort-U}, \textsc{Assort-I}},
    ytick={0,1,2,3,4,5,6,7,8,9,10},
    yticklabels={\textsc{SpaceSize}, \textsc{Shape}, \textsc{Density}, \textsc{Gini-U}, \textsc{Gini-I}, \textsc{AvgDegree-U}, \textsc{AvgDegree-I}, \textsc{AvgClustC-U}, \textsc{AvgClustC-I}, \textsc{Assort-U}, \textsc{Assort-I}},
    enlargelimits={abs=0.5}, 
    scatter/@pre marker code/.append code={
      \pgfplotstransformcoordinatex{sqrt(abs(\pgfplotspointmeta))}
      \scope[mark size=6, fill=mapped color]
    },
    scatter/@post marker code/.append code={%
      \endscope%
    },
    scale=0.8 
]
\addplot +[
    point meta=explicit,
    mark=square*,
    only marks,
    ] table [
    x expr={int(mod(\coordindex+0.01,11))},
    y expr={int((\coordindex+0.01)/11))},
    meta=value
] {
X   Y   value

0	0	1.000000
0	1	-0.212315
0	2	-0.954520
0	3	0.180507
0	4	0.134404
0	5	0.255242
0	6	0.254337
0	7	-0.408143
0	8	-0.377079
0	9	-0.328141
0	10	0.143009
1	0	-0.212315
1	1	1.000000
1	2	-0.021688
1	3	-0.524974
1	4	-0.254467
1	5	-0.794564
1	6	-0.767826
1	7	0.783785
1	8	0.765346
1	9	0.469329
1	10	0.394994
2	0	-0.954520
2	1	-0.021688
2	2	1.000000
2	3	-0.031705
2	4	-0.027120
2	5	0.044572
2	6	0.045499
2	7	0.118263
2	8	0.084351
2	9	0.163524
2	10	-0.325265
3	0	0.180507
3	1	-0.524974
3	2	-0.031705
3	3	1.000000
3	4	0.371105
3	5	0.506309
3	6	0.497189
3	7	-0.497250
3	8	-0.503362
3	9	-0.415233
3	10	-0.228037
4	0	0.134404
4	1	-0.254467
4	2	-0.027120
4	3	0.371105
4	4	1.000000
4	5	0.361340
4	6	0.363311
4	7	-0.362226
4	8	-0.355434
4	9	-0.111091
4	10	-0.620275
5	0	0.255242
5	1	-0.794564
5	2	0.044572
5	3	0.506309
5	4	0.361340
5	5	1.000000
5	6	0.999082
5	7	-0.984254
5	8	-0.989383
5	9	-0.569820
5	10	-0.573792
6	0	0.254337
6	1	-0.767826
6	2	0.045499
6	3	0.497189
6	4	0.363311
6	5	0.999082
6	6	1.000000
6	7	-0.983229
6	8	-0.989941
6	9	-0.568128
6	10	-0.577563
7	0	-0.408143
7	1	0.783785
7	2	0.118263
7	3	-0.497250
7	4	-0.362226
7	5	-0.984254
7	6	-0.983229
7	7	1.000000
7	8	0.998075
7	9	0.598835
7	10	0.511454
8	0	-0.377079
8	1	0.765346
8	2	0.084351
8	3	-0.503362
8	4	-0.355434
8	5	-0.989383
8	6	-0.989941
8	7	0.998075
8	8	1.000000
8	9	0.594505
8	10	0.523383
9	0	-0.328141
9	1	0.469329
9	2	0.163524
9	3	-0.415233
9	4	-0.111091
9	5	-0.569820
9	6	-0.568128
9	7	0.598835
9	8	0.594505
9	9	1.000000
9	10	-0.013674
10	0	0.143009
10	1	0.394994
10	2	-0.325265
10	3	-0.228037
10	4	-0.620275
10	5	-0.573792
10	6	-0.577563
10	7	0.511454
10	8	0.523383
10	9	-0.013674
10	10	1.000000
};
\end{axis}
\end{tikzpicture}
\caption{Pearson correlation of the selected characteristics and datasets in the study by \cite{DBLP:conf/recsys/MalitestaPAMNS24}. Many values in $[-0.5, 0.5]$ indicate loose correlation.}
\label{fig:correlation}
\end{figure}

These characteristics are first analyzed through their empirical correlation structure, in order to identify redundant signals and complementary patterns (see~\Cref{fig:correlation}). Moreover, characteristic values and distributions are computed both on the original datasets and on the collection of generated samples (\Cref{tab:dataset-chars}). As reported in the bottom part of the table, the aggregated statistics obtained from all generated samples span a considerably wider range than those computed on the original datasets alone (top part of the table). This observation highlights the increased topological diversity introduced by the sampling procedure and justifies the generation of multiple dataset samples from each original dataset to cover a broader spectrum of topological settings and recommendation scenarios. The explanatory model is then fitted to relate these characteristics to recommendation performance, and the resulting coefficients are examined across datasets and architectures.

This application illustrates how the generic interpretation guidelines proposed in this section are translated into concrete insights. The table of characteristics defines the feature space, the correlation analysis clarifies how these features interact, and the numerical regression results quantify their impact on performance. The reader is referred to that case study for a data-driven instantiation of the framework.

\begin{table}[!t]
\caption{Dataset overall characteristics and aggregated characteristics (minimum and maximum values, mean, and standard deviation) on the sampled sub-datasets. These values refer to the experimental setting adopted in the study proposed by \citet{DBLP:conf/recsys/MalitestaPAMNS24}.}
\label{tab:dataset-chars}
\begin{adjustbox}{width=1\textwidth,center}
    \begin{tabular}{l|cccc|cccc|cccc}
    \toprule
        \multicolumn{1}{l}{\textbf{Overall Chars.}} & \multicolumn{4}{c}{\textbf{Yelp2018}} & \multicolumn{4}{c}{\textbf{Amazon-Book}} & \multicolumn{4}{c}{\textbf{Gowalla}} \\ \midrule
        \textsc{SpaceSize} & \multicolumn{4}{c|}{1.411} & \multicolumn{4}{c|}{1.623} & \multicolumn{4}{c}{1.544} \\
        \textsc{Shape} & \multicolumn{4}{c|}{-0.002} & \multicolumn{4}{c|}{0.453} & \multicolumn{4}{c}{-0.138} \\
        \textsc{Density} & \multicolumn{4}{c|}{-2.957} & \multicolumn{4}{c|}{-3.318} & \multicolumn{4}{c}{-3.076} \\
        \textsc{Gini-U} & \multicolumn{4}{c|}{0.458} & \multicolumn{4}{c|}{0.445} & \multicolumn{4}{c}{0.467} \\
        \textsc{Gini-I} & \multicolumn{4}{c|}{0.604} & \multicolumn{4}{c|}{0.603} & \multicolumn{4}{c}{0.435} \\
        \textsc{AvgDegree-U} & \multicolumn{4}{c|}{1.434} & \multicolumn{4}{c|}{1.078} & \multicolumn{4}{c}{1.537} \\
        \textsc{AvgDegree-I} & \multicolumn{4}{c|}{1.431} & \multicolumn{4}{c|}{1.531} & \multicolumn{4}{c}{1.399} \\
        \textsc{AvgClustC-U} & \multicolumn{4}{c|}{-1.561} & \multicolumn{4}{c|}{-1.247} & \multicolumn{4}{c}{-1.577} \\
        \textsc{AvgClustC-I} & \multicolumn{4}{c|}{-1.491} & \multicolumn{4}{c|}{-1.606} & \multicolumn{4}{c}{-1.496} \\
        \textsc{Assort-U} & \multicolumn{4}{c|}{0.003} & \multicolumn{4}{c|}{-0.044} & \multicolumn{4}{c}{0.068} \\
        \textsc{Assort-I} & \multicolumn{4}{c|}{-0.017} & \multicolumn{4}{c|}{0.211} & \multicolumn{4}{c}{-0.030} \\
        \midrule
        \multicolumn{1}{c}{\makecell[l]{\textbf{Sample Chars.}\\(600 samples/dataset)}} & \textbf{Min} & \textbf{Max} & \textbf{Mean} & \multicolumn{1}{c}{\textbf{Std}} & \textbf{Min} & \multicolumn{1}{c}{\textbf{Max}} & \textbf{Mean} & \multicolumn{1}{c}{\textbf{Std}} & \textbf{Min} & \multicolumn{1}{c}{\textbf{Max}} & \textbf{Mean} & \textbf{Std} \\
        \cmidrule{1-13}  
        \textsc{SpaceSize} & 0.256 & 1.393 & 1.000 & 0.379 & 0.405 & 1.593 & 1.176 & 0.384 & 0.430 & 1.541 & 1.161 & 0.375 \\
        \textsc{Shape} & 0.019 & 0.105 & 0.045 & 0.014 & 0.325 & 0.443 & 0.407 & 0.021 & -0.149 & -0.097 & -0.129 & 0.008 \\ 
        \textsc{Density} & -3.699 & -2.693 & -3.219 & 0.358 & -3.902 & -2.896 & -3.497 & 0.365 & -3.889 & -2.876 & -3.380 & 0.363 \\ 
        \textsc{Gini-U} & 0.443 & 0.508 & 0.486 & 0.008 & 0.384 & 0.499 & 0.459 & 0.023 & 0.462 & 0.512 & 0.491 & 0.007 \\ 
        \textsc{Gini-I} & 0.500 & 0.609 & 0.575 & 0.019 & 0.518 & 0.618 & 0.586 & 0.018 & 0.437 & 0.502 & 0.478 & 0.008 \\ 
        \textsc{AvgDegree-U} & 0.523 & 0.926 & 0.758 & 0.110 & 0.318 & 0.609 & 0.476 & 0.077 & 0.603 & 1.017 & 0.846 & 0.115 \\ 
        \textsc{AvgDegree-I} & 0.565 & 0.955 & 0.804 & 0.098 & 0.682 & 1.043 & 0.883 & 0.096 & 0.487 & 0.888 & 0.717 & 0.109 \\ 
        \textsc{AvgClustC-U} & -1.144 & -0.662 & -0.947 & 0.126 & -0.757 & -0.407 & -0.602 & 0.095 & -1.211 & -0.741 & -1.013 & 0.122 \\ 
        \textsc{AvgClustC-I} & -1.092 & -0.652 & -0.922 & 0.105 & -1.124 & -0.751 & -0.967 & 0.099 & -1.080 & -0.614 & -0.881 & 0.124 \\ 
        \textsc{Assort-U} & -0.051 & 0.235 & 0.021 & 0.035 & -0.041 & 0.533 & 0.052 & 0.074 & 0.042 & 0.544 & 0.188 & 0.071 \\ 
        \textsc{Assort-I} & -0.002 & 0.237 & 0.067 & 0.037 & 0.000 & 0.842 & 0.443 & 0.264 & -0.037 & 0.161 & 0.021 & 0.028 \\ \bottomrule
    \end{tabular}
    \end{adjustbox}
\end{table}

\begin{figure*}[!t]
\centering

\subfloat[LightGCN]{
    \includegraphics[width=0.9\textwidth]{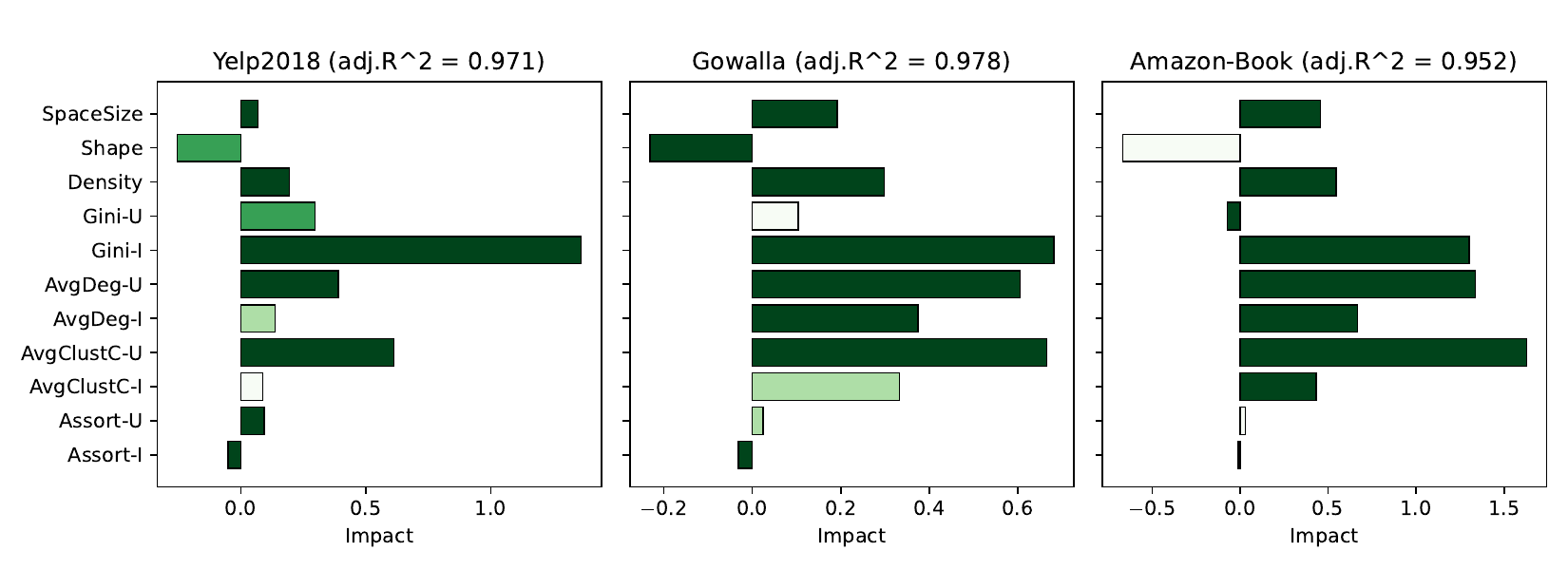}}

\subfloat[DGCF]{
    \includegraphics[width=0.9\textwidth]{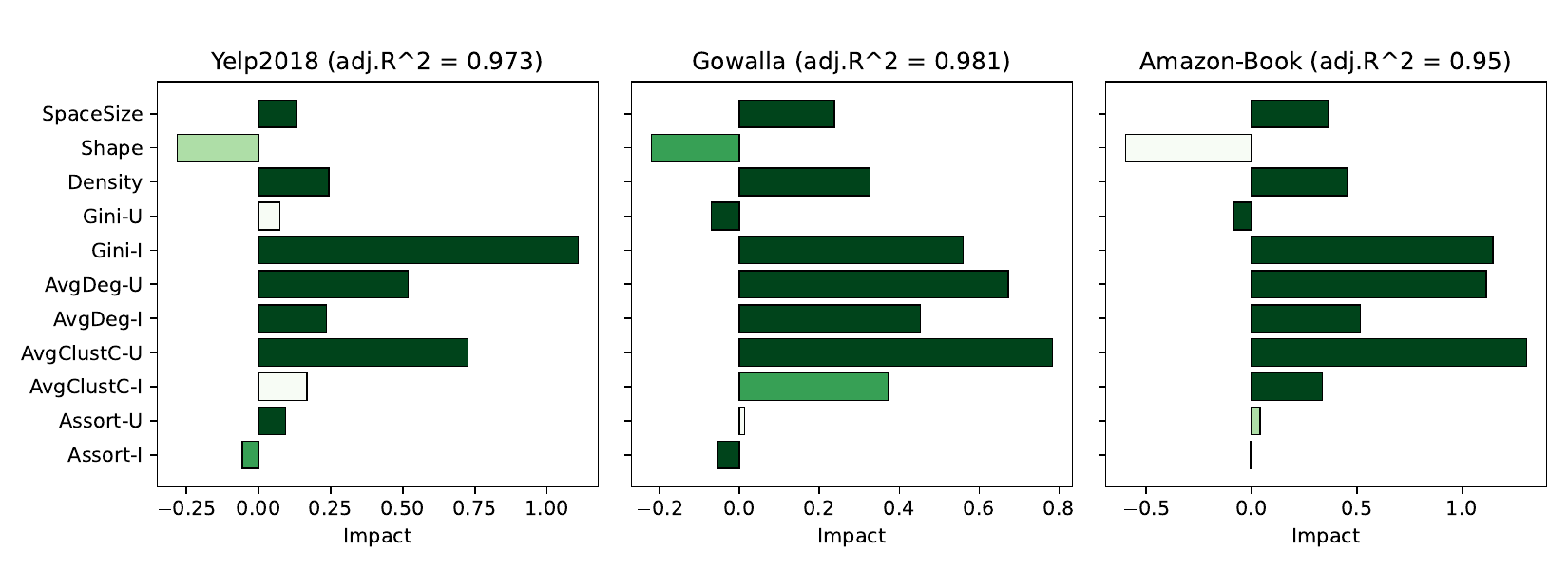}}

\subfloat[SVD-GCN]{
    \includegraphics[width=0.9\textwidth]{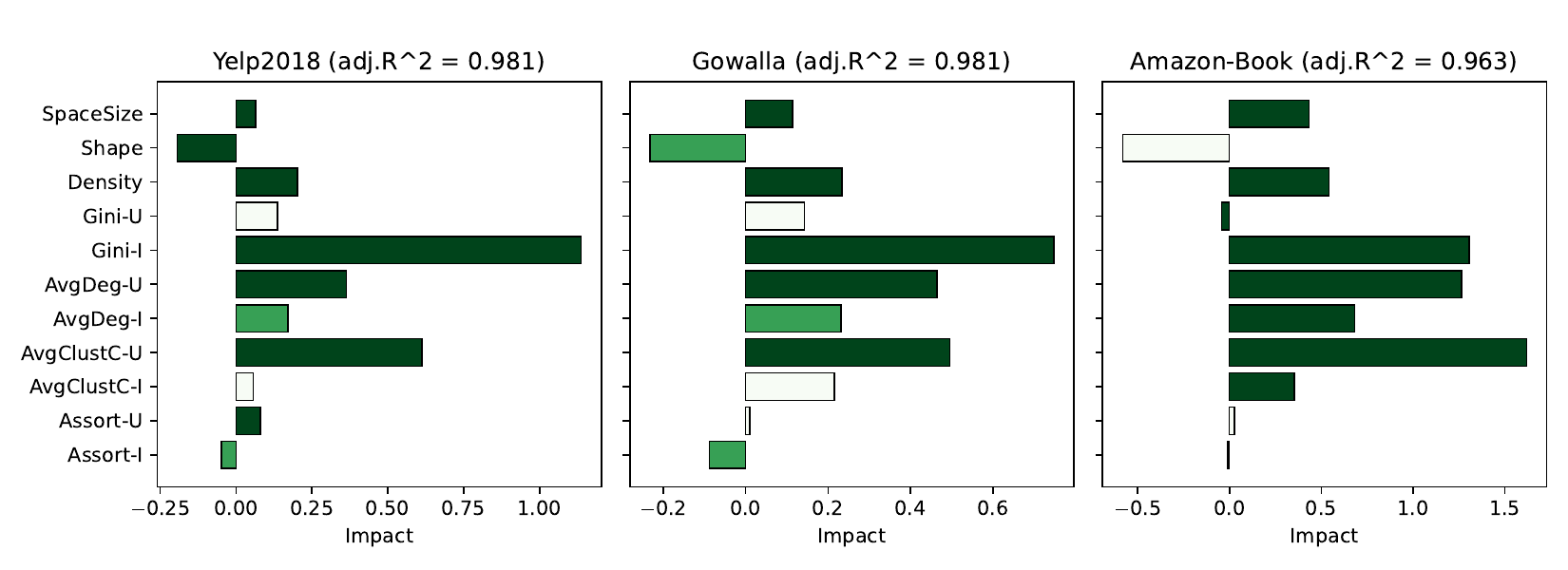}}

\subfloat{
    \includegraphics[width=0.85\textwidth]{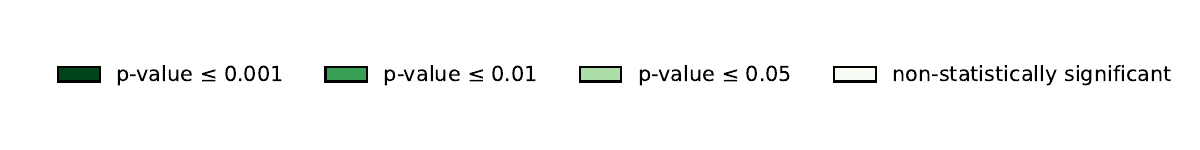}}

\caption{Visual representation of the impact of dataset characteristics on the recommendation performance (Recall@20) of GNNs-based recommender systems, for each dataset/model setting in \citep{DBLP:conf/recsys/MalitestaPAMNS24}. Bar plot length and direction represent the impact magnitude and whether there is a direct/inverse correspondence between characteristic and performance. Finally, the darker the bar plots, the higher their statistical significance.}\label{fig:rq1}
\end{figure*}

In particular, the results reported in~\Cref{fig:rq1} offer an illustrative example of this process. When viewed through the lens of our framework, they reveal that degree-related characteristics consistently exert a positive influence on performance, while more complex topological quantities, such as clustering coefficients and assortativity, display model-dependent patterns. At the same time, classical characteristics like density and item-side inequality retain a substantial role, reflecting the interplay between factorization-style effects and neighborhood aggregation in GNN-based recommenders. Rather than focusing on exact numerical values,~\Cref{fig:rq1} should thus be interpreted as a qualitative map of how different groups of characteristics jointly shape the behavior of topology-aware recommendation models.

\chapter{Recent trends in GNNs-based recommendation}\label{ch:recent_trends}

In the previous sections of this monograph, we have primarily focused on GNN-based recommender systems that learn from the user-item interaction graph in its original spatial form, typically without relying on additional sources of information. More recent research has progressively moved beyond this traditional setting by proposing methodologies that either exploit the graph from alternative perspectives or explicitly modify its topology to improve recommendation quality, robustness, or scalability. This section provides a non-exhaustive overview of these emerging directions. We first discuss spectral graph methodologies, highlighting how spectral information can be exploited to partition the interaction graph or filter its structural signals. We then review multimodal and LLM-enhanced recommender systems, showing how external semantic knowledge can be integrated into graph representations and, in the case of LLMs, even used to reshape the topology of the user-item interaction graph itself.

\section{Spectral graph methodologies}

While most GNN-based recommender systems operate in the spatial domain, an alternative line of research adopts a spectral perspective to analyze and exploit the structural properties of the user-item graph. Rather than focusing solely on message passing over graph neighborhoods, these methods leverage spectral information to improve either the scalability of recommendation algorithms or the quality of the learned representations. In the following, we review two representative research directions, namely, graph partitioning techniques for scalable recommendation and spectral denoising approaches for learning robust user-item representations. Moreover, we present recent studies that empirically investigate how such model families address the topological structure of the user-item graph.

\subsection{Partition-based recommendation and partitions' topology}

The first line of research in spectral-based recommendation addresses the scalability limitations of graph-based recommender systems by explicitly decomposing the recommendation problem into a collection of smaller and more manageable subproblems. Rather than processing the entire item graph as a single structure, these approaches partition it into multiple groups of closely related items, so that similarities can be learned independently within each partition. This divide-and-conquer strategy substantially reduces both computational complexity and memory requirements, making it possible to scale graph-based recommendation to much larger item collections.

In this respect, FPSR \citep{DBLP:conf/www/WeiMC23} follows this principle by first partitioning the item graph into nearly independent subgraphs using graph partitioning techniques. Each partition is processed separately to estimate local item similarities, replacing a single large optimization problem with several smaller ones that can be solved more efficiently and, if desired, in parallel. Although this decomposition greatly improves scalability, partitioning the graph may also remove useful dependencies between items belonging to different subgraphs. To alleviate this issue, FPSR complements the locally learned similarities with a global component derived from the overall structure of the interaction graph, allowing information that spans multiple partitions to be incorporated into the final recommendation model. In this way, the method aims to preserve the benefits of local optimization while retaining a global view of user-item relationships.

FPSR$+$ \citep{DBLP:journals/tkde/WeiCM24} further extends this idea by explicitly addressing situations in which the graph decomposition produces very small or highly unbalanced partitions. In these cases, the limited amount of interaction data available within individual partitions may lead to unreliable similarity estimates. To improve robustness, the method introduces a shared set of representative items that act as bridges across different partitions (defined by the authors as hub items). By allowing these representative items to convey information between otherwise disconnected subgraphs, the model can recover part of the global interaction structure that is lost during partitioning, leading to more stable similarity estimation and improved recommendation.

With reference to these two approaches, the authors in \citep{DBLP:conf/ecir/GioiaPBN26} have recently investigated the sensitivity of the FPSR family to the graph partitioning strategy by analyzing the impact of the partition's topology (in terms of its granularity) on recommendation performance. Particular attention is devoted to the partition-size parameter, which determines how granular the item graph is decomposed. Since the effectiveness of graph partitioning depends on balancing computational efficiency with the preservation of meaningful structural information, the authors examine how different partition sizes affect the behavior of both the original FPSR and other variants of the FPSR+ version, namely, FPSR+$_D$ and FPSR+$_F$.

The results reported in Table \ref{tab:fpsr} on the Douban dataset~\citep{DBLP:conf/cikm/ShiZLYYW15} show that the partition size (indicated as $\tau$) has a substantial influence on the performance of the original FPSR model. When the graph is divided into excessively small partitions, recommendation accuracy deteriorates noticeably because each partition contains only limited interaction information and captures fewer relationships with items belonging to other partitions. As the partition size increases reaching the best found partition size ($\tau_{\text{best}}$), this effect is progressively mitigated, leading to significantly more accurate recommendations.

FPSR+$_D$ and FPSR+$_F$ exhibit a considerably more stable behavior across different partitioning settings. By introducing representative items that are shared across partitions, these models preserve part of the global interaction structure that would otherwise be lost during graph decomposition. This mechanism reduces the sensitivity of local similarity estimation to small or unbalanced partitions, resulting in more consistent recommendation performance even under aggressive partitioning strategies. Among the two variants, FPSR+$_D$ demonstrates the highest degree of robustness, whereas FPSR+$_F$ provides a more moderate but still noticeable improvement over the original FPSR.

Overall, the analysis highlights that the effectiveness of partition-based recommendation depends not only on the graph decomposition itself but also on how information is propagated across partition boundaries. Appropriate tuning of the partition granularity can substantially improve the performance of the original FPSR model, while the introduction of hub items in FPSR+ offers an additional level of robustness by preserving global structural information when the graph is partitioned into many small subgraphs.

\begin{table}[!t]
\caption{Recommendation results of FPSR, FPSR$+_D$, and FPSR$+_F$ with varying values of $\tau$ (i.e., partition size) on Douban, as presented in the work by \citet{DBLP:conf/ecir/GioiaPBN26}. The value $\tau_{\text{best}}$ stands for the best configuration found in the authors' analysis.}
\label{tab:fpsr}
\centering
\footnotesize
\setlength{\tabcolsep}{6pt}

\begin{tabular}{lcccccc}
\toprule
& \multicolumn{2}{c}{FPSR ($\tau_{\text{best}}=0.5$)}
& \multicolumn{2}{c}{FPSR$+_D$ ($\tau_{\text{best}}=0.4$)}
& \multicolumn{2}{c}{FPSR$+_F$ ($\tau_{\text{best}}=0.4$)} \\
\cmidrule(r){2-3}\cmidrule(r){4-5}\cmidrule(r){6-7}
& R@20 & N@20 & R@20 & N@20 & R@20 & N@20 \\
\midrule
$\tau = 0.05$        & 0.1631 & 0.1561 & 0.2068 & 0.1882 & 0.1731 & 0.1602 \\
$\tau = 0.15$        & 0.1918 & 0.1767 & 0.2097 & 0.1908 & 0.1953 & 0.1771 \\
$\tau = 0.25$        & 0.1930 & 0.1780 & 0.2109 & 0.1912 & 0.1957 & 0.1780 \\
\midrule
$\tau_{\text{best}}$ & 0.2034 & 0.1855 & 0.2132 & 0.1928 & 0.2041 & 0.1841 \\
\bottomrule
\end{tabular}
\end{table}

\subsection{Robustness of spectral denoising methods to data sparsity}

Rather than processing the interaction graph exclusively in the spatial domain, spectral denoising methods analyze its spectral structure to identify the most informative patterns while attenuating noisy or redundant signals. The underlying intuition is that user preferences often exhibit an inherent low-dimensional organization, meaning that much of the relevant information can be captured by a limited number of dominant latent components.

Building on this intuition, these methods first derive a compact spectral representation of the interaction graph and then exploit it in different ways. For instance, SVD-GCN~\citep{DBLP:conf/cikm/PengSM22}, already discussed in this monograph, constructs an initial low-rank representation of the user-item graph and subsequently refines it through graph convolutional layers. Other models further extend this idea by explicitly manipulating the spectral representation. SGDE and RSGDE~\citep{DBLP:journals/tois/PengSM24} selectively suppress or reweight less informative spectral components to reduce the influence of noise and alleviate issues such as over-smoothing, while CSGDE~\citep{DBLP:journals/tois/PengSM24} complements this strategy with a contrastive learning objective aimed at preserving discriminative information in the learned embeddings. A different direction is followed by SGFCF~\citep{DBLP:conf/kdd/PengLSM24}, which applies a predefined spectral filter to the interaction graph, producing recommendation scores through a closed-form solution without requiring iterative optimization.

Although these models differ in their architectural choices and training strategies, they all share a common view of recommendation as a spectral filtering problem. Their main objective is to preserve the dominant structural patterns of the user-item interaction graph while suppressing components that are more likely to reflect noise or spurious correlations. Consequently, the recommendation process can be interpreted as enhancing the most informative spectral signals rather than treating all graph components as equally relevant.

The effectiveness of this family of methods relies on the assumption that the dominant spectral components of the interaction graph encode the essential preference patterns shared across users and items, whereas weaker components predominantly capture noise or less informative variations. Under this perspective, emphasizing the most significant spectral structures leads to cleaner representations and, ultimately, to more reliable recommendations even in data sparse scenarios.

To investigate whether spectral denoising methods are intrinsically more robust to sparse interaction data, the authors in \citep{DBLP:conf/sigir/GioiaPBN26} have recently conducted a controlled study on the Douban dataset~\citep{DBLP:conf/cikm/ShiZLYYW15}, where the amount of training information is progressively reduced while keeping the evaluation protocol unchanged. This setting provides a convenient testbed for analyzing the behavior of spectral denoising models under increasing levels of sparsity without introducing additional sources of variability.

As illustrated in \Cref{fig:douban_sparsity}, the analysis focuses on representative methods from the spectral denoising family, namely SVD-GCN, SVD-GCN-S (a variant of SVD-GCN), GDE \citep{DBLP:conf/sigir/PengSM22}, and SGFCF. Their performance is compared against two well-established neighborhood-based recommenders, SLIM \citep{DBLP:conf/icdm/NingK11} and ItemkNN \citep{DBLP:journals/tois/DeshpandeK04}, which serve as strong non-spectral baselines.

To simulate increasingly sparse scenarios, the test set is kept fixed while only the amount of training data is reduced. After partitioning the dataset into training, validation, and test sets, progressively smaller training sets are generated by retaining only 60\%, 40\%, and 20\% of the original training interactions. This protocol isolates the effect of diminishing training signal, ensuring that any variation in performance can be attributed solely to the increasing sparsity of the available interaction data.

\begin{figure}[t]
\centering

\begin{minipage}{0.48\columnwidth}
\centering
\begin{tikzpicture}
\begin{axis}[
    width=\linewidth,
    height=5.2cm,
    xlabel={Density (\%)},
    ylabel={Recall@20},
    xmin=15, xmax=65,
    ymin=0.035, ymax=0.095,
    xtick={20,40,60},
    ytick={0.03, 0.04, 0.05,0.06,0.07,0.08,0.09},
    xmajorgrids=true,
    ymajorgrids=true,
    grid style={dashed,gray!40},
    scaled y ticks=false,
    yticklabel style={
        /pgf/number format/fixed,
        /pgf/number format/precision=2
    },
    tick label style={font=\small},
    label style={font=\small},
    legend to name=sharedlegend,
    legend columns=3,
    legend style={font=\small, draw=none, column sep=0.5em, row sep=0.5em},
    cycle list={
    {blue, thick, mark=o},
    {red, thick, mark=square},
    {green!60!black, thick, mark=triangle},
    {orange, thick, mark=diamond},
    {purple, thick, mark=pentagon},
    {black, thick, mark=x}
    }
]

\addplot coordinates {(20,0.0592) (40,0.0662) (60,0.0785)};
\addlegendentry{ItemkNN}

\addplot coordinates {(20,0.0576) (40,0.0850) (60,0.0886)};
\addlegendentry{SLIM}

\addplot coordinates {(20,0.0684) (40,0.0787) (60,0.0726)};
\addlegendentry{SVD-GCN}

\addplot coordinates {(20,0.0700) (40,0.0750) (60,0.0809)};
\addlegendentry{SVD-GCN-S}

\addplot coordinates {(20,0.0586) (40,0.0679) (60,0.0701)};
\addlegendentry{GDE}

\addplot coordinates {(20,0.0654) (40,0.0649) (60,0.0570)};
\addlegendentry{SGFCF}

\end{axis}
\end{tikzpicture}
\end{minipage}
\hfill
\begin{minipage}{0.48\columnwidth}
\centering
\begin{tikzpicture}
\begin{axis}[
    width=\linewidth,
    height=5.2cm,
    xlabel={Density (\%)},
    ylabel={nDCG@20},
    xmin=15, xmax=65,
    ymin=0.035, ymax=0.095,
    xtick={20,40,60},
    ytick={0.03, 0.04, 0.05,0.06,0.07,0.08,0.09,0.09},
    xmajorgrids=true,
    ymajorgrids=true,
    grid style={dashed,gray!40},
    scaled y ticks=false,
    yticklabel style={
        /pgf/number format/fixed,
        /pgf/number format/precision=2
    },
    tick label style={font=\small},
    label style={font=\small},
    cycle list={
    {blue, thick, mark=o},
    {red, thick, mark=square},
    {green!60!black, thick, mark=triangle},
    {orange, thick, mark=diamond},
    {purple, thick, mark=pentagon},
    {black, thick, mark=x}
    }
]

\addplot coordinates {(20,0.0504) (40,0.0565) (60,0.0643)};
\addplot coordinates {(20,0.0520) (40,0.0731) (60,0.0735)};
\addplot coordinates {(20,0.0564) (40,0.0612) (60,0.0538)};
\addplot coordinates {(20,0.0581) (40,0.0573) (60,0.0645)};
\addplot coordinates {(20,0.0465) (40,0.0555) (60,0.0524)};
\addplot coordinates {(20,0.0539) (40,0.0500) (60,0.0403)};

\end{axis}
\end{tikzpicture}
\end{minipage}

\vspace{2mm}
\pgfplotslegendfromname{sharedlegend}
\caption{Recommendation performance on the Douban dataset when considering different density levels (20\%, 40\% and 60\%) on the training set. Results are taken from the work by \citet{DBLP:conf/sigir/GioiaPBN26}.}
\label{fig:douban_sparsity}
\end{figure}

The results reported in \Cref{fig:douban_sparsity} indicate that the performance of nearly all methods deteriorates as fewer training interactions become available. Among the spectral approaches, SGFCF exhibits a distinctive behavior, maintaining stable performance and even showing a slight improvement under the sparsest setting, suggesting that its fixed spectral filtering strategy is less sensitive to the reduction of training data than the mechanisms adopted by the other models. In contrast, SVD-GCN, SVD-GCN-S, and GDE follow a degradation trend comparable to that observed for the non-spectral baselines. Although SVD-GCN and SVD-GCN-S remain competitive in the most challenging setting, they do not consistently outperform SLIM or ItemkNN, while GDE does not display any evident robustness advantage.

Overall, these findings suggest that emphasizing the dominant spectral components of the interaction graph does not automatically translate into greater robustness to data sparsity. Instead, the benefits appear to depend on the specific filtering strategy adopted by each model, with SGFCF representing a notable exception. More generally, the experimental evidence indicates that robustness to sparse interaction data should be regarded as an empirical property of individual model-dataset combinations rather than as an inherent consequence of spectral denoising itself.

\section{Multimodality and large language models}

Recent advances in graph-based recommender systems have progressively moved beyond the sole exploitation of user-item interactions by incorporating complementary sources of information. In particular, multimodal data and large language models (LLMs) have emerged as two major research directions for enriching graph representations with semantic knowledge that is not directly observable from interaction data alone. In the following, we first provide an overview of multimodal GNN-based recommendation and then focus on LLM-enhanced recommender systems, highlighting how LLMs can explicitly reshape the topology of the user-item interaction graph. We deepen into this aspect through a recent empirical analysis that investigates how LLMRec \citep{DBLP:conf/wsdm/WeiRTWSCWYH24}, a notable example in the literature of multimodal LLM-based recommendation, modifies the topology of the user-item graph.

\subsection{Multimodal GNNs-based recommendation: an overview}

The integration of multimodal information into recommender systems~\citep{DBLP:journals/tors/MalitestaCPMNS25, DBLP:journals/csur/LiuHXZGWLT25, DBLP:journals/corr/abs-2302-04473, DBLP:journals/corr/abs-2502-15711, DBLP:conf/kdd/LiuZYDD0ZZD24, DBLP:journals/inffus/PanPCC26} has demonstrated substantial benefits across a wide range of application domains, including e-commerce~\citep{DBLP:conf/kdd/Xv0GGLDXZ23}, fashion~\citep{DBLP:conf/sigir/ChenCXZ0QZ19}, music~\citep{DBLP:conf/recsys/OramasNSS17}, food~\citep{DBLP:journals/tmm/MinJJ20}, and micro-video recommendation~\citep{DBLP:journals/tmm/CaiQFX22}. By leveraging multiple data modalities associated with items, these systems are able to learn richer representations that improve recommendation quality. Recent advances in multimodal learning~\citep{DBLP:books/acm/18/BaltrusaitisAM18} have further accelerated the development of multimodal recommender systems, enabling them to tackle long-standing issues such as data sparsity and the cold-start problem~\citep{DBLP:conf/aaai/HeM16,DBLP:conf/recsys/OramasNSS17}. Moreover, multimodal information facilitates a deeper understanding of user-item interactions by exploiting attention mechanisms, which can reveal users' preferences and intentions while also improving the interpretability of recommendation models~\citep{DBLP:conf/sigir/ChenCXZ0QZ19}.

The rapid adoption of graph neural networks in recommender systems~\citep{DBLP:conf/sigir/0001DWLZ020} has encouraged the development of methods that incorporate multimodal information into both user-item bipartite graphs and knowledge graphs~\citep{DBLP:journals/tmm/WangWYWSN23}. These approaches typically refine user and item representations through message-passing mechanisms that effectively propagate multimodal information across the graph structure. Early graph-based methods primarily relied on injecting multimodal item features into the recommendation pipeline~\citep{DBLP:conf/mm/WeiWN0HC19}. More recent solutions, however, construct dedicated graph representations for each modality and explicitly disentangle user preferences across different modalities~\citep{DBLP:conf/cikm/KimLSK22}, while simultaneously modeling both global and local user interests~\citep{DBLP:conf/aaai/GuoL0WSR24}. Other studies focus on capturing the structural relationships among multimodal item representations within the catalog~\citep{DBLP:conf/mm/Zhang00WWW21,DBLP:conf/mm/LiuYLWTZSM21, DBLP:conf/mm/ZhouS23, DBLP:conf/ecai/Zhou0Z023}, often by incorporating self-supervised~\citep{DBLP:conf/www/ZhouZLZMWYJ23, DBLP:conf/mm/Yu0LB23, DBLP:conf/www/WeiHXZ23, DBLP:conf/aaai/00030000N25, DBLP:journals/tmm/TaoLXWYHC23} and contrastive~\citep{DBLP:conf/sigir/Yi0OM22, DBLP:conf/mm/Su0L0024} learning paradigms. More recently, researchers have explored the integration of advanced generative models, including variational autoencoders~\citep{DBLP:journals/tmm/ZhouM24, DBLP:journals/tmm/YiC24} and diffusion models~\citep{DBLP:conf/mm/JiangX0LLH24}. Additional research directions include transforming multimodal representations into the frequency domain~\citep{DBLP:conf/wsdm/OngK25, DBLP:conf/aaai/Yu0LB25}, transferring multimodal knowledge from large-scale architectures to more compact models~\citep{DBLP:conf/apweb/WangYCJYKWHL24, DBLP:conf/www/WeiTXJH24}, and improving recommendations for long-tail items in the catalog~\citep{DBLP:conf/cikm/LinMWLZ024}.

\subsection{Multimodal LLMs \& GNNs: the notable example of LLMRec}

A representative example of the integration of large language models (LLMs) and multimodal information into graph-based recommender systems is LLMRec~\citep{DBLP:conf/wsdm/WeiRTWSCWYH24}, one of the earliest frameworks to combine these paradigms within a unified recommendation architecture. LLMRec enhances graph-based recommendation through LLM-driven graph augmentation and dedicated mechanisms for denoising and feature refinement. By exploiting the semantic reasoning capabilities and extensive knowledge encoded in LLMs, LLMRec alleviates the limitations imposed by sparse implicit feedback, ultimately improving the quality of user-item representations. More specifically, the interaction graph is enriched by augmenting three complementary components: (i) user-item interaction edges; (ii) item node attributes; and (iii) user node profiles. For the sake of this monograph, we only deepen into the description of the graph and node augmentation procedures, leaving aside the presentation of the denoising mechanisms; the interested reader may refer to the original work for a complete presentation of the approach.

To augment the user-item interaction graph, LLMRec employs the LLM as a knowledge-aware sampler that generates informative pairwise user-item training instances. This strategy aims to alleviate the effects of noisy interactions, reducing both false-positive user-item relationships and false-negative non-interactions. For each user, the LLM is prompted with the historical interaction sequence, the corresponding side information (e.g., year and genre), and a candidate item set produced by a backbone recommender such as LATTICE~\citep{DBLP:conf/mm/Zhang00WWW21} or MMSSL~\citep{DBLP:conf/www/WeiHXZ23}. Based on this information, the LLM identifies the items that the user is most likely and least likely to interact with. These predictions are then adopted as positive and negative samples during the BPR optimization~\citep{DBLP:conf/uai/RendleFGS09}.

The framework also exploits the LLM to augment node attributes. Specifically, user profiles are generated by summarizing each user's historical interactions, whereas enriched item descriptions are produced from prompts containing the available item metadata. The resulting textual information is subsequently encoded into embedding representations and incorporated into the recommendation model.

Once the augmented user and item information has been generated, LLMRec incorporates it into the original graph through a three-stage pipeline. First, an augmented semantic projection module reduces the dimensionality of the LLM-enhanced feature vectors. Next, a collaborative context injection stage exploits a lightweight Graph Neural Network (GNN) encoder to infuse higher-order collaborative signals into the projected user and item features. Finally, a 
semantic feature fusion module combines the enhanced semantic representations with the corresponding collaborative embeddings using a weighting coefficient and $L_2$ normalization, yielding the final user and item representations. Notably, the model is optimized using triplets sampled from the union of the original user-item graph set and a subset of the LLM-augmented interactions.

\subsection{A topological analysis of LLMRec's augmentation strategy}

One of the main strategies representing the key novelty of LLMRec is the augmentation of the historical user-item interaction graph through LLMs to refine noisy recorded edges among users and items. In this respect, it would be interesting to understand how the same graph changes after the LLM intervention from a \textit{topological} viewpoint. This is, in fact, an open research question, not necessarily referred to the graph-based recommendation domain. Other works, where the most prominent is the one by \citet{DBLP:conf/iclr/FatemiHP24}, have already questioned how and to what extent LLMs are able to understand and act over graph data. In this sense, the recent work by \citet{DBLP:conf/recsys/FiorettiLPMPNN25} analyzes this research dimension for the prominent example of LLMRec, studying the topological impact of the LLM's augmentation on the user-item interaction graph.

Table~\ref{tab:llmrec} summarizes the topological analysis conducted by \citet{DBLP:conf/recsys/FiorettiLPMPNN25} on the Amazon-Music \citep{DBLP:conf/www/HeM16, DBLP:conf/sigir/McAuleyTSH15} and the Netflix dataset\footnote{Shortened link to the dataset: \url{https://tinyurl.com/3wk3xchb}.}; as for the Netflix dataset, the authors consider two versions (indicated in the table with Netflix$^{(1)}$ and Netflix$^{(2)}$), namely, with standard and a custom training splitting. The authors' rationale is to compare the original interaction graphs with those obtained through LLM-based augmentation. The experiments consider two graph-based recommenders, LATTICE and MMSSL, as candidate generators, and three LLMs (\emph{gpt-3.5-turbo-16k}\footnote{\url{https://developers.openai.com/api/docs/models/gpt-3.5-turbo}.}, \emph{gpt-4-turbo}\footnote{\url{https://developers.openai.com/api/docs/models/gpt-4-turbo}.}, and \emph{Llama-3-405B-Instruct}\footnote{\url{https://huggingface.co/meta-llama/Llama-3.1-405B-Instruct}.}) to refine the candidate items.

Overall, the reported results consistently indicate that LLM-based augmentation substantially reshapes the topology of the user-item graph. The most evident effect is the increase in both Density and Average Degree across all datasets and experimental configurations, suggesting that the augmented graphs become more connected and contain a richer set of interactions. Such an increase is observed for both user and item partitions, indicating that users are connected to a larger number of items while items receive interactions from a broader set of users. According to the authors, this denser connectivity may facilitate information propagation through the graph and ultimately provide a richer space for recommendation.

\begin{table}[!t]
    \centering
    \caption{Topological analysis of LLMRec's augmentation strategy on the user-item graph, considering different LLMs and recommendation backbones. Results are taken from the study by \citep{DBLP:conf/recsys/FiorettiLPMPNN25}.}
    \label{tab:llmrec}
    \resizebox{\textwidth}{!}{%
    \begin{tabular}{ccccccccccc}
       \toprule
        \textbf{LLM} & \textbf{Baseline} & \textbf{Density} & \textbf{Gini-U} & \textbf{Gini-I} & \textbf{AvgDeg-U} & \textbf{AvgDeg-I} & \textbf{AvgClustC-U} & \textbf{AvgClustC-I} & \textbf{Assort-U} & \textbf{Assort-I} \\
        \midrule
        \multicolumn{11}{c}{\textit{Amazon-Music}} \\ \midrule
        \textbf{\emph{Original}} & -- & 0.0028 & 0.4929 & 0.5012 & 5.6015 & 14.4962 & 0.1184 & 0.0496 & -0.0884 & 0.0533 \\
        \midrule
        \multirow{2}{*}{gpt-3.5-turbo-16k} & LATTICE & 0.0038 & 0.3632 & 0.5497 & 6.3730 & 16.4927 & 0.1117 & 0.0470 & -0.0371 & 0.0322 \\
        & MMSSL & 0.0038 & 0.3632 & 0.5780 & 6.6823 & 17.2932 & 0.1114 & 0.0457 & -0.0164 & 0.0333 \\
        \midrule
        \multicolumn{11}{c}{\textit{Netflix\textsuperscript{(1)}}} \\ \midrule
        \textbf{\emph{Original}} & -- & 0.0006 & 0.1187 & 0.7493 & 5.2280 & 8.3996 & 0.1102 & 0.1474 & 0.2250 & -0.0357 \\
        \midrule
        \multirow{2}{*}{gpt-3.5-turbo-16k} & LATTICE & 0.0009 & 0.0859 & 0.7839 & 6.3641 & 10.2006 & 0.0963 & 0.1387 & 0.1671\textsuperscript{†} & -0.1542 \\
        & MMSSL & 0.0009 & 0.0859 & 0.8076 & 6.3907 & 10.2554 & 0.0956 & 0.1285 & 0.2529\textsuperscript{†} & -0.1857 \\
        \midrule
        \multicolumn{11}{c}{\textit{Netflix\textsuperscript{(2)}}} \\ \midrule
        \textbf{\emph{Original}} & -- & 0.0006 & 0.1467 & 0.7341 & 4.2291 & 7.4407 & 0.1380 & 0.1605 & 0.2502 & -0.0054 \\
        \midrule
        \multirow{2}{*}{Llama-3-405B-Instruct} & LATTICE &  0.0008 & 0.0997 & 0.7994 & 6.2271 & 10.9053 & 0.1019 & 0.1261 & 0.2001\textsuperscript{†} & -0.1710 \\
        & MMSSL & 0.0008 & 0.0997 & 0.8147 & 6.2265 & 10.9042 & 0.1026 & 0.1101 & 0.2738\textsuperscript{†} & -0.2408 \\
        \midrule
        \multirow{2}{*}{gpt-4-turbo} & LATTICE  & 0.0008 & 0.1037 & 0.7977 & 6.1347 & 10.7434 & 0.1037 & 0.1276 & 0.1506\textsuperscript{†} & -0.1643 \\
        & MMSSL & 0.0008 & 0.1033 & 0.8120 & 6.0657 & 10.6227 & 0.1044 & 0.1143 & 0.2048\textsuperscript{†} & -0.2425 \\
        \bottomrule
        \multicolumn{11}{l}{\footnotesize \textsuperscript{(1)} With the original training set by the authors.}\\
        \multicolumn{11}{l}{\footnotesize\textsuperscript{(2)} With a 90-10 random split of the original training set.}\\
        \multicolumn{11}{l}{\footnotesize\textsuperscript{†} These values represent a trend divergence with respect to the original dataset version.}   
    \end{tabular}}
\end{table}

The degree distributions also exhibit noticeable changes after augmentation. In particular, the Gini coefficient computed on the user partition consistently decreases, indicating that user activity becomes more evenly distributed rather than being concentrated on a limited number of highly active users. Conversely, the Gini coefficient on the item partition generally shows a slight increase, especially when MMSSL is adopted as the backbone recommender, suggesting that item popularity remains relatively concentrated and may even become slightly more skewed despite the richer interaction graph.

The structural organization of the graph is further characterized by a moderate reduction in the Average Clustering Coefficient for both user and item partitions, with the decrease being more pronounced on the item side. Since this metric captures the tendency of nodes to share common neighborhoods, the observed reduction suggests that the newly introduced interactions are distributed across a broader portion of the graph instead of reinforcing existing local communities. The authors interpret this as an indication of increased interaction diversity, which could positively affect the diversity of the generated recommendations.

Finally, the analysis of Degree Assortativity reveals more dataset-dependent trends. On the Amazon-Music dataset, User Assortativity increases while remaining negative, indicating a partial reduction of the graph's disassortative structure, whereas Item Assortativity decreases while remaining positive, implying that highly popular items become less likely to be associated with other popular items. On both 
Netflix dataset versions, User Assortativity remains positive, although the direction and magnitude of the variation depend on the specific combination of backbone recommender and LLM. This is the only topological property for which the authors observe inconsistent trends across all experimental settings, a behavior highlighted in Table~\ref{tab:llmrec} through the dagger ($\dagger$) notation. In contrast, Item Assortativity consistently becomes more negative, suggesting stronger interactions between highly popular and less popular items. Moreover, employing MMSSL as the generator generally produces larger absolute assortativity values than LATTICE.
\chapter{Take-home messages}\label{ch:message}

The power of GNN-based recommender systems stems from their ability to directly leverage the user-item interaction graph. However, this power is not monolithic; it is shaped by a complex interplay between model architecture and data characteristics. Throughout this monograph, we have systematically deconstructed these systems by establishing a formal pipeline (Section~\ref{ch:pipeline}), defining the classical and topological properties of their data (Section~\ref{ch:char_class}-\ref{ch:topo_gnn}), and empirically measuring their interplay through a rigorous explanatory framework (Section~\ref{ch:framework}).
This comprehensive analysis now allows us to distill our findings into a series of actionable take-home messages. We begin by focusing on the GNNs themselves, examining how fundamental architectural choices dictate their performance and behavior. Subsequently, we will delve into the properties of the data, providing guidance on how to study and connect them to a model's expected performance. These messages are designed to bridge the gap between theoretical analysis and practical application, providing clear guidelines for building and evaluating the next generation of GNN-based recommender systems.

\section{Architectural choices drive performance and behavior}

The specific design of a GNN is not a trivial implementation detail; it is the primary determinant of its performance, biases, and limitations within recommender systems. As our model taxonomy (Section~\ref{ch:pipeline}) and empirical analysis (Section~\ref{ch:framework}) have shown, architectural choices create fundamental trade-offs between complexity, accuracy, and beyond-accuracy objectives. The following messages distill these critical lessons learned from the evolution of GNN architectures:

\begin{itemize}
    \item \textbf{Simplicity often outperforms complexity.}
    As detailed in our model taxonomy, the evolution from complex early models like NGCF to simpler architectures like LightGCN marked a pivotal moment. Our analysis confirms that removing non-linearities and feature transformations often leads to superior accuracy and efficiency, demonstrating that the core strength of GNNs in this domain is effective neighborhood aggregation, not convoluted transformations~\citep{DBLP:conf/sigir/0001DWLZ020}.
    \item \textbf{Over-smoothing is the primary architectural bottleneck.}
    The tendency for node embeddings to become indistinguishable after a few message-passing steps is a fundamental limitation discussed in Section~\ref{sec:exploration_depth}. This not only restricts most models to shallow depths but also negatively impacts beyond-accuracy metrics like novelty and diversity~\citep{DBLP:conf/recsys/AnelliDNSFMP22}. This is why models like SVD-GCN, which we analyzed, avoid iterative message-passing to capture higher-order signals without degradation.

    \item \textbf{Standard GNNs can amplify popularity bias.}
    The message-passing mechanism inherently favors nodes with high degrees. Our own experimental results in Section~\ref{ch:framework} (\Cref{fig:rq1}) support this, showing a consistent positive correlation between the item Gini coefficient (a measure of popularity concentration) and model performance. This confirms that GNNs tend to reward popular items, a critical consideration for fairness~\citep{DBLP:conf/ecir/AnelliDNMPP23}. 

    \item \textbf{Contrastive learning is a powerful tool for data sparsity.}
    To address the data sparsity defined in~\Cref{sec:density}, architectures like SGL and SimGCL have proven effective. By creating augmented views of the graph or embeddings, these models learn more robust representations that are less dependent on the local neighborhood structure, improving performance in sparse data regimes~\citep{DBLP:conf/sigir/WuWF0CLX21,DBLP:conf/sigir/YuY00CN22}.
    
\end{itemize}

\section{Relationship of classical dataset characteristics with performance}
Even though classical dataset characteristics like shape or density (Section~\ref{ch:char_class}) have been investigated previously for more simple recommendation algorithms, the relationship between such characteristics and GNN-based recommenders is a novel contribution of our work, as presented in Section~\ref{ch:framework}.
The most remarkable insights, as evidenced in \Cref{fig:rq1}, can be summarized as:
\begin{itemize}
    \item \textbf{GNN models benefit from more (dense) data available.} Consistent positive correlation between density and performance, throughout datasets and GNN models, evidence that when the GNN has access to less sparse information, it can create more meaningful representations of users and items, hence producing more accurate recommendations.
    
    \item \textbf{Too many users may be detrimental.} By the interplay between space size and shape properties, we observed that the raw number of potential interactions (space size) is not a consistent predictor of performance, whereas the relationship between the number of users and items (shape) is always inversely related to performance. This implies that, for a fixed number of items, increasing the number of users will negatively impact performance; symmetrically, for a given number of users, having more items will produce a lower shape, which has been shown to be related to better performance.
    
    \item \textbf{Skewed item distributions produce better models.} While this is not true for user distributions (as measured by Gini user), since correlations are either negligible or not significant, for item distributions (Gini item), the correlation values are very high in every case. This means that GNN models need item distributions where popular and non-popular items are easily discriminated from each other, because of their tendency to promote popular items (hence, if the distributions are too flat, the promoted items may not be the most appropriate ones).
\end{itemize}

\section{Topological characterization for performance prediction}

While classical dataset characteristics, such as sparsity and density (Section~\ref{ch:char_class}), provide a necessary high-level overview, our work demonstrates that they tell only part of the story. To understand the potential and challenges of a dataset for a GNN-based recommender, practitioners must examine the user-item graph's structure in greater detail. The most potent insights come from its topological properties.
As we defined and re-interpreted in Section~\ref{ch:char_topo}, these properties translate abstract graph theory into tangible recommendation concepts:
\begin{itemize}
    \item \textbf{Node degree} is the most direct measure of user activity and item popularity, immediately quantifying the extent of the cold-start problem.
    \item \textbf{Clustering coefficient} captures the strength of the collaborative filtering signal, revealing the density of shared preferences within local neighborhoods. A high clustering coefficient implies a strong foundation for neighborhood-based methods.
    \item \textbf{Degree assortativity} uncovers the homophily within the graph. It tells us whether active users tend to engage with items that are also popular among other active users, a deeper, ``look-ahead'' property that classical metrics cannot capture.
    \item \textbf{Degree distribution} suggests the scale-free tendency of the user-item graph, where very few nodes have high degree (i.e., active users and popular items), while the majority of nodes present low degrees.
\end{itemize}

This is not merely a theoretical exercise. The explanatory framework in Section~\ref{ch:framework} provides strong empirical evidence that these topological features are powerful and often dominant predictors of GNN performance. Our regression analysis (\Cref{fig:rq1}) consistently showed that characteristics like average user degree and average user clustering coefficient have a statistically significant and positive impact on accuracy across different models and datasets.

Therefore, a critical take-home message is to make topological analysis a standard, non-negotiable step in any GNN-based recommendation project. This analysis moves beyond a simple sparsity diagnosis to a much richer understanding. It allows for the quantification of the inherent strengths and weaknesses of a dataset, sets realistic performance expectations, and guides the selection of an appropriate GNN architecture before investing in costly training and evaluation cycles.

\section{Unveiling topological properties within model formalizations} 

A GNN-based recommender does not learn from graph-structured data if not explicitly defined to do so; its mathematical formulation is explicitly or implicitly designed to embed the topological properties of that data. Understanding this connection is fundamental to moving beyond treating these models as black boxes. Our analysis in Section~\ref{ch:topo_gnn}, which dissected the equations of various GNNs, reveals that these properties are encoded in distinct ways, with significant consequences for model behavior.

\textbf{The most direct and explicit encoding is node degree.} As our review of model architectures showed, nearly every modern GNN for recommendation, from NGCF to LightGCN to SGL, incorporates node degree directly into its message-passing formulation via symmetric normalization. The $1/{(\sqrt{\sigma_u}\sqrt{\sigma_i})}$ term is not merely a mathematical trick for stabilization; it is a deliberate architectural choice that rescales the contribution of messages based on user activity and item popularity. This makes the model's learning process fundamentally dependent on the node degree distribution of the training graph.

\textbf{Other crucial properties, like the clustering coefficient and degree assortativity, are captured implicitly.} These characteristics are not represented by a specific variable in the GNN's equations. Instead, they are an emergent consequence of the multi-hop message-passing mechanism itself.
\begin{itemize}
    \item \textbf{Clustering} is implicitly learned through 2-hop paths (e.g., user-item-user). When two users have a high clustering coefficient, it means they share many common items. The message-passing process naturally aggregates information from these shared items, pulling the embeddings of these two users closer together in the latent space. Thus, the architecture inherently learns the collaborative filtering signal that the clustering coefficient measures.
    \item \textbf{Assortativity}, the tendency for nodes of similar degrees to connect, is captured over even deeper propagation paths. A multi-layer GNN can learn whether active users tend to interact with items that are also popular among other active users. This ``look-ahead'' property, as we termed it in Section~\ref{ch:char_topo}, is implicitly encoded as the model learns to navigate longer relationships in the graph.
\end{itemize}

The take-home message is that a model's architecture determines its topological sensitivity. The explicit reliance on node degree means that if the degree distribution of the training data is not representative, the model may be biased and fail to generalize. Furthermore, the depth of a GNN dictates its ability to implicitly capture more complex properties. A shallow, 1-layer model may only effectively leverage node degree, while a deeper model can begin to harness the signals from clustering and assortativity. This understanding helps explain the performance variations we observed in Section~\ref{ch:framework} and provides a powerful tool for predicting how a given model will behave on a new dataset with a unique topological signature.

\section{Connecting data topology to model performance} 

The ultimate goal of our analysis is to move beyond the costly and inefficient trial-and-error model selection paradigm. A recommender system's performance is not an intrinsic property of the model alone, but an emergent outcome of the dynamic interplay between a model's architectural sensitivities and the dataset's unique topological signature. Understanding and predicting this interplay is the fundamental step toward building more targeted and effective GNN-based systems.

Our explanatory framework, empirically validated in Section~\ref{ch:framework}, provides the tools to forge this connection. By first studying the topological properties of the data and then understanding how different GNNs embed these properties, we can predict how a given model will perform on a specific dataset. Our findings reveal several key principles:

\begin{itemize}
    \item \textbf{The factorization component and node degree are Foundational.} Against the common assumption that the message-passing mechanism is the sole source of GNNs' power, our empirical results suggest that the factorization component (i.e., the quality of the node embeddings) and its most direct topological influence (\textbf{node degree}) are the most consistent and powerful predictors of performance. This implies that the primary benefit of GNNs often comes from learning robust node representations that are effectively regularized by their immediate neighborhood, a task for which even shallow models are well-suited.
    
    \item \textbf{Higher-order topologies explain performance differences.} While node degree sets the baseline, the ability to leverage more complex properties like \textbf{clustering coefficient} and \textbf{degree assortativity} is what often differentiates the performance of various GNN architectures. As our analysis revealed, models with more complex message-passing schemes are better equipped to implicitly capture these signals. The stronger the collaborative signal in a dataset (high clustering) or the more defined its homophily patterns (high assortativity), the greater the potential advantage of a GNN that can effectively explore multi-hop neighborhoods.

    \item \textbf{Use topology as a predictive tool to guide model selection.} The most practical application of this framework is as a diagnostic and predictive tool. Before launching extensive experiments, a topological analysis can guide resource allocation. For instance, if a dataset exhibits a low clustering coefficient and weak assortativity, it is highly unlikely that a complex, deep GNN will outperform a simpler, more efficient baseline. This allows practitioners to avoid wasting experimental cycles on architectures that are topologically mismatched with the data. The obtained results and explanatory framework can help prevent situations where there is insufficient variability or richness in a topological property for a complex model to exploit, guiding the community toward more focused and justified model evaluation.
\end{itemize}

\chapter{Challenges and future directions}\label{ch:future}

Throughout this monograph, we have provided a comprehensive re-interpretation of GNN-based recommender systems from a topological perspective. By modeling the user-item interaction space as a bipartite graph, we have moved beyond treating GNNs as black-box predictors and instead have shown how their performance is fundamentally linked to the topological characteristics of the data. Our explanatory framework has demonstrated that these properties are not just passively present but are actively, if sometimes implicitly, encoded within the models' formulations.

This deep dive into the interplay between graph topology and recommendation performance serves as a natural springboard into the next crucial discussion: the challenges and future directions for the field. The very success of GNNs in this domain opens up a new frontier of complex questions. \textit{If topology is so critical, how can we design models that exploit it more deliberately? If our models are sensitive to these properties, what are the implications of noisy, sparse, and dynamic real-world data? And how must our evaluation methodologies evolve to capture this deeper understanding?}

This section addresses these questions by outlining the key challenges and promising future directions in GNN research, with a particular focus on their application in recommender systems. We categorize these into four main areas: \textbf{theoretical and model-centric challenges} that address the intrinsic limitations of GNN architectures; \textbf{data-centric challenges} concerning the quality and characteristics of graph data; \textbf{task-specific challenges} related to applying GNNs in complex domains like recommender systems; and \textbf{methodological challenges} that call for more rigorous and reproducible evaluation practices.

\section{Theoretical and model-centric challenges}

These challenges are inherent to the architectural and theoretical foundations of GNNs themselves. Addressing them is crucial for building more powerful, robust, and scalable models.

\vspace{0.5em}
\noindent \textbf{Expressiveness and isomorphism:} A well-documented limitation of many GNNs is their inability to distinguish between certain non-isomorphic graph structures. The expressive power of most message-passing GNNs is bounded by the Weisfeiler-Lehman (WL) test of graph isomorphism. This means that if two graphs are indistinguishable by the 1-WL test, a standard GNN will learn the same representation for them, limiting their ability to capture fine-grained topological details~\citep{DBLP:conf/iclr/XuHLJ19}. Research is actively exploring more expressive GNNs, such as those based on higher-order WL tests or subgraph-based message passing. The key challenge is to increase expressive power without incurring prohibitive computational costs, striking a balance between theoretical capability and practical feasibility. This remains an active area of research, with recent surveys mapping the landscape of GNN expressiveness~\citep{DBLP:journals/tkde/ZhangFLHZHL25}.

\vspace{0.5em}
\noindent \textbf{Oversmoothing and oversquashing:} These two related issues represent a major obstacle to building deep GNNs. \textbf{Oversmoothing} is the phenomenon where, as the number of GNN layers increases, the representations of all nodes converge to a single, uninformative point, losing their distinctiveness. \textbf{Oversquashing} occurs when information from distant nodes gets ``squashed'' into a fixed-size vector during message passing, creating an information bottleneck that prevents the model from effectively capturing long-range dependencies~\citep{DBLP:conf/iclr/ToppingGC0B22}. Potential solutions include architectural innovations like skip connections, graph rewiring to reduce bottlenecks, and moving beyond the standard message-passing paradigm altogether. Exploring techniques from differential geometry to better understand graph curvature and its impact on information flow is a particularly promising avenue.

\vspace{0.5em}
\noindent \textbf{Scalability:} Applying GNNs to massive, web-scale graphs with billions of nodes and edges remains a major computational hurdle. The ``neighborhood explosion'' problem, where the number of neighbors to consider grows exponentially with the number of layers, makes full-batch training intractable.
While various sampling techniques have been developed, future work should focus on improving their efficiency and reducing bias. Additionally, research into decentralized and parallel training algorithms, as well as hardware-software co-design for GNN acceleration, will be critical for scaling to industrial-sized problems.

\section{Data-centric challenges}
The performance of GNNs is intrinsically tied to the quality and characteristics of the underlying graph data. Real-world graphs, however, present numerous challenges that are often abstracted away in benchmark datasets.

\vspace{0.5em}
\noindent \textbf{Data scarcity and quality:} The field is hampered by a scarcity of large, high-quality, and diverse public datasets. Many real-world graphs are proprietary, forcing researchers to rely on a small set of well-worn benchmarks that may not be representative of real-world complexity~\citep{DBLP:journals/jmlr/DwivediJL0BB23}. Furthermore, real-world graphs are often noisy, imbalanced, and contain missing information, which can severely degrade GNN performance~\citep{DBLP:journals/corr/abs-2403-04468}.

\vspace{0.5em}
\noindent \textbf{Heterophily vs. homophily:} A core assumption of many GNNs is \textit{homophily}, that connected nodes are similar. However, many real-world graphs exhibit \textit{heterophily}, where connected nodes are dissimilar. Standard GNNs often fail in such scenarios. While specialized architectures for heterophilous graphs exist, a more unified understanding is needed. Future models should be able to adaptively learn whether to smooth or sharpen features based on the local graph structure, moving beyond a one-size-fits-all assumption about homophily~\citep{DBLP:conf/nips/ZhuYZHAK20}.

\vspace{0.5em}
\noindent \textbf{Generative models and topology modification:} The static nature of most graph datasets is a limitation. The topology of the user-item graph, for instance, could be modified to improve performance or efficiency.
A promising direction is the use of recent generative models, such as diffusion models, to leverage and modify the graph topology. These models could potentially reduce the noise and sparsity of the dataset by generating plausible, yet unobserved, edges, thereby enhancing the performance of downstream GNNs. Recent advances have demonstrated the effectiveness of diffusion models across a broad spectrum of graph-based recommendation tasks, including sequential recommendation \citep{DBLP:journals/tkde/WangZWZLYT24}, out-of-distribution generalization \citep{DBLP:conf/www/ZhaoYLL0ZG025, DBLP:conf/www/ZhaoYLZG025}, knowledge-aware recommendation \citep{DBLP:conf/kdd/ZhengZSTK26, DBLP:conf/wsdm/JiangYXH24}, graph contrastive learning \citep{DBLP:conf/www/HuangCZC26, DBLP:conf/wsdm/GeMCQ0BZ26, DBLP:journals/tkde/JingZWYT26}, and conditional or guided recommendation \citep{DBLP:journals/eswa/ZhuT26, DBLP:journals/ipm/WenYM26, DBLP:journals/tkde/ZhangCZZW25}. In a related vein, recent studies are also exploring how Large Language Models (LLMs) can be leveraged to augment or modify the user-item graph topology, further blurring the lines between structured and unstructured data modeling. Beyond the pioneering work of \citet{DBLP:conf/wsdm/WeiRTWSCWYH24}, LLMs have been successfully applied to a variety of graph-based recommendation tasks, including graph reasoning \citep{DBLP:conf/aaai/WangCOWHSGXZCLZ24, DBLP:conf/ecir/SakuraiTOH25, DBLP:conf/acl/WangCFS25}, graph retrieval-augmented generation for recommendation \citep{DBLP:conf/acl/0002F0LMWY25}, sequential recommendation \citep{DBLP:conf/acl/KimNBCL26}, and multimodal recommendation \citep{DBLP:conf/aaai/MengMJLW25}. Another emerging research direction exploits LLMs to generate augmented versions of the user-item graph for tasks such as contrastive learning and hard negative sampling, as demonstrated by several recent studies \citep{DBLP:conf/emnlp/ZhengWCZWSYLL25, DBLP:conf/aaai/SongCL26, DBLP:journals/corr/abs-2505-12396, DBLP:journals/corr/abs-2602-17410}. Despite the growing body of work leveraging diffusion models and LLMs to enhance the performance of GNN-based recommender systems, little attention has been paid to understanding the topological implications of these approaches, with only a few notable exceptions \citep{DBLP:conf/recsys/FiorettiLPMPNN25}. We argue that investigating these methods through a topological perspective is crucial for uncovering how they reshape the structure of the user-item graph and, ultimately, why such structural modifications translate into improved recommendation performance.

\section{Task-specific and application-oriented challenges}
As GNNs are applied to more complex tasks, new challenges emerge related to the specific demands of each domain. Recommender systems provide a clear example of this.

\vspace{0.5em}
\noindent \textbf{Exploiting topology in GNN-based recommendation:} While GNNs are a natural fit for recommendation, there is still a limited understanding of how they exploit the topological properties of the user-item graph. Current research often focuses on accuracy, neglecting the rich, interpretable insights that a topological perspective can offer.
There is a need for more research that explicitly investigates the link between topological characteristics (i.e., node degree, clustering coefficients, assortativity, degree distribution) and recommendation outcomes like novelty, diversity, and fairness. This involves designing GNNs that are not just powerful predictors but are also tailored to leverage specific, desirable topological properties of the data~\citep{DBLP:conf/aaai/HuangQ0025,DBLP:journals/kais/ZhangSQWWYC25,DBLP:journals/asc/MoPZ25}.

\vspace{0.5em}
\noindent \textbf{Integrating multimodal and sequential data:} Recommender systems and other applications often involve rich side information that is not easily captured in a simple graph structure. For example, user behavior is often sequential, and items may have multimodal features (e.g., text, images) \citep{DBLP:journals/tors/MalitestaCPMNS25}. A key challenge is to develop principled ways to integrate this information. This could involve treating multimodal attributes as node features or viewing sequential data as a way of re-structuring the user-item bipartite graph. The goal is to create a unified model that can reason over the graph topology and the rich semantics of its components simultaneously.

\section{Evaluation and methodological challenges}
The way GNN models are evaluated is crucial for driving meaningful progress. Current benchmarking and reproducibility practices, however, face significant criticism.

\vspace{0.5em}
\noindent \textbf{Benchmarking and reproducibility:} The GNN field has been criticized for its lack of standardized and rigorous benchmarking practices, a topic of intense debate within the community \citep{DBLP:journals/jmlr/DwivediJL0BB23,DBLP:conf/iclr/DongSHWL25}. Different studies often use different datasets, data splits, and experimental settings, making fair and reproducible comparisons nearly impossible. This can lead to a ``reproducibility crisis'' where reported gains are difficult to verify.
There is an urgent need for the community to rally around high-quality, standardized benchmarks that cover a diverse range of tasks and graph types. This includes not just node classification but also graph classification, link prediction, and applications in science and industry. Platforms that facilitate easy and fair comparison, with open-sourcing code/data, are essential for fostering robust and reproducible research~\citep{DBLP:journals/corr/abs-2502-14546}.

\vspace{0.5em}
\noindent \textbf{Beyond accuracy:} Evaluation often focuses narrowly on a single metric like accuracy, ignoring other crucial aspects of model performance such as fairness, robustness, interpretability, and efficiency.
Future work must adopt a more holistic approach to evaluation. This means extending the analysis of GNNs to measures that go beyond accuracy, such as diversity, novelty, bias, and fairness, as it has been done for other families of recommendation algorithms~\citep{DBLP:journals/ipm/DeldjooBN21}. A multi-faceted evaluation framework will lead to the development of GNNs that are not just accurate but also reliable, equitable, and trustworthy.

\backmatter 

\printbibliography

\end{document}